\documentclass[prx,aps,twocolumn,footinbib,superscriptaddress,notitlepage,10pt]{revtex4-2}
\usepackage[T1]{fontenc}
\usepackage{amsmath}
\usepackage{amssymb}
\usepackage{graphicx}
\usepackage{dcolumn}
\usepackage{enumerate}
\usepackage{bm}
\usepackage{xcolor}
\usepackage{comment}
\usepackage{subfigure}
\usepackage{breqn}
\usepackage{braket}
\usepackage{hyperref}
\hypersetup{colorlinks=true, citecolor=blue, urlcolor=blue, linkcolor=blue}

\makeatletter
\let\cat@comma@active\@empty
\makeatother

\begin{document}
\title{Reconfigurable dissipative entanglement between many spin ensembles: \\from robust quantum sensing to many-body state engineering}

\author{Anjun Chu}
\email{anjunchu@uchicago.edu}
\affiliation{Pritzker School of Molecular Engineering, University of Chicago, Chicago, Illinois 60637, USA}
\author{Mikhail Mamaev}
\affiliation{Pritzker School of Molecular Engineering, University of Chicago, Chicago, Illinois 60637, USA}
\author{Martin Koppenh\"{o}fer}
\affiliation{Fraunhofer Institute for Applied Solid State Physics IAF, Tullastr.~72, 79108 Freiburg, Germany}
\author{Ming Yuan}
\affiliation{Pritzker School of Molecular Engineering, University of Chicago, Chicago, Illinois 60637, USA}
\author{Aashish A. Clerk}
\email{aaclerk@uchicago.edu}
\affiliation{Pritzker School of Molecular Engineering, University of Chicago, Chicago, Illinois 60637, USA}
\date{\today}

\begin{abstract}
An attractive approach for stabilizing entangled many-body spin states is to employ engineered dissipation. 
Most existing proposals either target relatively simple collective spin states, or require numerous independent and complex dissipative processes. 
Here, we show a surprisingly versatile scheme for many-body reservoir engineering that relies solely on fully collective single-excitation decay, augmented with local Hamiltonian terms.  Crucially, all these ingredients are readily available in cavity QED setups.  
Our method is based on splitting the spin system into groups of sub-ensembles, and provides an easily tunable setup for stabilizing a broad family of pure, highly entangled states with closed-form analytic descriptions.  
Our results have immediate application to multi-ensemble quantum metrology, enabling Heisenberg-limited sensing of field gradients and curvatures.  Notably, our approach solves an important challenge in differential quantum sensing by providing the first example of Heisenberg-limited differential sensing immune to common-mode noise and accessible with only simple one-body measurements.
The same setup also allows the stabilization of an entire family of entangled states in a 1D chain of spin ensembles with symmetry-protected topological (SPT) order, and have a direct connection to the outputs of sequential unitary circuits.  
A special case of our protocol efficiently stabilizes the celebrated Affleck–Kennedy–Lieb–Tasaki (AKLT) state.
\end{abstract}

\maketitle

\section{Introduction}
Among the major achievements of modern quantum science is the ability to control systems of many qubits or spins with increasing levels of complexity. 
This progress is driven both by applications to quantum technologies and by the fundamental pursuit of understanding many-body physics.  
Cavity QED systems, where multiple atoms interact through their couplings to common cavity modes, have proven particularly fruitful.  
In these systems, photons commonly serve as mediators of effective Hamiltonian spin–spin interactions, giving rise to rich and nontrivial dynamical behavior.  
This standard approach has been widely used to generate highly entangled spin-squeezed states for quantum-enhanced metrology \cite{Pezze2018}, and has also served as a powerful tool for exploring a variety of many-body effects \cite{Mivehvar2021}.  

Cavity QED platforms also naturally provide an alternative route for generating nontrivial spin dynamics and entangled states by harnessing cavity dissipation.
A generic situation in experiments is the limit where cavity loss induces fully collective dissipation on the spins, a single dissipative process where each atom is essentially indistinguishable. 
This setting underlies the celebrated phenomenon of Dicke superradiance \cite{Dicke1954,Gross1982}, characterized by a collectively enhanced atomic decay and the generation of a photonic burst.
Beyond superradiance, the interplay between collective decay and coherent Rabi driving can give rise to dissipative phase transitions \cite{puri_exact_1979,Carmichael1980} and even time-crystal–like phenomena \cite{Iemini2018}. 
Moreover, the combination of collective decay and tailored external driving enables the generation of metrologically useful spin-squeezed states \cite{Agarwal1990,DallaTorre2013,Bai2021,Groszkowski2022,Guti2023}, as well as extensions to two-ensemble configurations \cite{Parkins2006,Krauter2011,Mamaev2025,Kaubruegger2025}.

\begin{figure}[t]
    \centering
    \includegraphics[width=1.0\columnwidth]{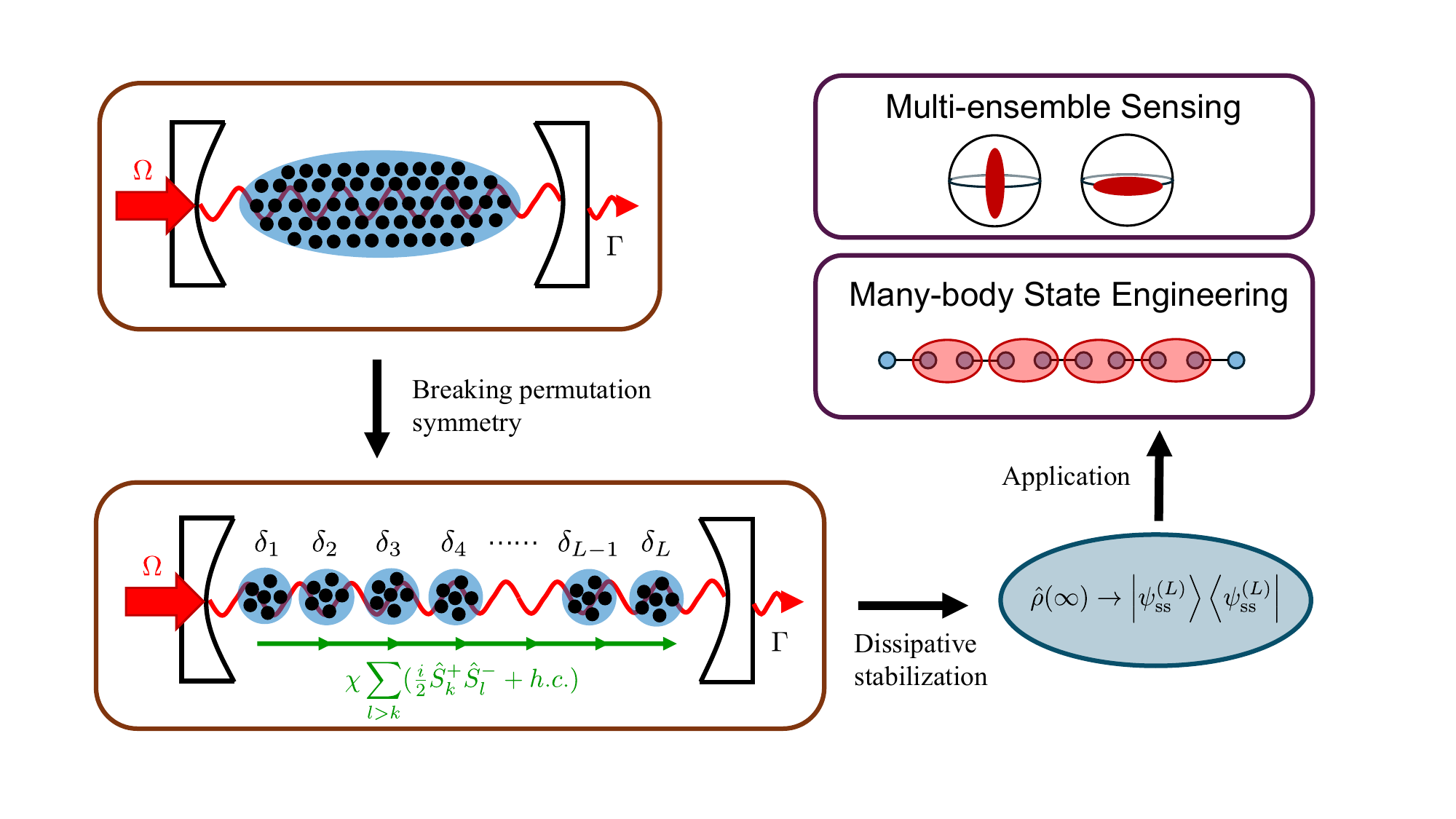}
    \caption{Schematic of a cavity QED platform for reconfigurable many-body reservoir engineering.  We start with a fully collective setup:  $N_{\rm tot}$ spins are subject to Rabi drives $\Omega$, and a collective, cavity-mediated superradiant decay $\Gamma$.  We break the full permutation symmetry by assigning different detunings $\delta_l$ to sub-ensembles of spins, and/or adding chiral spin-exchange interactions $\chi$ between sub-ensembles.  This setup stabilizes a variety of non-trivial pure, entangled states, and is easily reconfigured by changing the detuning pattern $\delta_l$ and/or other system parameters ($\chi$ and $\Omega$).}
    \label{fig:schematic}
\end{figure}

While permutation-symmetric spin models with collective cavity-induced decay are appealing for their experimental relevance, a broader goal is to realize more general forms of reservoir engineering \cite{Poyatos1996}, where controlled dissipation can stabilize a wide variety of nontrivial many-body spin states beyond the permutation-symmetric case. 
Many-body dissipative state-preparation schemes are of interest both for their practical utility, such as the potential to design protocols that are more robust than unitary approaches, and for their fundamental implications including the characterization of dissipative phases of matter \cite{Rakovszky2024,Sang2024}.
There are by now a host of theoretically proposed protocols for reservoir-engineering (see, e.g., Ref.~\onlinecite{Harrington2022} for a recent review). 
However, these protocols typically involve a considerable amount of complexity for near-term experiments: Instead of a single collective cavity-induced loss process, they often require engineering a large, even extensive, number of independent and highly tailored dissipative processes.

Given this context, a natural question is whether one can bridge these two extremes: is it possible to {\it only} rely on the ubiquitous ingredient of collective loss in cavity QED systems, but still dissipatively stabilize complex, non-collective many-body states?
One might even be more ambitious and ask whether such an approach could be made {\it reconfigurable}, allowing the same setup to stabilize a wide variety of target states. 
In this work, surprisingly we show that this is indeed possible.
We introduce and analyze a general method for dissipative stabilization of a broad class of non-collective many-body entangled states using only a single collective loss dissipator. 
Our approach breaks the permutation symmetry of $N_{\rm tot}$ atoms only through Hamiltonian terms, namely a pattern of drive detunings and a structured cavity-mediated spin-exchange interaction. 
The Hamiltonian terms reduce the symmetry of the system and lead to $L$ distinct spin ensembles, each with the same collective spin $S = N_{\rm tot}/2L$.
By varying the detuning pattern, we show that one can dissipatively stabilize a wide range of pure entangled states of these $L$ ensembles.
We can further derive closed-form analytic descriptions of the corresponding steady states. 
Note that in the extreme limit $L = N_{\rm tot}$, where the steady states attain a greatly simplified structure, our construction coincides with Refs.~\onlinecite{Stannigel2012,Ramos2014,Pichler2015}.
Our work has immediate relevance to recent cavity QED experiments involving two or more spin ensembles \cite{Periwal2021,Malia2022,Cooper2024,Young2024}.

While the space of states accessible with our approach is vast, we focus on two particularly intriguing classes. 
In Secs.~\ref{sec:two} and \ref{sec:many}, we demonstrate how multi-ensemble entangled states with exceptional metrological properties for field gradients and curvatures can be naturally stabilized.
Our approach solves an important challenge in differential quantum sensing \cite{Corgier2025,Kaubruegger2025} by providing the first example of Heisenberg-limited differential sensing immune to common-mode noise and accessible with only simple one-body measurements - the standard measurement primitive in essentially all atomic sensors.
In Sec.~\ref{sec:1dchain}, we discuss how our approach can stabilize states of fundamental interest in many-body physics: an entire family of entangled states in a 1D chain of spin-$S$ ensembles exhibiting symmetry-protected topological (SPT) order \cite{Chen2011,Pollmann2012} and connecting directly to recent ideas for exploiting sequential unitary circuits \cite{Chen2024}.
A special case of our protocol efficiently stabilizes the celebrated Affleck–Kennedy–Lieb–Tasaki (AKLT) state \cite{Affleck1987,Affleck1988}.
Together, these examples highlight both the power and versatility of our general approach.

\section{Setup for reconfigurable reservoir engineering}
\label{sec:GeneralSetup}

Our approach is sketched in Fig.~\ref{fig:schematic}.  
We start with a simple, fully-collective system where $N_{\rm tot}$ two-level atoms experience a single collective decay process (induced by resonant interaction with a lossy cavity), and are also subject to uniform and resonant Rabi drives.  This realizes the well-known cooperative Rabi fluorescence (CRF) model \cite{puri_exact_1979,Carmichael1980}.  
Working in the rotating frame of Rabi drives and adiabatically eliminating the cavity, the spin dynamics follows a Gorini–Kossakowski–Sudarshan–Lindblad (Lindblad) master equation:
\begin{equation}
    \frac{d}{dt}\hat{\rho} = -i[\Omega \hat{S}^x,\hat{\rho}] + \Gamma \mathcal{D}[\hat{S}^-]\hat{\rho},
    \label{eq:crf}
\end{equation}
where $\mathcal{D}[\hat{K}](\cdot) = \hat{K}(\cdot)\hat{K}^{\dag}-\{\hat{K}^{\dag}\hat{K},(\cdot)\}/2$ is the Lindblad superoperator describing dynamics induced by a jump operator $\hat{K}$,
$\Omega$ is the Rabi frequency, $\Gamma$ is the rate of superradiant emission, and $\hat{S}^x$, $\hat{S}^-$ are collective spin operators for the $N_{\rm tot}$ atoms.  
Despite its seeming simplicity, the CRF model exhibits a variety of non-trivial non-equilibrium physics, including dissipative phase transitions \cite{puri_exact_1979,Carmichael1980}, dissipative spin squeezing \cite{Lee2014,Barberena2019}, boundary time crystals \cite{Iemini2018}, and hidden time-reversal symmetry \cite{Roberts2023}. 
The dissipative steady state of this model can be found exactly \cite{Carmichael1980}, even if one adds single-spin loss and disorder \cite{Roberts2023}.
Dissipative phase transitions related to this model have recently been observed experimentally \cite{Ferioli2023,Song2025}. 

The CRF model is fully collective (i.e., total angular momentum is conserved), severely limiting the complexity of the resulting dynamics and steady states.  
As promised, we will achieve a richer dynamics by perturbing the system Hamiltonian, while keeping the dissipative part of the dynamics unchanged
(see Fig.~\ref{fig:schematic}).  
We first partition our full set of $N_{\rm tot}$ atoms equally into $L$ sub-ensembles (indexed by $l=1, \dots, L$, with $L$ an even number).
Then we break the permutation symmetry by assigning detuning $\delta_l$ for atoms in sub-ensemble $l$, and introducing an all-to-all spin-exchange interaction (amplitude $\chi$) with a ``chiral'' structure \cite{Stannigel2012,Ramos2014,Pichler2015}.  
The modified master equation describes a system of $L$ ensembles with the same collective spin $S = N_{\rm tot}/ 2L$, and has the form:
\begin{equation}
    \begin{gathered}
    \frac{d}{dt}\hat{\rho} = -i[\hat{H},\hat{\rho}] + \Gamma \mathcal{D}\bigg[\sum_{l=1}^L\hat{S}_{l}^{-}\bigg]\hat{\rho},\\
    \hat{H}=\sum_{l=1}^L \Omega\hat{S}^x_l + \sum_{l=1}^L \delta_{l}\hat{S}^z_l + i\frac{\chi}{2}\sum_{l>k}(\hat{S}^{+}_k\hat{S}^{-}_l-\hat{S}^{-}_k\hat{S}^{+}_l).
    \end{gathered}
    \label{eq:model}
\end{equation}
Here, all the spin operators are now spin-$S$ operators acting on a particular sub-ensemble.  
As we will see, even without the spin-exchange interactions $\chi$ we can generate interesting states, but including them for the case of $L>2$ gives access to an even richer set.  
The imaginary amplitude of these interactions will be crucial, since they provide an effective 1D ordering of the ensembles (as the sign of spin exchange amplitude between $k$ and $l$ depends on the sign of $(k-l)$).  

All the new terms introduced in $\hat{H}$ are compatible with the experimental tools of cavity QED.
One can implement Eq.~(\ref{eq:model}) by trapping many spin ensembles in a cavity as shown in Fig.~\ref{fig:schematic}.
The detunings $\delta_l$ can be realized by applying separate Rabi drives with different frequencies that selectively address individual sub-ensembles, or by using a single global drive while engineering effective detunings for each ensemble via local frequency shifts (e.g. AC Stark shifts, Zeeman shifts).
Similar to the setup in Ref.~\onlinecite{Periwal2021}, the chiral spin-exchange couplings $\chi$ can be engineered by applying a magnetic-field gradient along the cavity, which allows for engineering cavity-assisted Raman processes between ensembles separated by different distances independently via frequency selection.
Alternatively, one can also engineer $\chi$ using a chiral quantum network \cite{Stannigel2012,Ramos2014,Pichler2015,Lodahl2017,Suarez-Forero2025}, which functions by coupling each spin ensemble to an optical cavity, and then connecting all these cavities via chiral waveguides.
We discuss the details of the implementation of chiral spin-exchange couplings $\chi$ in App.~\ref{sec:experiment}. 

As we increase $L$, we systematically reduce the permutation symmetry of the model, yielding dynamics that is far richer than the simple CRF model.  
We will consider the full range of this explicit symmetry breaking, from the minimal case of $L=2$, to the maximal case $L = N_{\rm tot}$ (i.e., each sub-ensemble consists of a single atom).  
Note that the fully-permutation-symmetric limit $L=1$ (i.e., the bare CRF model) does not admit pure dissipative steady states except in the trivial case where there is no drive, $\Omega = 0$.  
As we show below, the situation is very different (and richer) when permutation symmetry is broken.  
For an even number $L$, we will show that Eq.~(\ref{eq:model}) has a {\it unique pure} steady state $|\psi_{\rm ss}^{(L)}\rangle$ for any $\Omega > 0$ as long as the following two conditions on the drive detunings are satisfied:
\begin{itemize}
    \item For each spin ensemble $k$, there exists another spin ensemble $l$ such that $\delta_k = -\delta_l$.
    \item If $\chi=0$, all the detunings $\delta_l$ are nonzero and different from each other.
\end{itemize}
At a heuristic level, these conditions suggest a kind of pairing structure in the steady state between ensembles with equal-in-magnitude, opposite-signed detunings.
We will see explicitly how this manifests itself.

Note that our approach is connected to (but distinct from) previous work exploiting non-reciprocal interactions for entanglement stabilization.  In the limit where the chiral interaction is tuned to exactly $\chi = \Gamma$, Eq.~(\ref{eq:model}) has the form of a cascaded master equation \cite{Gardiner1993,Carmichael1993,Stannigel2012}, and mimics a setup where the $L$ ensembles are all coupled to a common unidirectional waveguide (meaning that ensemble $l$ is only influenced by ensembles $k < l$).  In the case $L=2$, dissipative pure-state entanglement in this fully directional limit can be understood as an example of the coherent quantum absorber method \cite{Stannigel2012} and a consequence of hidden time-reversal symmetry \cite{Roberts2021}; applications to two spin ensembles was studied in Ref.~\onlinecite{Roberts2023,Cabot2024}.  
Our work shows that for $L=2$, entanglement generation survives beyond the perfectly directional limit, indicating a kind of adiabatic continuity of the steady state as one tunes $\chi$.

In the other extreme case $L = N_{\rm tot}$ (i.e., each ensemble is just a single qubit with $S=1/2$), our work reduces to the entanglement stabilization scheme of  Refs.~\onlinecite{Stannigel2012,Ramos2014,Pichler2015}.  
Note that, in this case, the complexity of the stabilized states is reduced, and there is no obvious metrological utility to the states that are generated.
Although the possibility of states with SPT order exists for the case of $S=1/2$, it was not discussed by the previous works.   
Our work non-trivially generalizes this scheme to the case where each subsystem has an {\it arbitrary} spin $S > 1/2$.

\section{Entangling two spin ensembles}
\label{sec:two}

\subsection{Exact solution for pure steady states}

\begin{figure}[t]
    \centering
    \includegraphics[width=1.0\columnwidth]{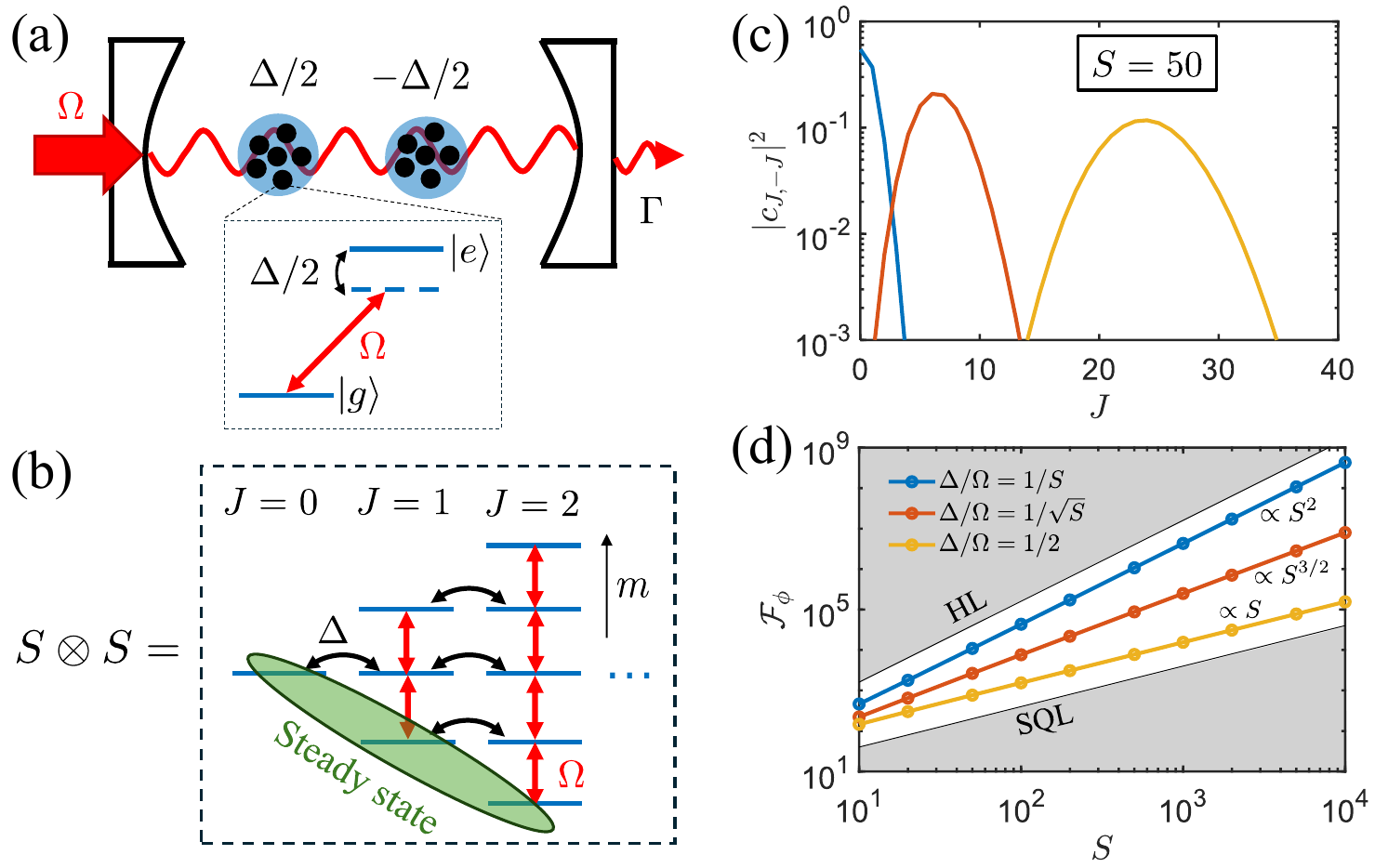}
    \caption{(a) Schematic of the $L=2$ setup: two spin-$S$ ensembles coupled to a cavity, each driven by Rabi fields with amplitudes $\Omega$ and opposite detunings $\pm \Delta/2$, and subject to a cavity-mediated collective decay $\Gamma$.
    (b) In the total-angular-momentum basis $|J,m\rangle$, the Rabi drive $\Omega$ couples $|J,m\rangle \rightarrow |J,m\pm 1\rangle$, and the detuning $\Delta$ couples $|J,m\rangle \rightarrow |J\pm 1,m\rangle$. The collective decay ensures the pure steady state is a linear combination of $|J,-J\rangle$ states, with coefficients $c_{J,-J}$ determined by destructive interference.
    (c) Steady-state wave function coefficients $c_{J,-J}$ with $S=50$. For $\Delta/\Omega=1/S$, the wave function is peaked at $J=0$; increasing $\Delta/\Omega$ shifts weight to larger $J$.
    (d) Steady-state quantum Fisher information (QFI) for measuring a differential phase $\phi$ (see Eq.~(\ref{eq:qfipure})). In the case of $\Delta/\Omega=1/S$, the QFI achieves Heisenberg scaling. 
    We compare the QFI for different $\Delta/\Omega$ scalings with the Heisenberg limit (HL) and the standard quantum limit (SQL).}
    \label{fig:twoensemble}
\end{figure}

We start by considering the case with just two spin ensembles (i.e., $L=2$), showing that the resulting dynamics can stabilize entangled states with remarkable metrological properties.  
As we will show later, the chiral interaction $\chi$ is unnecessary in this case, and the 
general conditions in Sec.~\ref{sec:GeneralSetup} for a pure steady-state only require that the two drive detunings are non-zero and the same up to a sign: $\delta_1 = - \delta_2 \equiv \Delta/2$.
Setting $\chi=0$, Eq.~(\ref{eq:model}) becomes:
\begin{equation}
    \begin{gathered}
    \frac{d}{dt}\hat{\rho} = -i[\hat{H},\hat{\rho}] + \Gamma\mathcal{D}[\hat{S}_1^{-}+\hat{S}_2^{-}]\hat{\rho},\\
    \hat{H} = \Omega (\hat{S}^x_1+\hat{S}^x_2) +\frac{\Delta}{2}(\hat{S}^z_1-\hat{S}^z_2).
    \end{gathered}
    \label{eq:twospin}
\end{equation}
As shown in Fig.~\ref{fig:twoensemble}(a), one can directly implement Eq.~(\ref{eq:twospin}) in a cavity QED setup similar to the realization of the CRF model \cite{Song2025}.

We now show that this dynamics admits a pure steady state.  
In general, such a state must be both an eigenstate of the Hamiltonian and annihilated by every Lindblad jump operator \cite{Kraus2008}.
In our case, the only jump operator is $\hat{S}_1^{-}+\hat{S}_2^{-}$, which describes the collective decay of the two ensembles.  
If such a steady state exists, it can be written in the basis of total angular momentum states $|J,m \rangle$, which are eigenstates of the total angular momentum operator $\hat{\mathbf{J}}\cdot \hat{\mathbf{J}}\ket{J,m} = J(J+1)\ket{J,m}$, with $\hat{\mathbf{J}}=(\hat{J}^x, \hat{J}^y, \hat{J}^z)$ and $\hat{J}^{\alpha}=\hat{S}^{\alpha}_1 + \hat{S}^{\alpha}_2$. Here $J\in \{0,\dots, 2S\}$ is the total angular momentum, and $m \in \{-J, \dots, J\}$ is the angular momentum projection along the quantization axis with $\hat{J}^{z}\ket{J,m} = m \ket{J,m}$. The jump operator $\hat{S}_1^{-} + \hat{S}_{2}^{-}$ annihilates the ground state $|J,-J \rangle$ of each total-angular-momentum sector; hence any pure steady state should take the form
\begin{equation}
    \left|\psi_{\rm ss}^{(2)}\right\rangle = \sum_{J=0}^{2S} c_{J,-J} |J,-J\rangle,
    \label{eq:assu}
\end{equation}
with wavefunction coefficients $c_{J,-J}$.

In the total angular momentum basis, one can interpret the Hamiltonian in Eq.~(\ref{eq:twospin}) as generating nearest-neighbor hopping in an effective 2D lattice formed by the $|J,m\rangle$ states (see Fig.~\ref{fig:twoensemble}(b)),  
\begin{equation}
    \begin{aligned}
    \hat{H} &=  \frac{\Omega}{2}\sum_{Jm}A_{J,m}\Big(|J,m+1\rangle\langle J,m|+h.c.\Big)\\
    &+\frac{\Delta}{2}\sum_{Jm} B_{J,m}\Big( |J+1,m\rangle\langle J,m|+h.c.\Big), 
    \end{aligned}
    \label{eq:hop}
\end{equation}
where $A_{J,m} = \langle J,m+1|\hat{S}^{x}_1+\hat{S}^{x}_2|J,m\rangle$, $B_{J,m} = \langle J+1,m|\hat{S}^{z}_1-\hat{S}^{z}_2|J,m\rangle$. 
As shown in App.~\ref{sec:analytic}, these coefficients can be calculated analytically, yielding:
\begin{equation}
    \begin{aligned}
    A_{J,m} = \sqrt{(J-m)(J+m+1)},
    \end{aligned}   
    \label{eq:ACoeff}
\end{equation}
\begin{equation}
    \begin{aligned}
    B_{J,m}=\sqrt{\frac{(2S-J)(2S+J+2)(J+m+1)(J-m+1)}{(2J+1)(2J+3)}}.
    \end{aligned}
    \label{eq:BCoeff}
\end{equation}

Applying Eq.~(\ref{eq:hop}) to Eq.~(\ref{eq:assu}), one sees that $\hat{H}|\psi_{\rm ss}^{(2)}\rangle$ is a linear combination of states $|J+1,-J\rangle$, and hence is necessarily orthogonal to $|\psi_{\rm ss}^{(2)}\rangle$.
Therefore, the pure steady state in our case must satisfy $\hat{H}|\psi_{\rm ss}^{(2)}\rangle=0$, which is equivalent to destructive interference of the hopping terms due to Rabi frequency $\Omega$ and detuning $\Delta$ in the $|J,m\rangle$ basis. 
One can thus obtain a recurrence relation for the coefficients of $|\psi_{\rm ss}^{(2)}\rangle$,
\begin{equation}
    \frac{c_{J+1,-J-1}}{c_{J,-J}} = -\frac{\Delta}{\Omega}\frac{B_{J,-J}}{A_{J+1,-J-1}}.
    \label{eq:twodetune}
\end{equation}

We find that $|\psi_{\rm ss}^{(2)}\rangle$ is the unique steady state for any $\Delta \neq 0$. 
Note that for $\Delta=0$ (CRF model with two ensembles), one can obtain $(2S+1)$ distinct steady states, as in this case there are no couplings between subspaces of different total angular momentum $J$. In App.~\ref{sec:unique} we show explicitly that the introduction of a perturbatively-small non-zero $\Delta$ completely breaks this degeneracy, and leads to a unique, pure steady state.  

Fig.~\ref{fig:twoensemble}(c) shows the amplitudes $c_{J,-J}$ of the pure steady state as a function of $J$. The probability $|c_{J,-J}|^2$ is localized in $J$ and the peak position can be tuned by varying the ratio $\Delta/\Omega$. 
When $\Delta/\Omega\gg 1/S$, the peak position of $|c_{J,-J}|^2$ is near $|c_{J+1,-J-1}/c_{J,-J}|\,\sim 1$, leading to $J\sim S\Delta/\Omega$.
In contrast, in the large drive regime 
$\Delta/\Omega\sim 1/S$, $|c_{J,-J}|^2$ is centered at $J=0$ (as in this regime we always have $|c_{J+1,-J-1}/c_{J,-J}|\,<1$).
The upshot is that by varying the detuning-to-drive ratio, we have a flexible tool for controlling the angular-momentum distribution of the system's unique, pure steady state.  
The large-Rabi-drive limit $\Omega\rightarrow\infty$ is of particular interest, as in this limit $|\psi_{\rm ss}^{(2)}\rangle$ approaches the $|J=0,m=0\rangle$ state.  
This is a maximally entangled state between the two spin ensembles, and has been discussed in several recent works as enabling robust, quantum-enhanced differential sensing \cite{Mamaev2025,Kaubruegger2025}.
As we show below, our state continues to be metrologically useful even away from the large-drive limit. 

Before turning to a discussion of metrology, we briefly comment on what happens if (still for $L=2$) one includes a non-zero chiral interaction amplitude $\chi$ in Eq.~(\ref{eq:model}).  We still obtain a pure steady state of the form in Eq.~(\ref{eq:assu}), but the recursion relation in Eq.~(\ref{eq:twodetune}) determining the coefficients is modified in a simple way: 
\begin{equation}
    \frac{c_{J+1,-J-1}}{c_{J,-J}} = -\frac{\Delta-i\chi (J+1)}{\Omega}\frac{B_{J,-J}}{A_{J+1,-J-1}}.
    \label{eq:twogeneral}
\end{equation}
This result is based on the relation $C_{J,m}=-(J+1)B_{J,m}$ proved in App.~\ref{sec:analytic}, where $C_{J,m}$ is defined as
$C_{J,m}=\langle J+1,m|\hat{S}_1^{+}\hat{S}_2^{-}-\hat{S}_1^{-}\hat{S}_2^{+}|J,m\rangle$.  
Thanks to this relation, one can simply replace $\Delta$ by $\Delta-i\chi(J+1)$ in Eq.~(\ref{eq:twodetune}).
For the rest of our discussion of the two-ensemble ($L=2$) case, we will return to setting $\chi=0$.  
However, in later sections, when we explore the $L>2$ multi-ensemble versions of our setup, a non-zero $\chi$ will be important, and Eq.~(\ref{eq:twogeneral}) will be an important ingredient in building up our full solution.

\subsection{Quantum enhancement for differential sensing}
\label{subsec:DifferentialSensing}

\begin{figure}[t]
    \centering
    \includegraphics[width=1.0\columnwidth]{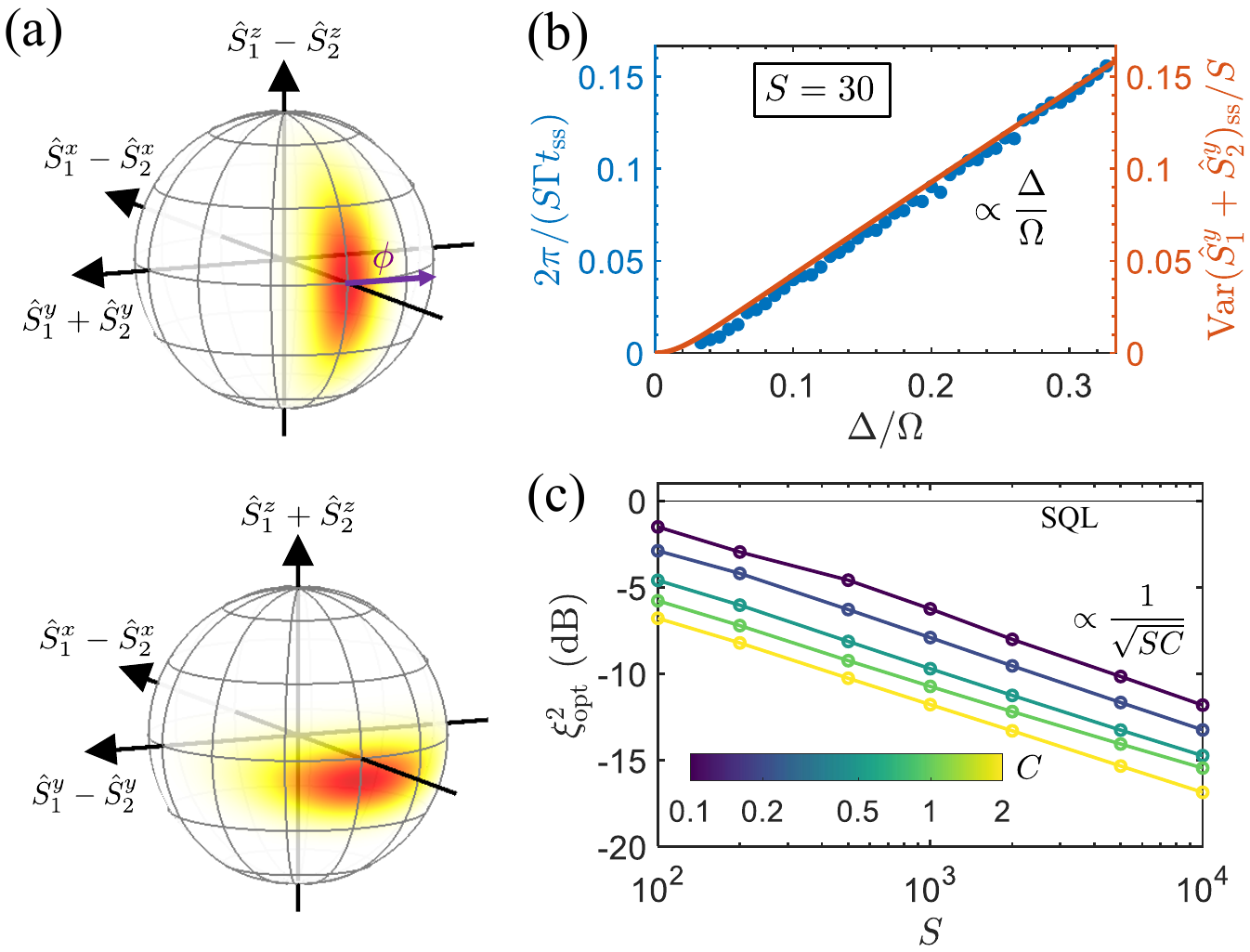}
    \caption{(a) Generalized Bloch sphere representation of the $L=2$ steady state. It can be interpreted as a two-mode spin squeezed state with squeezing along the $\hat{S}^y_1+\hat{S}^y_2$ and $\hat{S}^z_1+\hat{S}^z_2$ axes, and anti-squeezing along the $\hat{S}^z_1-\hat{S}^z_2$ and $\hat{S}^y_1-\hat{S}^y_2$ axes. Applying a differential phase $\phi$ between the two ensembles is equivalent to rotations about the $\hat{S}^z_1-\hat{S}^z_2$ axis in the first Bloch sphere. (b) Comparison between the optimal time scale to reach the steady state ($t_{\rm ss}$) and the steady-state squeezing (described by $\mathrm{Var}(\hat{S}^y_1+\hat{S}^y_2)_{\rm ss}$), with $S=30$. Note that $\mathrm{Var}(\hat{S}^y_1+\hat{S}^y_2)_{\rm SQL}=S$. We consider the initial state where all the spins are pointing down, and define $t_{\rm ss}$ as the optimal evolution time for the infidelity to the steady state to reach $10^{-3}$ with fixed $\Gamma$ and $\Delta/\Omega$. The same initial state is used in (c). (c) The scaling of optimal two-mode Wineland spin-squeezing parameter $\xi^2_{\rm opt}$ with spin $S$ for each ensemble and single-atom cooperativity $C$ (obtained from a second-order cumulant expansion). The scaling $\xi^2_{\rm opt}\propto 1/\sqrt{SC}$ is determined via numerical fitting.}
    \label{fig:twomodesqueezing}
\end{figure}

Having established the form of the pure steady state of our two-ensemble setup, we now explore its remarkable metrological utility for differential sensing.  
We consider the measurement of a differential phase $\phi$ between two spin ensembles, encoded by the unitary evolution $\hat{U}_{\phi} = e^{-i\phi (\hat{S}^z_1-\hat{S}^z_2)}$. 
The quantum Fisher information (QFI) for measuring this phase is given by the variance of $\hat{S}^z_1-\hat{S}^z_2$ in $|\psi_{\rm ss}^{(2)}\rangle$
(c.f.~Eqs.~\eqref{eq:assu}, \eqref{eq:BCoeff}, and \eqref{eq:twodetune})
\begin{equation}
    \mathcal{F}_{\phi} = 4\mathrm{Var}(\hat{S}^z_1-\hat{S}^z_2)_{\rm ss} = 4\sum_{J=0}^{2S} |c_{J,-J}|^2|B_{J,-J}|^2.
    \label{eq:qfipure}
\end{equation}

A central question is how this QFI scales with the number of atoms, i.e., with $S$.  
Fig.~\ref{fig:twoensemble}(d) shows this scaling for different choices of the parameter $\Delta/\Omega$.
As discussed in the previous section, away from the large drive regime, i.e., when $\Delta / \Omega \gg 1/S$, the steady state's angular-momentum distribution is localized near $J\sim S\Delta/\Omega$.  
It then follows from Eq.~(\ref{eq:qfipure}) and the form of the $B_{J,-J}$ coefficients that $\mathcal{F}_{\phi}\propto S\Omega/\Delta$.
This immediately suggests how to achieve different canonical scalings of the QFI.  
For $\Delta/\Omega\sim 1$, we have a scaling near the standard quantum limit (SQL), $\mathcal{F}_{\phi}\propto S$.  For a Rabi drive that increases with system size, $\Delta/\Omega\sim 1/\sqrt{S}$, we obtain a scaling that exceeds the SQL, $\mathcal{F}_{\phi}\propto S^{3/2}$.  

For even larger Rabi drives, $\Delta/\Omega\sim 1/S$, we obtain a QFI that exhibits Heisenberg-limited (HL) scaling, i.e., $\mathcal{F}_{\phi} \propto S^2$.  
This is consistent with our earlier observation that our state approaches the $|J=0,m=0\rangle$ state in the large-Rabi-drive limit ($\Omega\rightarrow\infty$), a state which has a maximal QFI $[\mathcal{F}_{\phi}]_{\rm max} = 16S(S+1)/3$.
The same maximal QFI is also reported in Refs.~\onlinecite{Mamaev2025,Kaubruegger2025}.
Since the $|J=0,m=0\rangle$ state is invariant under global rotations, in the large-Rabi-drive limit, one can obtain the same HL-scaling QFI for parameter estimation corresponding to generators $\hat{S}^x_1-\hat{S}^x_2$, $\hat{S}^y_1-\hat{S}^y_2$, and $\hat{S}^z_1-\hat{S}^z_2$, and vanishing variances for the operators $\hat{S}^x_1+\hat{S}^x_2$, $\hat{S}^y_1+\hat{S}^y_2$, and $\hat{S}^z_1+\hat{S}^z_2$. 
The squeezing and anti-squeezing of these operators can be extended to the case of a finite $\Omega$ as we show later, although the rotation symmetry of the three directions no longer holds.

While the $|J=0,m=0\rangle$ has an optimally-large QFI, one must also ask whether this metrological enhancement can be achieved with experimentally feasible measurements. 
The $|J=0,m=0\rangle$ is problematic in this regard, as the expectation values of all one-body operators vanish, and hence achieving the QFI necessarily requires more complicated, multi-body measurements.  
Surprisingly, our more general class of states provides a way of evading this issue while still maintaining HL scaling.  
The trick is to use a large enough value of $\Omega$ to yield HL scaling, but not so large as to cause the average spin length in the two ensembles to be zero (as it is the case for the $|J=0,m=0\rangle$ state).  
Having a non-zero spin length means that a simple Ramsey-style measurement, i.e., a linear measurement of collective spin variables, can potentially achieve the sensitivity predicted by the QFI.  

To make this idea more quantitative, we first derive a relation between the differential $x$ polarization of our spin ensembles and the QFI.  
One can show through an explicit calculation (see App.~\ref{sec:analytic}) that
\begin{equation}
    \langle\hat{S}^x_1-\hat{S}^x_2\rangle_{\rm ss}=\sum_J c^{*}_{J+1,-J-1}c_{J,-J} B^{-}_{J,-J} = -\frac{\Delta}{4\Omega}\mathcal{F}_{\phi},
    \label{eq:spinlength}
\end{equation}
where $B_{J,m}^{-} = \langle J+1,m-1|\hat{S}^{-}_1-\hat{S}^{-}_2|J,m\rangle$. 
Since $\mathcal{F}_{\phi}\propto S\Omega/\Delta$, as discussed previously, we have $\langle\hat{S}^x_1-\hat{S}^x_2\rangle_{\rm ss}\propto S$ for a wide range of $\Delta/\Omega$, with corrections when decreasing $\Delta/\Omega$ below $\sim 1/S$. 
This result suggests the possibility of having the QFI exhibit Heisenberg-scaling while still having a non-zero average spin length of order $S$, e.g., by having  $\Delta/\Omega \sim 1/S$.  
We indeed find that this is true through an explicit evaluation of Eq.~(\ref{eq:spinlength}).  
Note that having a spin length of order $S$ not only ensures that a Ramsey-type protocol is applicable, but also enhances the robustness against detection noise via large measurable signals.

Based on the large spin length $\langle\hat{S}^x_1-\hat{S}^x_2\rangle_{\rm ss}\propto S$ in our case, as shown in Fig.~\ref{fig:twomodesqueezing}(a), one can further interpret our steady state $|\psi_{\rm ss}^{(2)}\rangle$ as a two-mode spin squeezed state (see Refs.~\onlinecite{Kitzinger2020,Sundar2023,Mamaev2025} for other types of two-mode spin squeezed states).
The squeezed quadratures are along the $\hat{S}^y_1+\hat{S}^y_2$ and $\hat{S}^z_1+\hat{S}^z_2$ axes, and the anti-squeezed quadratures are along the $\hat{S}^z_1-\hat{S}^z_2$ and $\hat{S}^y_1-\hat{S}^y_2$ axes.
Both of the Bloch spheres share the same spin length, which is along the $\hat{S}^x_1-\hat{S}^x_2$ axis.
In particular, the differential phase $\phi$ leads to rotations along the $\hat{S}^z_1-\hat{S}^z_2$ axis, resulting in a non-zero average value of $\hat{S}^y_1+\hat{S}^y_2$.
This suggests a simple Ramsey-style measurement to measure the average collective spin component $\hat{S}^y_1+\hat{S}^y_2$. 
The quantum enhancement of this Ramsey-type measurement can be characterized by the two-mode generalization of the Wineland spin squeezing parameter \cite{Mamaev2025}.
Letting $(\Delta \phi)_{\rm est}$ denote the error in estimating $\phi$ from the Ramsey measurement, we have
\begin{equation}
    \xi_{\rm ss}^2 = 4S
    \left(\Delta \phi \right)^2_{\rm est} 
    =\frac{4S\mathrm{Var}(\hat{S}^y_1+\hat{S}^y_2)_{\rm ss}}{\langle\hat{S}^x_1-\hat{S}^x_2\rangle_{\rm ss}^2}.
    \label{eq:squeeze}
\end{equation} 
Note that $\mathrm{Var}(\hat{S}^y_1+\hat{S}^y_2)_{\rm ss} = \sum_{J=0}^{2S} |c_{J,-J}|^2 J/2$. 
Based on the previous discussions of $|c_{J,-J}|^2$, one can obtain $\xi_{\rm ss}^2\propto \Delta/\Omega$ when $\Delta/\Omega\gg 1/S$. 
Thus, by setting $\Delta/\Omega\sim 1/S$, one finds that Eq.~(\ref{eq:squeeze}) is approaching HL scaling, $\xi_{\rm ss}^2\propto 1/S$.
We thus have established a major advantage of our scheme:  HL scaling can be achieved for the differential phase $\phi$ using a simple Ramsey-type measurement with a signal of size $O(S)$.
Based on the Bloch sphere representation, one can also extract the phase imprinted by $\hat{S}^y_1-\hat{S}^y_2$ with HL scaling by measuring $\hat{S}^z_1+\hat{S}^z_2$.

We stress that the quantum enhancement for differential sensing does not require exactly the same atom number for the two spin ensembles (see App.~\ref{sec:imperfection} for detailed analysis).
For constant population imbalance, the HL scaling $\xi_{\rm ss}^2 \propto 1/S$ is maintained.
Even with $O(\sqrt{S})$ population imbalance, one can still obtain steady state spin squeezing $\xi_{\rm ss}^2 \propto 1/\sqrt{S}$.

Our approach and the resulting steady state offer several distinct advantages over previous dissipative preparation schemes for enhanced differential sensing based on entanglement between two spin ensembles.
Ref.~\onlinecite{Kaubruegger2025} stabilizes a state with vanishing spin length, precluding its use in Ramsey-type measurements. 
Ref.~\onlinecite{Mamaev2025} stabilizes a similar (but distinct) two-mode squeezed state, while it relies on engineering two independent dissipators, making the setup more demanding experimentally. 
In contrast, our scheme is naturally compatible with a simple Ramsey-type protocol and achieves HL scaling with a simpler experimental design.

\subsection{Impact of single-spin dissipation during state preparation}
\label{subsec:singlespindiss}

While our ideal scheme produces entangled states with exceptional metrological properties, experimental utility also requires that we have robustness to inevitable imperfections.  
A key issue here is the presence of single spin relaxation.  Each of our two spin ensembles
is formed by $2S$ spin-$1/2$ particles, $\hat{S}_j^{-}=\sum_{l=1}^{2S}\hat{s}_{j,l}^{-}$, where $\hat{s}_{j,l}^{-}$ are spin-$1/2$ operators. Single-particle spontaneous emission is then described by adding additional dissipative processes with jump operators $\sqrt{\gamma}\hat{s}^{-}_{j,l}$ to Eq.~(\ref{eq:model}).
We define the single-atom cooperativity $C=\Gamma/\gamma$, which characterizes the ratio between emission into the cavity and unwanted emission into free space.

In the presence of such unwanted dissipation, the relaxation timescale $t_{\rm ss}$ for the ideal dynamics to reach the steady state $|\psi_{\rm ss}^{(2)}\rangle$ must not be too large, otherwise the steady-state spin squeezing will be strongly degraded once we introduce $\gamma$.  
We define $t_{\rm ss}$ such that the ideal ($\gamma = 0$) evolution yields a high fidelity with the steady state for $t> t_{\rm ss}$.  
More specifically ~$F(t)=\mathrm{tr}\Big(|\psi_{\rm ss}^{(2)}\rangle\langle\psi_{\rm ss}^{(2)}|\,\hat{\rho}(t)\Big)$ satisfies $F(t)>1-\epsilon$ for $t>t_{\rm ss}$, where $\epsilon$ is a small positive number (we choose $\epsilon=10^{-3}$).
Here, $\hat{\rho}(0)$ can be any specific state easy to initialize in an experiment.
One finds that there is a fundamental tradeoff between the speed of the dynamics and the amount of steady-state squeezing and entanglement.  
Similar to Refs.~\onlinecite{Pocklington2024,Pocklington2025}, one can quantify this with a lower bound on the relaxation time (see App.~\ref{sec:relax}),
\begin{equation}
    \Gamma t_{\rm ss} \geq \frac{1-\epsilon-F(0)}{4\mathrm{Var}(\hat{S}^y_1+\hat{S}^y_2)_{\rm ss}}.
    \label{eq:bound}
\end{equation}
In Fig.~\ref{fig:twomodesqueezing}(b), we compute $t_{\rm ss}$ numerically using the quantum-jump method \cite{Plenio1998} and compare with $\mathrm{Var}(\hat{S}^y_1+\hat{S}^y_2)_{\rm ss}$.
We use an initial state $|J=2S,m=-2S\rangle$, i.e., all spins pointing down, and then minimize $t_{\rm ss}$ by varying the ratio $\Omega/\Gamma$ while keeping $\Gamma$ and $\Delta/\Omega$ fixed.
The optimal $t_{\rm ss}$ can be achieved when $\Omega\gtrsim S\Gamma$.
We stress that the required strong Rabi drive ($\Omega\gtrsim S\Gamma$) has been realized in cavity QED experiments with a single spin ensemble (up to $10^4$ atoms) \cite{Song2025}.
Based on this choice of $\Omega$, we find $\Gamma t_{\rm ss}/(2\pi) \approx 1/\mathrm{Var}(\hat{S}^y_1+\hat{S}^y_2)_{\rm ss}$, which is a constant factor away from the lower bound (see Eq.~(\ref{eq:bound})).
As we discussed in Sec.~\ref{subsec:DifferentialSensing}, $\mathrm{Var}(\hat{S}^y_1+\hat{S}^y_2)_{\rm ss}\sim S\Delta/\Omega$, we thus have $t_{\rm ss}\sim \Omega/(S\Gamma\Delta)$.

Having understood the basic speed-squeezing tradeoff in our dynamics, we now explicitly introduce single-spin decay.
At a heuristic level, this non-collective decay will degrade the spin-squeezing because of the effective shortening of the collective spin polarization.  
To the lowest order, this increases the Wineland parameter by an amount $\sim \gamma t$.
Considering evolution over a time scale $t\sim t_{\rm ss}$, one can thus approximate the spin-squeezing parameter by
\begin{equation}
    \xi^2 \sim \xi_{\rm ss}^2 + \gamma t_{\rm ss} \sim \frac{\Delta}{\Omega} + \frac{\Omega}{S C\Delta}.
\end{equation}
where we have made use of the tradeoff between speed and spin squeezing discussed above.
Optimizing over the ratio $\Delta/\Omega$, one finds an optimal spin squeezing parameter $\xi^2_{\rm opt}\propto 1/\sqrt{SC}$, occurring for an optimal parameter ratio $\Delta/\Omega\sim 1/\sqrt{SC}$.
In Fig.~\ref{fig:twomodesqueezing}(d), we explore the scaling of $\xi^2_{\rm opt}$ with spin $S$ for each ensemble and single-atom cooperativity $C$, using a second-order cumulant expansion to numerically simulate the system dynamics (see App.~\ref{sec:cumulant} for details). 
We vary parameters $\Delta$ and $\Omega$ as well as the evolution time $t$ while fixing $\Gamma$.
The numerical results show good agreement with our analytically predicted scaling $\xi^2_{\rm opt}\propto 1/\sqrt{SC}$.
Note that single-mode spin squeezing due to the iconic one-axis twisting interaction \cite{Kitagawa1993} shares the same scaling when considering dissipative processes such as single-particle spin flips \cite{Schleier-Smith2010a,Chu2021,Barberena2024}.

\subsection{Robustness against common-mode noise during interrogation}
\label{subsec:TwoEnsembleCommonModeNoise}

\begin{figure}[t]
    \centering
    \includegraphics[width=1.0\columnwidth]{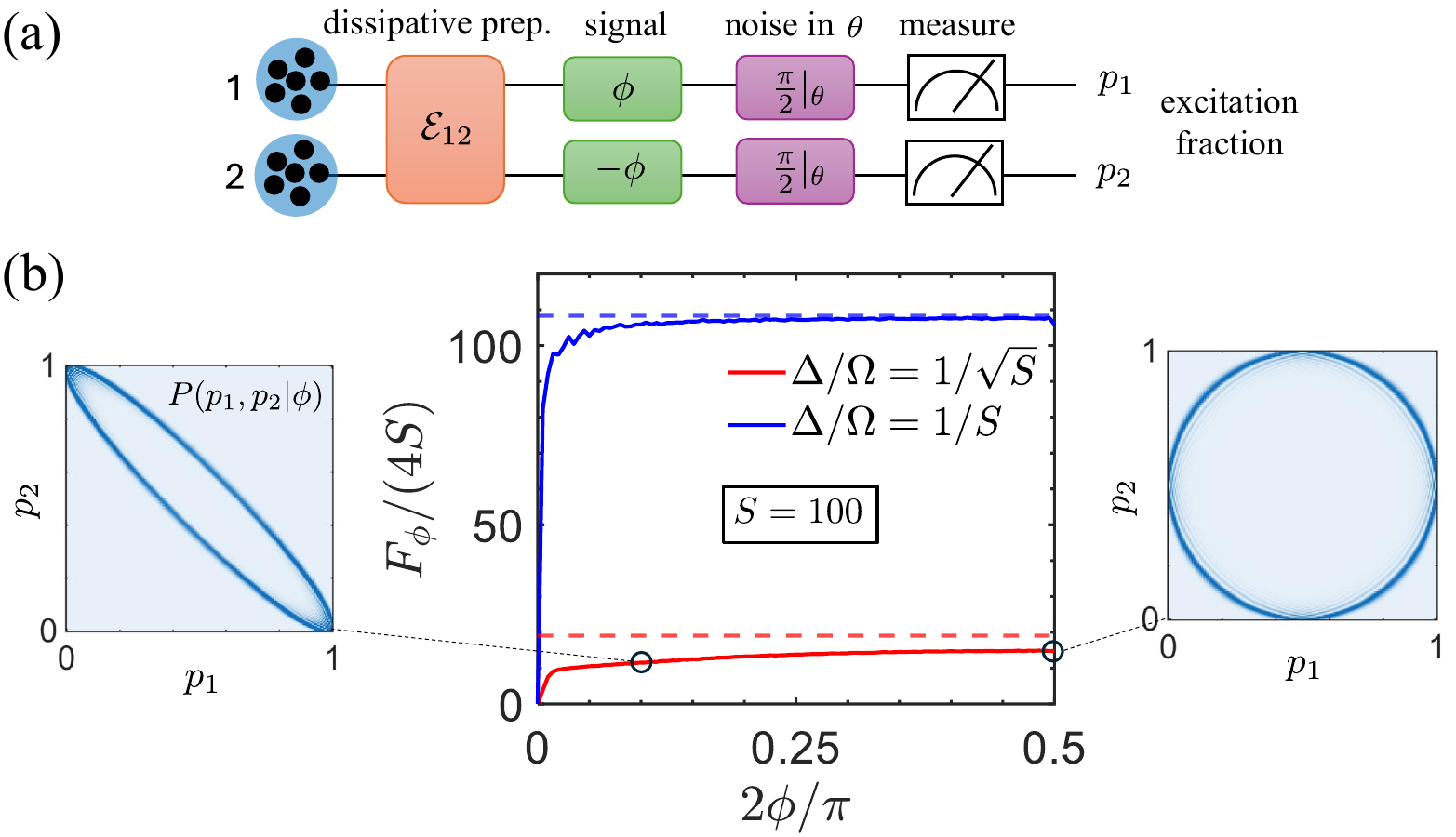}
    \caption{(a) Schematic for differential sensing in the presence of common phase noise. We first apply the quantum channel $\mathcal{E}_{12}$ (see Eq.~(\ref{eq:twospin})) 
    to stabilize the steady state $|\psi_{\rm ss}^{(2)}\rangle$.  The system then evolves under the differential phase $\phi$. We next apply a global $\pi/2$ pulse (random common phase $\theta$ due to laser noise) and then perform projective measurement of each ensemble in the $z$ basis to obtain the excitation fractions $p_1$ and $p_2$. (b) Classical Fisher information $F_{\phi}$ for ellipse fitting (solid lines). The dashed lines are the corresponding quantum Fisher information $\mathcal{F}_{\phi}$. $F_{\phi}$ is roughly approaching $\mathcal{F}_{\phi}$ except for the small region near $\phi=0$. The left and right insets are color-scale plots of the probability distribution for excitation fractions $(p_1,p_2)$. The points $(p_1,p_2)$ with peak probability lie on an ellipse whose form depends on $\phi$. }
    \label{fig:ellipsefitting}
\end{figure}

Common-mode noise can be a major practical limitation during the signal-acquisition (interrogation) part of a differential sensing protocol:  while one wants to estimate a small phase difference between the two ensembles, the {\it average} phase of the two ensembles could be completely random and uncontrolled.  
In AMO sensing platforms, common-mode noise can arise from laser-phase fluctuations in optical clocks \cite{Marti2018a,Zheng2024,Kim2025} and vibrations of reference platforms in atom interferometers \cite{Fixler2007,Rosi2015,Parker2018}.  
A standard method to mitigate this noise, without assuming it to be small, is the so-called ``ellipse fitting'' method \cite{Foster2002}, which is originally proposed for sensing with unentangled atoms.  
Recent work has sought to generalize this technique by entangling each spin ensemble separately \cite{Corgier2025}.  
An alternative is to use an entangled state between two spin ensembles that is intrinsically insensitive to common-mode noise. 
As discussed in Ref.~\onlinecite{Kaubruegger2025}, the $|J=0,m=0\rangle$ state (equivalent to our $\Omega \rightarrow \infty$ steady state) is such a state, as it is invariant under global rotations.  
Hence, in principle its HL scaling for differential sensing is robust to such noise.  
However, based on Sec.~\ref{subsec:DifferentialSensing}, the $|J=0,m=0\rangle$ state has a significant practical drawback: the lack of any net spin polarization means that metrology must employ complicated non-linear measurements.

Here, we demonstrate that our generalized class of states provide a means for combining the common-mode noise robustness of the $|J=0,m=0\rangle$ state while still allowing for a simple, linear measurement.  
The trick is to use our steady state $|\psi_{\rm ss}^{(2)}\rangle$ at finite values of $\Omega$, so that it retains a net spin length.
The loss of rotational symmetry can be compensated by ellipse fitting, such that we can still mitigate common-phase noise and maintain the same HL scaling while measuring one-body observables.
Our approach circumvents the complications of measuring two-body observables in Ref.~\onlinecite{Kaubruegger2025}.

We include the effects of the common-mode noise by applying a common phase rotation by angle $\theta$ to our steady state, encoded by the unitary evolution $\hat{U}_{\theta} = e^{-i\theta (\hat{S}^z_1+\hat{S}^z_2)}$.  We randomly sample $\theta$ with probability distribution $P(\theta)$.  As a result, the pure steady state is transformed to the impure state
\begin{equation}
    \hat{\rho}_{\rm c} = \int d\theta P(\theta) \hat{U}_{\theta}|\psi_{\rm ss}^{(2)}\rangle\langle \psi_{\rm ss}^{(2)}|\hat{U}^{\dag}_{\theta}.
\end{equation}
We can now calculate the sensitivity of this corrupted state to a differential phase $\phi$ by calculating its QFI $\mathcal{F}_{\phi}$:
\begin{equation}
    \begin{aligned}
    \mathcal{F}_{\phi} &= 2\sum_{\lambda_k+\lambda_l>0} \frac{(\lambda_k-\lambda_l)^2}{\lambda_k+\lambda_l}|\langle k|(\hat{S}^z_1-\hat{S}^z_2)|l\rangle|^2 \\
    &= 4 \sum_J |c_{J,-J}|^2 |B_{J,-J}|^2.
    \end{aligned}
    \label{eq:qfimix}
\end{equation}
Here $\lambda_k$ and $|k\rangle$ are the eigenvalues and eigenvectors of $\hat{\rho}_{\rm c}$. 
Comparing Eq.~(\ref{eq:qfipure}) with Eq.~(\ref{eq:qfimix}), we reach a surprising conclusion: the QFI for the differential phase $\phi$ remains unchanged for {\it any} probability distribution $P(\theta)$ describing the common-mode phase noise.
This result can be understood intuitively by considering the generalized Ramsey measurement scheme of the previous subsection, characterized by the two-mode spin-squeezing parameter in Eq.~(\ref{eq:squeeze}).  
The robustness against common phase noise is the simple consequence of the fact that $\mathrm{Var}(\hat{S}^y_1+\hat{S}^y_2)_{\rm ss} = \mathrm{Var}(\hat{S}^x_1+\hat{S}^x_2)_{\rm ss}$.  
Hence, while random common phase rotations mix these collective variances, this does not alter the variance that limits the generalized Ramsey measurement. Further, as discussed in Ref.~\onlinecite{Corgier2025}, this equal-variance condition corresponds to the optimal situation for extracting a differential phase $\phi$ using ellipse fitting.

We now go beyond the QFI calculation, and provide an explicit strategy for differential sensing in the presence of common-mode noise, see  Fig.~\ref{fig:ellipsefitting}(a). 
We consider a Ramsey-type measurement with the following steps: 
1) Turn on laser drives for dissipative stabilization of the steady state $|\psi_{\rm ss}^{(2)}\rangle$. For simplicity, we assume perfect state preparation. 
2) Turn off laser drives and let the system evolve under the signal source, acquiring the differential phase $\phi$ of interest during this  
``dark time''. 
3) Apply a global $\pi/2$ laser pulse. Assuming the standard situation where the dark time duration is much longer than the coherence time of the laser, this second laser pulse has an effectively random phase $\theta$ (common for the two ensembles).  The pulse can be described by $\hat{R}_{\theta}(\pi/2)=\hat{U}^{\dag}_{\theta}e^{-i\pi(\hat{S}^x_1+\hat{S}^x_2)/2}\hat{U}_{\theta}$. We assume that the phase $\theta$ is uniformly distributed between $0$ and $2\pi$, i.e., $P(\theta)=1/(2\pi)$.  Note that even in an experiment where $P(\theta)$ is possibly more structured, one could simply add phase noise to ensure $P(\theta)=1/(2\pi)$.
4) Perform a projective measurement on each ensemble in the $z$ basis, obtaining the excitation fractions $p_j$ ($j=1,2$) as measurement outcomes for operators $1/2 + \hat{S}^z_j/ (2S)$.
The expectation values of $p_1$ and $p_2$ in a single shot are thus given by
\begin{equation}
    \begin{aligned}
    E[p_1|\phi,\theta] &= \frac{1}{4S}\langle\hat{S}^x_1-\hat{S}^x_2\rangle_{\rm ss}\sin(\phi+\theta)+\frac{1}{2}, \\
    E[p_2|\phi,\theta] &= \frac{1}{4S}\langle\hat{S}^x_1-\hat{S}^x_2\rangle_{\rm ss}\sin(\phi-\theta)+\frac{1}{2}.
    \end{aligned}
    \label{eq:ellipse}
\end{equation}
Notice that Eq.~(\ref{eq:ellipse}) forms an ellipse in the ($p_1,p_2$) plane when varying $\theta$ from $0$ to $2\pi$, so one can extract the differential phase $\phi$ by fitting the ellipse formed by the measurement outcomes.  
Note that Eq.~(\ref{eq:ellipse}) is a circle for $\phi=\pi/4$.

One can use the classical Fisher information (CFI) to characterize the sensitivity of extracting a differential phase $\phi$ from the ellipse fitting method \cite{Corgier2025}. We first define the probability of the measurement outcome $p_1$ and $p_2$ conditioned on $\phi$ and $\theta$ as
\begin{equation}
    P(p_1,p_2|\phi,\theta)=\bigg|\Big(\langle S,m_1|\otimes\langle S,m_2|\Big)\hat{R}_{\theta}(\pi/2)\hat{U}_{\phi}| \psi_{\rm ss}^{(2)}\rangle\bigg|^2,
    \label{eq:condprob}
\end{equation}
where $p_j=m_j/(2S)+1/2$ with $j=1,2$.
We then integrate the common phase $\theta$ over the random distribution, $P(p_1,p_2|\phi) = \int d\theta P(\theta)P(p_1,p_2|\phi,\theta)$, which is shown in Fig.~\ref{fig:ellipsefitting}(b).
The CFI is thus given by
\begin{equation}
    F_{\phi} = \sum_{p_1,p_2}P(p_1,p_2|\phi)\bigg(\frac{\partial \ln P(p_1,p_2|\phi)}{\partial \phi}\bigg)^2.
    \label{eq:cfi}
\end{equation}
We then numerically calculate the CFI $F_{\phi}$ for $S=100$.
For both $\Delta/\Omega=1/\sqrt{S}$ and $\Delta/\Omega=1/S$, the CFI $F_{\phi}$ is roughly approaching the QFI $\mathcal{F}_{\phi}$ except for the small region near $\phi=0$.
The best sensitivity is achieved near $\phi=\pi/4$, where Eq.~(\ref{eq:ellipse}) forms a circle.
Note that one can of course still measure a small differential phase $\phi$ by deliberately adding a fixed differential phase bias of $\pi/4$.

This result shows another key feature of our scheme and the resulting steady-state solution: even with large common-phase noise, it is possible to reach Heisenberg scaling using simple Ramsey-style measurements and the ellipse-fitting method.
It is also interesting to note that because of the inter-ensemble entanglement in our scheme, we achieve a much better scaling than what is possible by applying the ellipse-fitting technique to a tensor product of two single-ensemble squeezed states.
As shown in Ref.~\onlinecite{Corgier2025}, this only yields a scaling of CFI $F_{\phi}\sim S^{4/3}$, in contrast to the HL-type scaling $F_{\phi}\sim S^2$ we achieve. 

\section{Entangling many spin ensembles}
\label{sec:many}

\subsection{Exact solution for pure steady states}

\begin{figure}[t]
    \centering
    \includegraphics[width=1.0\columnwidth]{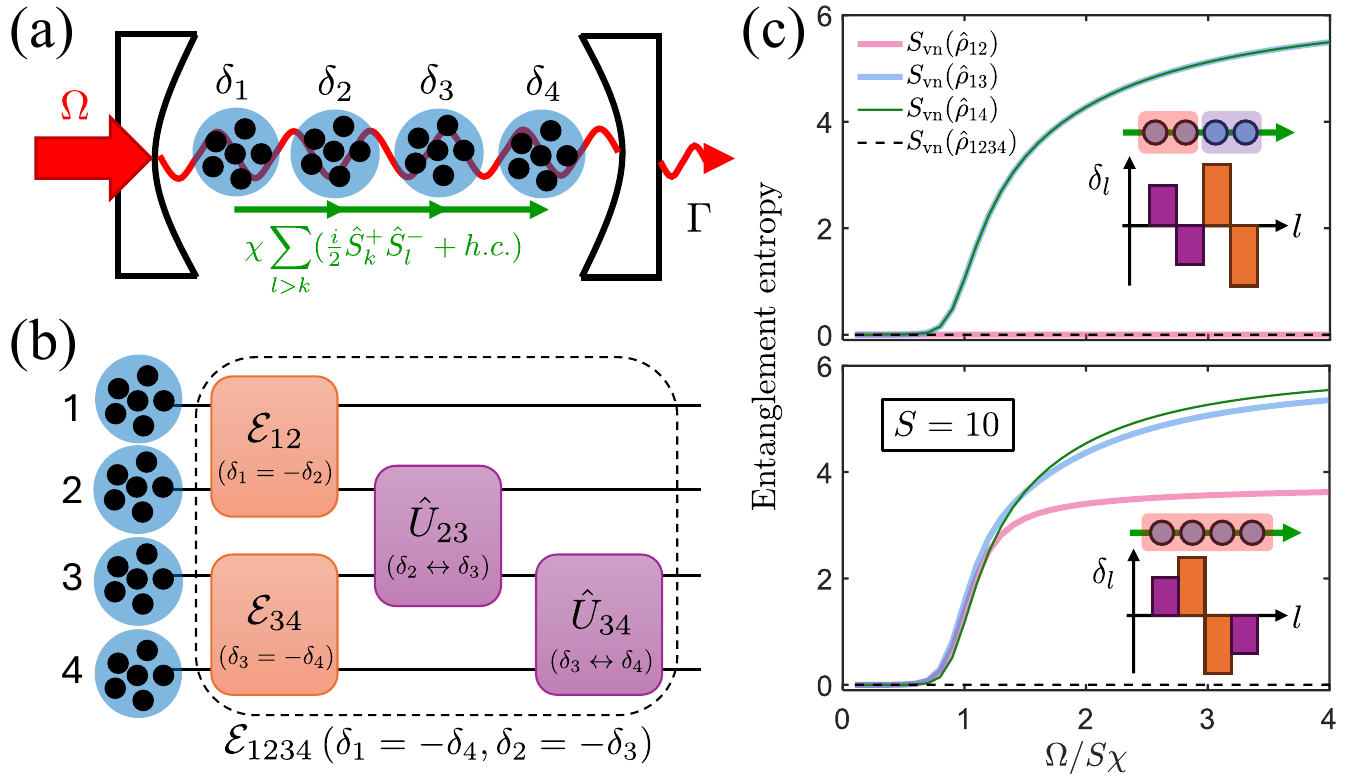}
    \caption{(a) Schematic of steady-state entanglement between four spin-$S$ ensembles in an optical cavity. Similar to Fig.~\ref{fig:twoensemble}(a), we consider Rabi drives with Rabi frequency $\Omega$ and detunings $\delta_l$ for each ensemble, as well as collective decay with rate $\Gamma$. We further engineer chiral spin-exchange couplings with rate $\chi$ between spin ensembles. (b) The quantum channel for relaxation to the steady state can be decomposed into quantum channels between ensemble pairs (initially setting $\delta_1=-\delta_2$ and $\delta_3=-\delta_4$) and unitary operators to swap the detunings to the final form (see text). (c) von Neumann entanglement entropy for the steady state in the case of $S=10$. The subscripts of $\hat{\rho}$ label the indices of spin ensembles included in the reduced density matrix. As shown in the insets, we fix $\delta_1=-\delta_2=2\chi$, $\delta_3=-\delta_4=3\chi$ in the top panel, and $\delta_1=-\delta_4=2\chi$, $\delta_2=-\delta_4=3\chi$ in the bottom panel. Multipartite entanglement between spin ensembles is achieved in the bottom panel.}
    \label{fig:fourensemble}
\end{figure}

We now build on our discussion of the two-spin-ensemble ($L=2$) case in the previous section, showing how a wide class of entangled states of $L \geq 2$ spin ensembles can be stabilized as pure steady states of the dissipative dynamics in Eq.~(\ref{eq:model}).
Fig.~\ref{fig:fourensemble}(a) shows a schematic for the case $L=4$.
Recall from Sec.~\ref{sec:GeneralSetup} that obtaining a pure steady state requires $L$ to be even and detunings to come in equal-in-magnitude, opposite-sign pairs.  We will also focus on the case where the chiral interaction $\chi$ is non-zero; as discussed, this interaction provides an effective 1D ordering of the ensembles.    

To show the existence of a pure steady state, consider first the special case where the detunings are ordered to form a dimerized pattern: each odd-numbered ensemble $l$ with detuning $\delta_l$ is immediately followed by an ensemble with the opposite detuning: $\delta_{2k-1}=\Delta_k/2$, $\delta_{2k}=-\Delta_k/2$ with $k=1,2,\cdots,L/2$. 
A direct calculation shows that in this case, Eq.~(\ref{eq:model}) has a pure steady state that is a product of entangled states for each adjacent pair of ensembles $(2k-1,2k)$:
\begin{equation}
    \left|\psi_{\rm ss}^{(L)} \right \rangle = \bigotimes_{k=1}^{L/2} 
    \left | \psi^{(2)}_{\rm ss}(\Delta_k,\chi,\Omega) \right \rangle_{2k-1,2k}.
    \label{eq:steady1}
\end{equation}
Here $|\psi^{(2)}_{\rm ss} \rangle_{k,l}$ is just the two-ensemble steady state of the last section 
(c.f.~Eq.~(\ref{eq:twogeneral})), realized using ensembles $k$ and $l$. 
One can further show that $|\psi^{(L)}_{\rm ss}\rangle$ is the {\it unique} steady state as long as permutation symmetry is fully broken, which can be achieved by setting $\chi\neq 0$ or making all $\Delta_k$ nonzero and distinct (see App.~\ref{sec:unique}). 
This dimerized form is easily understood in the non-reciprocal limit $\chi = \Gamma$, as the ensemble $2k$ is a perfect absorber \cite{Stannigel2012,Roberts2021,Ramos2014,Pichler2015} for the $2k-1$ ensemble.

The more interesting case is where the detuning pairs ($\pm\Delta_k/2$) are not perfectly ordered to appear in adjacent ensembles.  To address this general case, we first parametrize the Hamiltonian in Eq.~(\ref{eq:model}) as $\hat{H}(\vec{\delta})$, where $\vec{\delta}=(\delta_1,\delta_2,\cdots,\delta_L)$ is the ordered vector of drive detunings.
We also define a reordered vector of detunings 
$\vec \delta_{\rm init}$, where the elements of $\vec{\delta}$ have been permuted to yield a dimerized ordering:  $\vec{\delta}_{\rm init}=(\Delta_1/2,-\Delta_1/2,\Delta_2/2,-\Delta_2/2,\cdots)$, such that 
$\vec{\delta}=\mathcal{P}\vec{\delta}_{\rm init}$ with $\mathcal{P}$ a permutation.  Note that the non-uniqueness of $\vec{\delta}_{\rm init}$ plays no role in the following analysis.

If the system Hamiltonian were $\hat{H}(\vec{\delta}_{\rm init})$, then the system steady state would be given by the fully dimerized form of Eq.~(\ref{eq:steady1}).
Inspired by the discussion for the $S=1/2$ case  \cite{Stannigel2012,Ramos2014,Pichler2015}, we seek to design, for arbitrary $S$, a unitary operator $\hat{U}$ that transforms the steady-state solution of the simple ordering $\vec{\delta}_{\rm init}$ of detunings to the exact steady-state solution of the actual ordering of detunings $\vec{\delta}$.  This requires that $\hat U$ satisfy
\begin{equation}
    \begin{aligned}
    \hat{U}\hat{H}(\vec{\delta}_{\rm init})\hat{U}^{\dag} = \hat{H}(\vec{\delta}),
    \,\,\,\,\,\,\,\,
    \hat{U}\Big(\sum_l\hat{S}_l^{-}\Big)\hat{U}^{\dag} = \sum_l\hat{S}_l^{-}.
    \end{aligned}
    \label{eq:cond}
\end{equation}
If such a unitary operator exists, it immediately follows that the steady solution for the general case will be given by 
$\left|\psi_{\rm ss}^{(L)}(\vec{\delta}) \right \rangle=\hat{U}
\left |\psi_{\rm ss}^{(L)}(\vec{\delta}_{\rm init}) \right \rangle$.

The problem thus reduces to finding a $\hat{U}$ of the above form that corresponds to the needed permutation $\mathcal{P}$.  
Note that any permutation can be decomposed as a product of nearest-neighbor swap operations $\mathcal{P}_{l,l+1}$ that just permute two adjacent detunings:  $\mathcal{P}_{l,l+1}(\cdots,\delta_l,\delta_{l+1},\cdots)=(\cdots,\delta_{l+1},\delta_{l},\cdots)$. 
We can thus construct a $\hat{U}_{l,l+1}$ that corresponds to 
$\mathcal{P}_{l,l+1}$ (and satisfies Eq.~(\ref{eq:cond})), and then use it to build the full unitary $\hat{U}$ that relates $\vec \delta$ and $\vec \delta_{\rm init}$.  

We make the following SU(2)-symmetric ansatz for $\hat{U}_{l,l+1}$:
\begin{equation}
    \hat{U}_{l,l+1}(\delta_l,\delta_{l+1},\chi) = \exp\bigg(i\sum_{J,m}\theta_J|J,m\rangle\langle J,m|_{l,l+1} \bigg),
    \label{eq:gate}
\end{equation}
where $|J,m\rangle_{l,l+1}$ with $J=0,1,\cdots, 2S$ is the total-angular-momentum basis for the subsystem formed by ensembles $l$ and $l+1$, and $\theta_J$ are coefficients to be determined.  
This form ensures that any terms symmetric between ensembles $l$ and $l+1$ will remain unchanged when transformed by $\hat{U}_{l,l+1}$, so the second condition of Eq.~(\ref{eq:cond}) is automatically satisfied.  The first condition reduces to
\begin{equation}
    \begin{aligned}
    &\hat{U}_{l,l+1} \bigg(\frac{\delta_l-\delta_{l+1}}{2}(\hat{S}^{z}_l-\hat{S}^{z}_{l+1})+i\frac{\chi}{2}(\hat{S}_l^{+}\hat{S}_{l+1}^{-}-h.c.)\bigg)\hat{U}^{\dag}_{l,l+1}\\
    &= - \left( \frac{\delta_l-\delta_{l+1}}{2} \right)
    (\hat{S}^{z}_l-\hat{S}^{z}_{l+1})+i\frac{\chi}{2}(\hat{S}_l^{+}\hat{S}_{l+1}^{-}-h.c.).
    \end{aligned}
\end{equation}
Rewriting this in the total-angular-momentum basis, and applying the analytical results in 
App.~\ref{sec:analytic}, we find an equation that determines the coefficients $\theta_J$,
\begin{equation}
    e^{i(\theta_{J+1}-\theta_J)} = \frac{-(\delta_l-\delta_{l+1}) -i\chi (J+1)}{(\delta_{l}-\delta_{l+1}) -i\chi (J+1)}.
    \label{eq:gate1}
\end{equation}

To obtain intuition for this result, note first that $\hat{U}_{l,l+1}$ applies a different phase shift $\theta_J$ to each total-angular-momentum sector of the two-ensemble subsystem.  
When $\chi\gg |\delta_l-\delta_{l+1}|$, we have $\theta_{J+1}-\theta_J\rightarrow 0$, and $\hat{U}_{l,l+1}$ is close to the identity matrix.
In contrast, when $\chi\ll |\delta_l-\delta_{l+1}|$, we have $\theta_{J+1}-\theta_J\rightarrow \pm \pi$.
Due to the symmetry properties of Clebsch-Gordan coefficients, $\langle S,m_l;S,m_{l+1}|J,m\rangle = (-1)^{2S-J}\langle S,m_{l+1};S,m_l|J,m\rangle$, 
$\hat{U}_{l,l+1}$ in this limit is close to a many-body swap gate that swaps the quantum states for the entire ensemble $l$ with ensemble $l+1$. 
The general case in between these limits corresponds to a ``partial swap'' between the two ensembles, an operation that is entangling. 

Further insights follow from re-writing $\hat{U}_{l,l+1}$ in terms of an effective Hamiltonian generator for spin-spin interactions between the two ensembles.  We find:
\begin{equation}
    \hat{U}_{l,l+1}(\delta_l,\delta_{l+1},\chi) = \exp\bigg(i\sum_{q=0}^{2S} a_q \big(\hat{\bf S}_{l}\cdot\hat{\bf S}_{l+1}\big)^q\bigg),
    \label{eq:spinspin}
\end{equation}
where the coefficients $a_q$ can be obtained from the phase shifts $\theta_J$. In the previously studied $S=1/2$ case  \cite{Stannigel2012,Ramos2014,Pichler2015}, we have a massive simplification, as the Hamiltonian in Eq.~(\ref{eq:spinspin}) reduces to a simple Heisenberg interaction $\hat{\bf S}_{l}\cdot\hat{\bf S}_{l+1}$.
In contrast, for the more general case of arbitrary spin $S$ considered in our work, Eq.~(\ref{eq:spinspin}) is highly nontrivial, as it is the sum of many different multi-body interaction terms.   

Having understood how to construct a unitary of the required form corresponding to a nearest-neighbor permutation $\mathcal{P}_{l,l+1}$, we can return to the general case where the detunings in the Hamiltonian are obtained from a general permutation $\mathcal{P}$ acting on the dimerized ordering $\vec \delta_{\rm init}$.  This general permutation can be written as a product of nearest-neighbor permutations.  It follows that the needed unitary $\hat{U}(\vec{\delta}_{\rm init}\rightarrow\vec{\delta})$ can be obtained as product of nearest-neighbor operations $\hat{U}_{l,l+1}(\delta_l,\delta_{l+1},\chi)$, specified by the decomposition of $\mathcal{P}$ into nearest-neighbor swaps.  We thus obtain:
\begin{equation}
    \left |\psi_{\rm ss}^{(L)} \right\rangle = \hat{U}(\vec{\delta}_{\rm init}\rightarrow\vec{\delta}) \bigg[\bigotimes_{k=1}^{L/2} 
    \left|\psi_{\rm ss}^{(2)} \right \rangle_{2k-1,2k}\bigg].
    \label{eq:steadyfinal}
\end{equation}

We thus have a concrete way of understanding the broad class of steady states that can be generated through our reconfigurable dynamics.
Our result also provides a useful circuit decomposition for the quantum channel generated by infinite-time evolution under the Lindbladian in Eq.~(\ref{eq:model}) (with detunings satisfying the constraints in Sec.~\ref{sec:GeneralSetup}).  
We show this explicitly in Fig.~\ref{fig:fourensemble}(b) for an example with $L=4$. 
The infinite-time evolution of our Lindblad master equation is a quantum channel ($\mathcal{E}_{1234}$) that converts an arbitrary state into the steady state. 
For the case of $\delta_1=-\delta_4$, $\delta_2=-\delta_3$, our analysis decomposes this quantum channel into quantum channels for two spin ensembles ($\mathcal{E}_{12}$ and $\mathcal{E}_{34}$), and then applies $\hat{U}_{23}$ and $U_{34}$ to swap the detunings sequentially. 
In computer science, forming an ordered list of numbers via nearest-neighbor permutations is known as ``bubble sort'' and it requires $O(L^2)$ permutations in the worst case.
This sets an upper bound for the number of nearest-neighbor operations in Eq.~(\ref{eq:steadyfinal}). 

Eq.~(\ref{eq:steadyfinal}) shows that by using an ordering of detunings that is not simply dimerized, we can produce steady states with more complex entanglement properties.  
In Fig.~\ref{fig:fourensemble}(c) and (d), we numerically calculate the von Neumann entanglement entropy for the four-ensemble steady state in the case of $S=10$. 
Here we use subscripts to label the indices of spin ensembles included in a reduced density matrix.
For example, $\hat{\rho}_{12}$ is the reduced two-ensemble density matrix formed by tracing out ensembles $3$ and $4$, i.e., $\hat{\rho}_{12}=\mathrm{tr}_{34}(|\psi_{\rm ss}^{(4)}\rangle\langle \psi_{\rm ss}^{(4)}|)$.
The von Neumann entanglement entropy for $\hat{\rho}_{12}$ is given by $S_{\rm vn}(\hat{\rho}_{12}) = -\mathrm{tr}(\hat{\rho}_{12}\ln(\hat{\rho}_{12}))$.
The case of $\delta_1=-\delta_2$, $\delta_3=-\delta_4$ is shown in Fig.~\ref{fig:fourensemble}(c).  For this choice, we find $S_{\rm vn}(\hat{\rho}_{12})=0$, implying that $\hat{\rho}_{12}$ is a pure state.
This agrees with the dimerized structure predicted in Eq.~(\ref{eq:steady1}). 
The case of $\delta_1=-\delta_4$, $\delta_2=-\delta_3$ is shown in Fig.~\ref{fig:fourensemble}(d).  
Here, we instead find $S_{\rm vn}(\hat{\rho}_{12}),S_{\rm vn}(\hat{\rho}_{13}),S_{\rm vn}(\hat{\rho}_{14})\neq 0$, indicating the existence of true multipartite entanglement between all spin ensembles.
In both Fig.~\ref{fig:fourensemble}(c) and (d), the entanglement entropy vanishes below a threshold $\Omega<S\chi$, where the steady state is approaching a product state of the four spin ensembles, $|\psi_{\rm ss}^{(4)}\rangle\approx \bigotimes_l |S,-S\rangle_l$.

\subsection{Entanglement-enhanced metrology with multiple spin ensembles}
There is growing interest in extending quantum-enhanced metrology to 
setups with many spatially separated spin ensembles \cite{Cooper2024,Li2025}, opening new frontiers in nonlocal-field sensing, multi-parameter estimation, and environmental-noise mitigation. 
Here, we show that the unique entanglement properties of our many-ensemble steady state (see Eq.~(\ref{eq:steadyfinal})) can be directly harnessed as a metrological resource.
We focus on the case of four spin-$S$ ensembles with $\delta_1=-\delta_4=\Delta_A/2$, $\delta_2=-\delta_3=\Delta_B/2$ as a concrete example. 
Note that the general structure here can be exploited for even larger values of $L$. 

\begin{figure}[t]
    \centering
    \includegraphics[width=1.0\columnwidth]{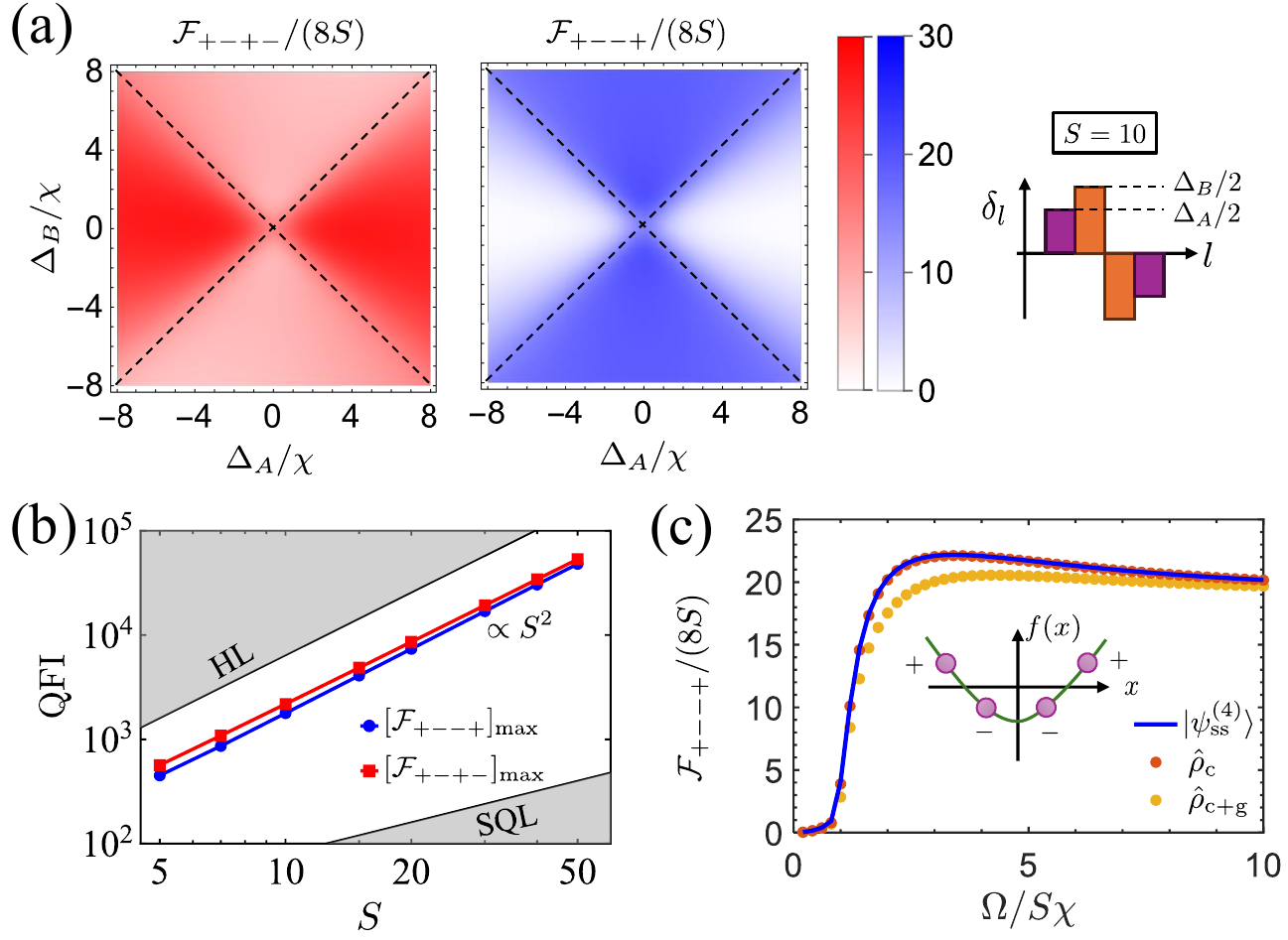}
    \caption{(a) Steady-state QFI $\mathcal{F}_{+-+-}$ and $\mathcal{F}_{+--+}$ (see text for definitions) associated with estimating different differential parameters in a four spin ensemble setup (with each ensemble having $S=10$).  The sensing state is the steady-state of our dynamics in the large-Rabi-drive limit, with drive detunings $\delta_j$ set to $\delta_1=-\delta_4=\Delta_A/2$, $\delta_2=-\delta_3=\Delta_B/2$. 
    The regime with $|\Delta_A|\;>|\Delta_B|$ favors $\mathcal{F}_{+-+-}$, while the regime with $|\Delta_B|\;>|\Delta_A|$ favors $\mathcal{F}_{+--+}$. 
    (b) Scaling of the maximum QFI for both composite parameters after optimizing over $\Omega$, $\Delta_A$ and $\Delta_B$. Heisenberg scaling can be reached for both cases. (c) $\mathcal{F}_{+--+}$ in the case of $S=10$ as a function of Rabi drive amplitude $\Omega$. We fix $\Delta_A=0$ and $\Delta_B=8\chi$. The inset shows that $\mathcal{F}_{+--+}$ captures the sensitivity for estimating the curvature of a spatially-varying field. The blue line, orange points, and yellow points describe the cases without noise ($|\psi_{\rm ss}^{(4)}\rangle$), with maximal common phase noise ($\hat{\rho}_{\rm c}$), and with both maximal common phase noise and gradient noise ($\hat{\rho}_{\rm c+g}$), respectively.}
    \label{fig:qfi}
\end{figure}

We first demonstrate that this parameterized family of four-ensemble entangled states can be tuned so as to change how quantum metrological enhancement is distributed between different ensembles.  
We consider the QFI matrix for operators $\hat{S}^z_l$, 
\begin{equation}
    \mathcal{F}_{kl} = 4\mathrm{Cov}(\hat{S}^z_k,\hat{S}^z_l)_{\rm ss}.
\end{equation}
In the large-Rabi-drive limit ($\Omega\rightarrow\infty$), it is possible to diagonalize $\mathcal{F}_{kl}$ analytically (see App.~\ref{sec:largedrive}).  
The matrix is diagonalized by the Hadamard matrix
\begin{equation}
    H_4=\begin{pmatrix}
        1 & 1 & 1 & 1\\
        1 & -1 & 1 & -1\\
        1 & 1 & -1 & -1\\
        1 & -1 & -1 & 1\\
    \end{pmatrix}.
\end{equation}
Here we consider the small detuning regime $\Delta_A,\Delta_B<S\chi$.  
In this regime, there are two dominant orthogonal modes (generators) that exhibit a quantum-metrological enhancement with HL-scaling in $S$:
\begin{equation}
     \begin{aligned}
     \mathcal{F}_{+-+-} &= 4\mathrm{Var}(\hat{S}^z_1-\hat{S}^z_2+\hat{S}^z_3-\hat{S}^z_4)_{\rm ss},\\
     \mathcal{F}_{+--+} &= 4\mathrm{Var}(\hat{S}^z_1-\hat{S}^z_2-\hat{S}^z_3+\hat{S}^z_4)_{\rm ss}.
     \end{aligned}
\end{equation}
In Fig.~\ref{fig:qfi}(a), we show both $\mathcal{F}_{+-+-}$ and $\mathcal{F}_{+-+-}$ in the large-Rabi-drive limit. 
$\mathcal{F}_{+-+-}$ dominates when $|\Delta_A|\;>|\Delta_B|$, while $\mathcal{F}_{+--+}$ dominates when $|\Delta_A|\;<|\Delta_B|$.
In Fig.~\ref{fig:qfi}(b), we optimize $\Omega$, $\Delta_A$ and $\Delta_B$ to calculate the maximal QFI.  
One sees clearly that both $[\mathcal{F}_{+-+-}]_{\rm max}$ and $[\mathcal{F}_{+--+}]_{\rm max}$ show Heisenberg scaling with increasing spin size $S$. 
It is worth mentioning that $\mathcal{F}_{+--+}$ reaches its maximum at a finite $\Omega$ (see Fig.~\ref{fig:qfi}(c)). 
Considering the prefactor of the Heisenberg scaling, we find that the maximal QFI for both cases approaches $64S(S+1)/3$.

We thus see that the specific family of stabilized four-ensemble entangled states analyzed here can provide a metrological advantage for distributed-sensing tasks, where one wishes to estimate parameters that couple in a structured manner to all four ensembles.  
We now discuss potential practical applications of this ability.
Using many spin ensembles, one can naturally consider the sensing of a spatially varying field $f(x)$ in 1D, where $x$ is discretely sampled by the
spins in ensemble $l$ which are localized near $x = x_l$. 
Suppose $f(x)$ is a slowly varying function near position $x_c$, one can approximate $f(x)$ using the Taylor expansion, $f(x)\approx f(x_c) + (x-x_c)f'(x_c) + \frac{1}{2}(x-x_c)^2 f''(x_c)$, where the first-order derivative $f'(x_c)$ and the second-order derivative $f''(x_c)$ are the gradient and curvature, respectively. 
If we place two spin ensembles at positions $x_l = x_c + (l-3/2)d$, where $l=1,2$ and $d$ is the separation distance, we can measure the gradient $f'(x_c)$ using the differential sensing protocol discussed in Sec.~\ref{subsec:DifferentialSensing}.
If, instead, we place four spin ensembles at positions $x_l = x_c + (l-5/2)d$ with $l=1,2,3,4$, the Hamiltonian for the spatially varying field is given by
\begin{equation}
    \begin{aligned}
    \hat{H}_{\rm field} &= \sum_{l=1}^4 f(x_l) \hat{S}^z_l\\
    &\approx \Big(f(x_c) + \frac{5d^2}{8}f''(x_c)\Big) \Big(\hat{S}^z_1+\hat{S}^z_2+\hat{S}^z_3+\hat{S}^z_4\Big)\\
    &-\frac{d}{2}f'(x_c)\Big(3\hat{S}^z_1+\hat{S}^z_2-\hat{S}^z_3-3\hat{S}^z_4\Big)\\
    &+\frac{d^2}{2}f''(x_c)\Big(\hat{S}^z_1-\hat{S}^z_2-\hat{S}^z_3+\hat{S}^z_4\Big).
    \end{aligned}
    \label{eq:field}
\end{equation}
Based on Eq.~(\ref{eq:field}), the effects of the curvature $f''(x_c)$ can be described by the following unitary evolution (removing the common phase), $\hat{U}_{\tilde{\phi}} = e^{-i\tilde{\phi}(\hat{S}^z_1-\hat{S}^z_2-\hat{S}^z_3+\hat{S}^z_4)}$.
Therefore, our steady-state solution can be directly applied to curvature measurements and the sensitivity is captured by the QFI $\mathcal{F}_{+--+}$. 

Our class of four-ensemble entangled states has another strong advantage: not only can it provide a quantum enhancement for estimating the curvature of a spatially-varying field, it can do this with robustness against various kinds of phase noise.  
We imagine a situation where the goal is to estimate the value of the curvature of a spatially varying field $f(x)$, despite possible noise and uncertainty in the average value and gradient of the field.  
For four spin ensembles with positions $x_l$ chosen as before, the unitary evolution due to a common phase $\theta$ is described by $\hat{U}_{\theta} = e^{-i\theta (\hat{S}^z_1+\hat{S}^z_2+\hat{S}^z_3+\hat{S}^z_4)}$.  In contrast, the evolution due to the gradient of the field $f(x)$ corresponds to $\hat{U}_{\phi} = e^{-i\phi (3\hat{S}^z_1+\hat{S}^z_2-\hat{S}^z_3-3\hat{S}^z_4)}$.
Similar to the two-ensemble case analyzed in Sec.~\ref{subsec:TwoEnsembleCommonModeNoise}, phase fluctuations of laser beams can lead to large common-phase noise, implying evolution under a random common phase $\theta$.  The resulting corrupted four-ensemble sensor state is then $\hat{\rho}_{\rm c} = \int \frac{d\theta}{2\pi} \hat{U}_{\theta}|\psi_{\rm ss}^{(4)}\rangle\langle \psi_{\rm ss}^{(4)}|\hat{U}^{\dag}_{\theta}$. 
One can further include large fluctuations of a field gradient and consider the density matrix, $\hat{\rho}_{\rm c+g} = \int \frac{d\theta}{2\pi}\frac{d\phi}{2\pi} \hat{U}_{\phi}\hat{U}_{\theta}|\psi_{\rm ss}^{(4)}\rangle\langle \psi_{\rm ss}^{(4)}|\hat{U}^{\dag}_{\theta}\hat{U}_{\phi}^{\dag}$.
Note that the uniform distribution over $2\pi$ is the worst-case scenario, since any probability distribution can be reduced to the uniform distribution by randomizing the common phase and the field gradient.
In Fig.~\ref{fig:qfi}(c), we numerically estimate the QFI of $\mathcal{F}_{+--+}$ for $|\psi_{\rm ss}^{(4)}\rangle$, $\hat{\rho}_{\rm c}$ and $\hat{\rho}_{\rm c+g}$. 
We see that, even with maximal amounts of common phase noise and gradient noise, the reduction of the QFI associated with estimating the field curvature is remarkably small.  

\section{Stabilizing SPT order in 1D chains of spin ensembles}
\label{sec:1dchain}

\subsection{Motivation and connection to sequential quantum circuits}
We now segue away from exploring the metrological utility of the entangled states stabilized by our scheme, and instead discuss how these states can possess more general many-body and topological features.
From a broad perspective, setting $\chi\neq 0$ (i.e., non-zero chiral spin-exchange interaction) leads to a 1D ordering of the ensembles, so steady states realized in our setup correspond to pure entangled states of an effective one-dimensional (1D) chain of spin-$S$ ensembles.

Depending on the choice of drive detunings in the system Hamiltonian, output states from a variety of quantum circuits can be obtained as the dissipative steady states (see Eq.~(\ref{eq:steadyfinal})).
In this section, we focus on a non-trivial subset of them known as sequential quantum circuits (SQC), i.e., a linear-depth circuit with each layer acting only on one subregion in the system. 
SQC play an important role in the preparation of matrix product states (MPS) and other tensor-product states \cite{Schon2005,Schon2007,Banuls2008,Wei2022}, as well as connecting different gapped quantum phases \cite{Chen2024}.
We show that the class of steady states as outputs of SQC has a natural MPS description with a low bond dimension, and, in appropriate limits, one can achieve symmetry-protected topological (SPT) order \cite{Chen2011,Pollmann2012}.

As a specific example, we show how to stabilize the spin-$1$ Affleck-Kennedy-Lieb-Tasaki (AKLT) state using $S=1/2$ ensembles.
Initially proposed to gain analytical insight into Haldane's conjecture \cite{Haldane1983} for ground states of the 1D integer-spin antiferromagnetic Heisenberg model, spin-$1$ AKLT states \cite{Affleck1987,Affleck1988} serve as a paradigmatic example of SPT order \cite{Chen2011,Pollmann2012} and are an important resource state for measurement-based quantum computation (MBQC) \cite{Verstraete2004,Gross2007}. 
We further discuss the stabilization of a continuum of states in the same SPT phase and the generalization to higher spin $S$.

It is also worth to emphasize that viewing the dissipative steady state as the output state of a quantum circuit (see Eq.~(\ref{eq:steadyfinal})) is a useful way of understanding the final steady state, but does not reflect the actual time-dynamics of the dissipative evolution (i.e., unlike Ref.~\onlinecite{Verstraete2009} we are not directly encoding a unitary circuit into a sequence of dissipators).     

\subsection{MPS representation}

\begin{figure}[t]
    \centering
    \includegraphics[width=1.0\columnwidth]{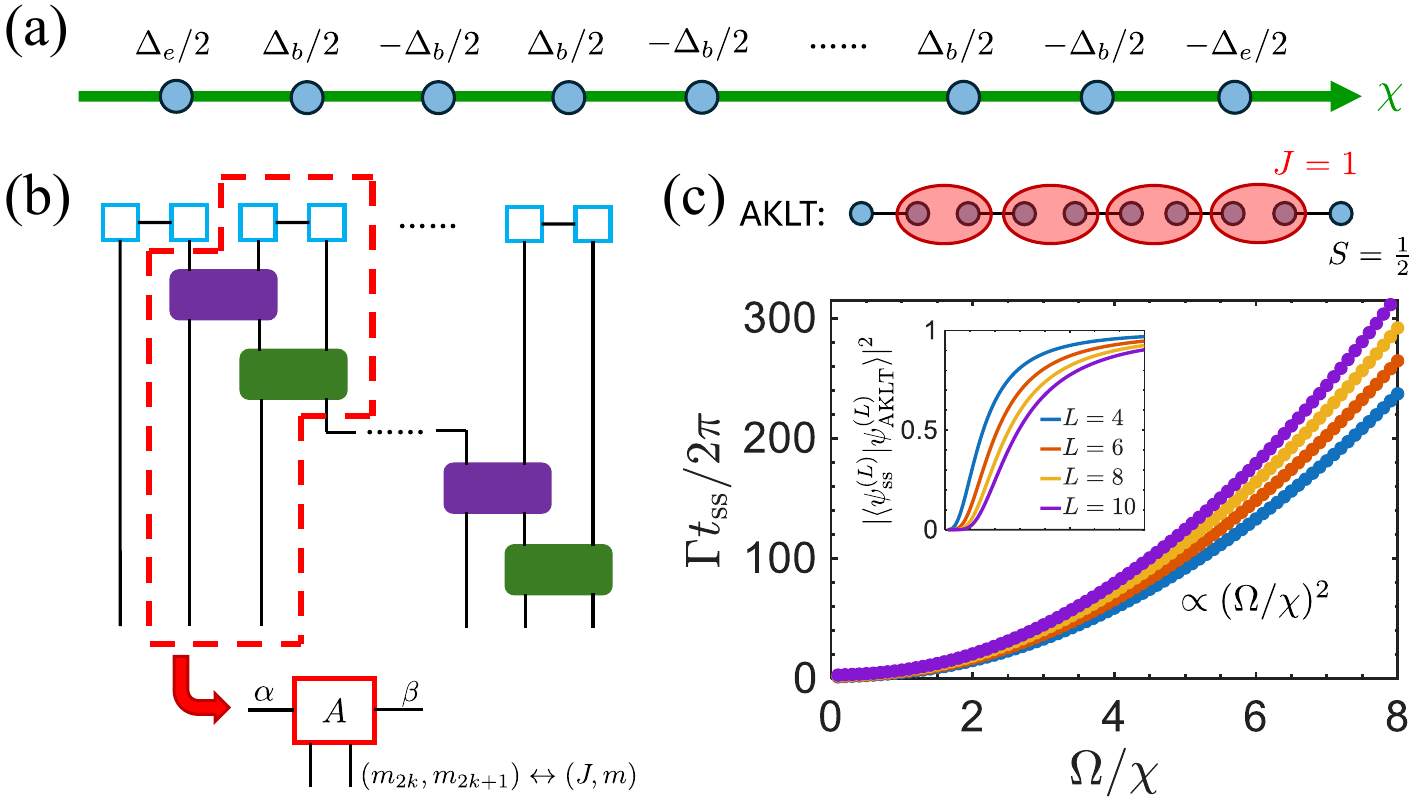}
    \caption{(a) We consider an even number $L$ of spin-$S$ ensembles placed in a 1D spin chain, and set their detunings as $\delta_1=\Delta_e/2$, $\delta_L=-\Delta_e/2$, $\delta_{2k}=\Delta_b/2$, $\delta_{2k+1}=-\Delta_b/2$, with $k=1,2,\cdots,L/2-1$. (b) The steady-state solution is the same as the output state of a ``staircase'' quantum circuit: Starting from the initial dimerized state (two blue boxes linked by a bond), we apply quantum gates $\hat{U}(-\Delta_e/2,\Delta_b/2,\chi)$ (purple boxes) and $\hat{U}(-\Delta_e/2,-\Delta_b/2,\chi)$ (green boxes) alternately. The red dashed line marks the translation-invariant structure for constructing a matrix product state (MPS). (c) The upper panel shows the schematic of the spin-$1$ AKLT state, which is formed by projecting every unit cell between the spin-$1/2$ dimers to total spin $1$. The bottom panel shows the relaxation timescale $t_{\rm ss}$ for reaching the steady state with $\chi=\Gamma$. We also set $\Delta_e=0$ and $\Delta_b=\sqrt{2}\,\chi$, which make the steady state equal to the spin-$1$ AKLT state in the large-Rabi-drive limit. The inset shows the fidelity between our steady-state solution $|\psi_{\rm ss}^{(L)}\rangle$ and the spin-$1$ AKLT state as a function of $\Omega/\chi$. The line colors specifying the system size $L$ in the inset also apply to the data points in the main plot. }
    \label{fig:spinchain}
\end{figure}

As shown in Fig.~\ref{fig:spinchain}(a), we consider an even number $L$ of spin-$S$ ensembles with 
$\chi\neq 0$.  
In our scheme, a particularly simple way of achieving a three-parameter $(\Delta_e,\Delta_b,\chi)$ family of SQC is to use a detuning pattern $\delta_1=\Delta_e/2$, $\delta_L=-\Delta_e/2$ on the edges, and $\delta_{2k}=\Delta_b/2$, $\delta_{2k+1}=-\Delta_b/2$ in the bulk, with $k=1,2,\cdots,L/2-1$. 
Notice that we have equal-in-magnitude, opposite-sign detuning pairs on adjacent sites {\it except} at the edges.  
This corresponds to starting with a perfectly dimerized ordering of detunings, and then swapping a single detuning $-\Delta_e/2$ across the whole 1D chain, moving it to the end.

Recall from Sec.~\ref{sec:many} that the permutation $\mathcal{P}$, which maps the perfectly dimerized configuration to the actual ordering of detunings, determines the form of the unitary operator $\hat{U}$ to construct the steady state $|\psi_{\rm ss}^{(L)}\rangle$.
For the detuning pattern specified above, Eq.~(\ref{eq:steadyfinal}) simplifies to a “staircase” circuit (see Fig.~\ref{fig:spinchain}(b)), in which we alternately apply quantum gates $\hat{U}(-\Delta_e/2,\Delta_b/2,\chi)$ (purple boxes) to ensemble pairs $(2k,2k+1)$ and $\hat{U}(-\Delta_e/2,-\Delta_b/2,\chi)$ (green boxes) to pairs $(2k+1,2k+2)$, starting from the dimerized initial state $\bigotimes_{k=1}^{L/2}|\psi_{\rm ss}^{(2)}\rangle_{2k-1,2k}$ (two blue boxes connected by a bond).
Here, the quantum gates are given by Eqs.~(\ref{eq:gate}) and (\ref{eq:gate1}), in which we are dropping the subscripts to emphasize the translationally invariant structure of quantum gates.

From the circuit description in Fig.~\ref{fig:spinchain}(b) of the resulting state, one can identify a translationally invariant structure based on a two-ensemble unit cell (depicted by red dashed lines).  This allows us to describe the final state (i.e., the steady state) as a translationally-invariant matrix-product state (MPS) with open boundary conditions. Using a product-state basis of $\hat{S}^z_{j}$ eigenstates,  
$|\mathbf{m}\rangle\equiv\otimes_{j=1}^L|S,m_j\rangle$, the MPS representation of our steady state can be written as
\begin{equation}
    |\psi_{\rm ss}^{(L)}\rangle = \sum_{\mathbf{m}}(\mathbf{v}_{\rm left}^{m_1})^{T}\bigg(\prod_{k=1}^{L/2-1} \mathbf{A}^{m_{2k}m_{2k+1}} \bigg) \mathbf{v}_{\rm right}^{m_L}|\mathbf{m}\rangle.
    \label{eq:mps}
\end{equation}
Here, $\mathbf{A}^{m_{2k}m_{2k+1}}$ is the translationally-invariant MPS matrix corresponding to the basis state $|S,m_{2k}\rangle|S,m_{2k+1}\rangle$ of a given two-ensemble unit cell.  From Fig.~\ref{fig:spinchain}(b), we see that the bond-dimension of this matrix is $2S+1$.  
The left and right boundary vectors $\mathbf{v}_{\rm left}^{m_1}$ and $\mathbf{v}_{\rm right}^{m_L}$ are uniquely determined by the form of the steady-state  and are defined in App.~\ref{sec:largemps}.

In the following, we mainly focus on the large-Rabi-drive limit ($\Omega\rightarrow\infty$), both to gain analytical insight and to highlight the resulting emergent symmetries.
In this limit, the initial dimerized state $\bigotimes_{k=1}^{L/2}|\psi_{\rm ss}^{(2)}\rangle_{2k-1,2k}$ in Eq.~(\ref{eq:steadyfinal}) becomes a tensor product of maximally entangled ensemble pairs $\bigotimes_{k=1}^{L/2}|J=0,m=0\rangle_{2k-1,2k}$.  This state is often termed the spin-$S$ dimer state of a 1D chain.
This state is the steady state of our scheme when we tune $\Delta_e=\Delta_b=0$, as, in this limit, all the quantum gates become identity matrices [as per Eq.~\eqref{eq:steadyfinal}].
For general choices of $\Delta_e$ and $\Delta_b$, it is more convenient to focus on the MPS matrices $\mathbf{A}^{m_{2k}m_{2k+1}}$, and transform to the total-angular-momentum basis in the two-ensemble unit cell, $\mathbf{A}^{Jm} = \sum_{m_{2k}m_{2k+1}} \langle S,m_{2k}; S,m_{2k+1}|J,m\rangle \mathbf{A}^{m_{2k}m_{2k+1}}$. Then one can obtain 
\begin{equation}
    \begin{aligned}
    \left[ \mathbf{A}^{Jm} \right]_{\alpha\beta} &= \frac{(-1)^{S+\beta}}{\sqrt{2S+1}}\langle S,\alpha;S,-\beta|J,m\rangle f_J(\Delta_e,\Delta_b,\chi),
    \end{aligned}
    \label{eq:vbs}
\end{equation}
where the subscripts $\alpha,\beta=-S,\cdots,S$ are the internal bond-dimension indices.  
Calculation details and the general analytical form of the amplitudes $f_J(\Delta_e,\Delta_b,\chi)$ are discussed in App.~\ref{sec:largemps}.

Equation~(\ref{eq:vbs}) shows that the MPS matrices factorize in the large-$\Omega$ limit: the matrix structure of $\mathbf{A}^{Jm}$ is only determined by Clebsch-Gordan coefficients, and the control parameters $\Delta_e,\Delta_b,\chi$ only enter through an overall $J$-dependent scalar prefactor.
The form of Eq.~(\ref{eq:vbs}) indicates that the MPS is invariant under a global on-site SO(3) symmetry \cite{Cirac2021}.
At a physical level, this symmetry emerges in the large-$\Omega$ limit as each dimer in the input to our circuit is in the $|J=0,m=0\rangle$ state (no preferred axis for rotation), and the unitaries in the form of Eq.~(\ref{eq:gate}) always preserve this symmetry.    
For a finite Rabi drive $\Omega$, the structure of the ``staircase'' quantum circuit remains the same with a general input state $\bigotimes_{k=1}^{L/2}|\psi_{\rm ss}^{(2)}\rangle_{2k-1,2k}$. 
In this case, it is more cumbersome to write down closed-form expressions for the MPS matrices, since they no longer exhibit the simple factorization in Eq.~(\ref{eq:vbs}).

We return to the large-$\Omega$ limit and consider some special, illustrative cases of Eq.~(\ref{eq:vbs}). 
To get a sense of the kinds of states described by this form, we start by taking the $f_J$ as arbitrary parameters.  
One simple case is $f_J=1$ for all $J$: this is just the spin-$S$ dimer state introduced earlier, where the dimers are located {\it between} the two-ensemble unit cells.
This choice of $f_J$ and the corresponding state are achievable in our setup by setting $\Delta_e=\Delta_b=0$.
Another special case is where only one value of $J=J_0$ is allowed, i.e., $f_J$ vanishes for $J\neq J_0$.  In this case, Eq.~(\ref{eq:vbs}) describes the spin-$J_0$ valence-bond-solid (VBS) state with virtual spin-$S$ particles \cite{Tu2009}.
For $J_0=2S$, the VBS state is also known as the spin-$2S$ AKLT state \cite{Affleck1987,Affleck1988}.
In our model, one can attempt to tune detunings and $\chi$ to achieve a desired pattern of the $f_J$ and a given target state.  
For general $S$, there is insufficient tunability to have only one $f_J$ be non-zero and hence stabilize a general VBS state.  
There is, however, an important and interesting exception: as we now show, for $S=1/2$, it is possible to set parameters such that the steady state exactly coincides with the spin-$1$ AKLT state.

\subsection{Stabilization of spin-$1$ AKLT states}

In the case of $S=1/2$ in our scheme, the tunability of $\Delta_e/\chi$ and $\Delta_b/\chi$ allows for the dissipative stabilization of the iconic spin-$1$ AKLT states.
For $S=1/2$, the MPS bond dimension is $2$, and Eq.~(\ref{eq:vbs}) becomes
\begin{equation}
    \begin{gathered}
    \mathbf{A}^{1,1} = \frac{f_1}{\sqrt{2}}\sigma^{+}, \quad \mathbf{A}^{1,0} = -\frac{f_1}{2}\sigma^z, \quad \mathbf{A}^{1,-1} = -\frac{f_1}{\sqrt{2}}\sigma^{-},\\
    \mathbf{A}^{0,0} = -\frac{f_0}{\sqrt{2}}I,
    \label{eq:mpsspin}
    \end{gathered}
\end{equation}
where $I$ is the $2\times 2$ identity matrix, $\sigma^{\pm}$ and $\sigma^z$ are Pauli matrices. The coefficients $f_1$ and $f_0$ are given by
\begin{equation}
    \begin{aligned}
    f_1 &= -\frac{\chi^2+i\chi \Delta_b}{\Big[(\Delta_e+\Delta_b)/2+i\chi\Big]\Big[(\Delta_e-\Delta_b)/2+i\chi\Big]},\\
    f_0 &= \frac{(\Delta_b^2-\Delta_e^2)/2-\chi^2+i\chi\Delta_e}{\Big[(\Delta_e+\Delta_b)/2+i\chi\Big]\Big[(\Delta_e-\Delta_b)/2+i\chi\Big]}.
    \end{aligned}
\end{equation}
We see that $f_0=0$ can be achieved by tuning $\Delta_e=0$ and $\Delta_b=\pm \sqrt{2}\,\chi$, thus realizing the exact MPS matrices of the spin-$1$ AKLT state.
As shown in the upper panel of Fig.~\ref{fig:spinchain}(c), one can understand the spin-$1$ AKLT state as projecting every unit cell between the spin-$1/2$ dimers (singlets) to total spin $1$.
It is worth mentioning that the AKLT state we dissipatively stabilized is unique (see App.~\ref{sec:largemps} and \ref{sec:unique}). 
Nevertheless, one can still access the 4 different AKLT states with open boundary conditions depending on the projective measurement outcome of the two edge qubits \cite{Smith2023}.

To understand whether this exact stabilization of the spin-$1$ AKLT state has practical relevance, we need to additionally analyze the relaxation time of the dynamics in Eq.~(\ref{eq:model}) when tuned to the AKLT parameters ($\Delta_e=0$ and $\Delta_b=\pm \sqrt{2}\,\chi$), and what happens when the drive amplitude $\Omega$ is finite. 
We will be interested in seeing how the relaxation time scales as we increase the system size $L$.  
Here we use two different metrics:  
1) $t_{\rm ss}$, the timescale for reaching the steady state $|\psi_{\rm ss}^{(L)}\rangle$ when $\Omega/\chi$ remains fixed as $L$ increases.
Note that, in this case, the fidelity with the ideal AKLT state $|\psi_{\rm AKLT}^{(L)}\rangle$ will decrease with increasing system size.  
2) $t_{\rm AKLT}$, the timescale for reaching the steady state $|\psi_{\rm ss}^{(L)}\rangle$ when fixing the fidelity overlap between the steady state and the ideal AKLT state, i.e., $|\langle \psi_{\rm ss}^{(L)}|\psi_{\rm AKLT}^{(L)}\rangle|^2$. 
In this case we need to increase $\Omega/\chi$ with increasing system size.

We start by considering the first case, i.e., $t_{\rm ss}$, with fixed system parameters. 
Similar to Sec.~\ref{subsec:singlespindiss}, we define $t_{\rm ss}$ in the following way: For $t>t_{\rm ss}$, the fidelity with the steady state, $F(t)=\mathrm{tr}\Big(|\psi_{\rm ss}^{(L)}\rangle\langle\psi_{\rm ss}^{(L)}|\,\hat{\rho}(t)\Big)$, satisfies $1-F(t)<10^{-3}$. 
We consider the initial state where all the spins are pointing down, numerically compute $t_{\rm ss}$ via exact evolution of Lindblad master equation.
In App.~\ref{sec:relax}, we show that optimal $t_{\rm ss}$ can be achieved near $\chi=\Gamma$ for fixed $\Gamma$ and $\Omega/\chi$, and the deviation from $\chi=\Gamma$ reduces as we increase $L$.
Therefore, we show $t_{\rm ss}$ with $\chi=\Gamma$ as an upper bound of the optimal relaxation time scale in the bottom panel of Fig.~\ref{fig:spinchain}(c).

Remarkably, although dissipative stabilization at $\chi=\Gamma$ (unidirectional propagation) share similarities with the SQC structure of our steady state, we find that $\Gamma t_{\rm ss}/(2\pi)\approx (\Omega/\chi)^2g(L)$, where $g(L)$ is a function that depends only weakly on the system size $L$, as suggested by numerical calculations up to $L=10$ (see App.~\ref{sec:relax} for further discussions).
This is in contrast to an SQC, where preparation time (circuit depth) is simply proportional to system size $L$.
As we already mentioned, the SQC with only nearest-neighbor quantum gates does not reflect the actual dissipative dynamics, since the collective dissipation and interactions in our system (see Eq.~(\ref{eq:model})) allow spins to interact with one another instantaneously, regardless of their distance.
Note that the weak dependence of system size for $t_{\rm ss}$ is also reported for dissipative stabilization of a different class of states in a related driven-dissipative setup \cite{Agusti2023}.

We next discuss the time scale $t_{\rm AKLT}$ relevant for the preparation of approximate AKLT states with a {\it fixed fidelity} with the ideal AKLT state. 
As shown in the inset of Fig.~\ref{fig:spinchain}(c), for a fixed value of $\Omega/\chi$, the fidelity with $|\psi_{\rm AKLT}^{(L)}\rangle$ decreases slowly as we increase the system size $L$.
To maintain the same fidelity, one has to compensate by increasing $\Omega/\chi$.
We further explore this tradeoff in App.~\ref{sec:largemps}, where we find that the preparation time for a fixed fidelity target scales as $t_{\rm AKLT}\propto Lg(L)$.
Our scheme serves as a promising protocol for dissipative stabilization of spin-$1$ AKLT states.
Compared to other dissipative stabilization schemes for such states \cite{Kraus2008,Sharma2021,Zhou2021a,Wang2023,Langbehn2024}, our scheme features a more favorable scaling of $t_{\rm AKLT}$ with system size, and requires only a single dissipator, which is more directly compatible with existing cavity-QED platforms.  
More importantly, unlike other approaches, our scheme is capable of stabilizing more than just the AKLT state.  
As we have stressed, our setup is easily reconfigurable (e.g., just change drive detunings), leading to a broad class of many-body entangled steady states.  
We will further explore the tunability in the following subsection.

\subsection{Tunability and string order for arbitrary spin-$S$}

\begin{figure}[t]
    \centering
    \includegraphics[width=1.0\columnwidth]{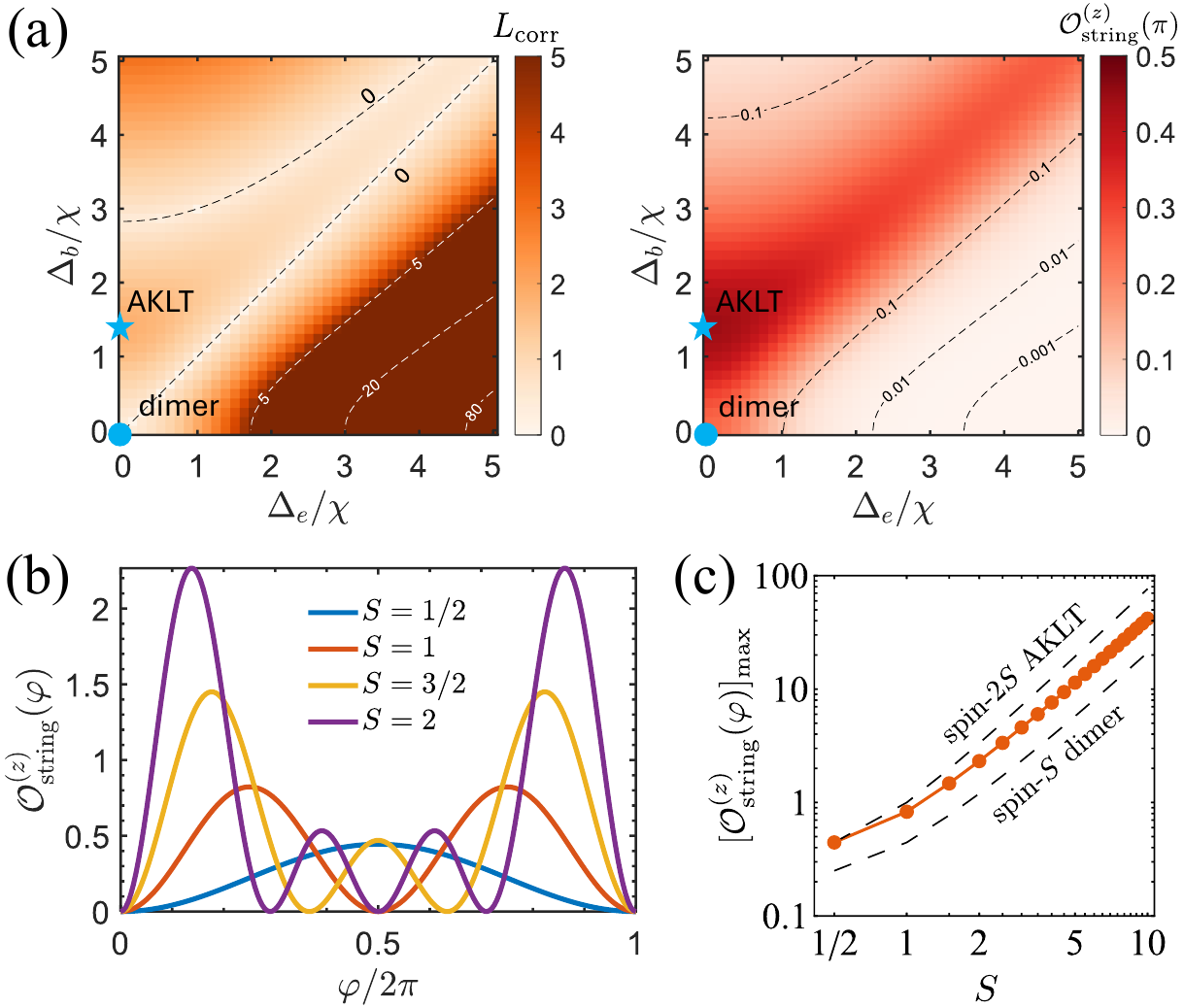}
    \caption{(a) Correlation length $L_{\rm corr}$ (left panel) and string order parameter $\mathcal{O}^{(z)}_{\rm string}(\pi)$ (right panel) in the dissipative steady state for $S=1/2$, as a function of $\Delta_e/\chi$ and $\Delta_b/\chi$ (and in the large-Rabi-drive limit). The blue star (circle) marks detuning choices that yield the spin-$1$ AKLT state (spin-$1/2$ dimer state). (b) Dependence of the string order parameter $\mathcal{O}^{(z)}_{\rm string}(\varphi)$ on the angle $\varphi$. We set $\Delta_e=0$ and $\Delta_b=\sqrt{2}\,\chi$ (parameters for spin-$1$ AKLT state in the case of $S=1/2$, blue star in panel (a)) and extend the calculation to $S>1/2$. (c) Scaling of the maximum value of the string order parameter as a function of $S$. We optimize over system parameters $\Delta_e/\chi$ and $\Delta_b/\chi$ and angle $\varphi$, and compare with spin-$2S$ AKLT states and spin-$S$ dimer states.}
    \label{fig:spinchain1}
\end{figure}

Apart from the spin-$1$ AKLT state, the reconfigurable nature of our scheme allows one to stabilize a class of entangled MPS steady states simply by changing detunings (see Eq.~(\ref{eq:vbs})).
To get an idea of the richness of steady states, we use the the same specific parameterized detuning pattern in Fig.~\ref{fig:spinchain}(a) with arbitrary $\Delta_e$ and $\Delta_b$, and again take the large-$\Omega$ limit.
We characterize the resulting two-parameter family of states, in the limit of infinite system size and arbitrary spin $S$.  

Similar to the spin-$1$ AKLT state, the states we obtain in this limit 
(described by Eq.~(\ref{eq:vbs})) do not exhibit any long-range order when looking at conventional spin-spin correlation functions, e.g.:
\begin{equation}
    \langle \hat{S}^z_k\hat{S}^z_l\rangle - \langle \hat{S}^z_k\rangle\langle\hat{S}^z_l\rangle \sim e^{-|l-k|/L_{\rm corr}},
    \label{eq:corr}
\end{equation}
with $L_{\rm corr}$ the spin-spin correlation length. In contrast, the string order parameter \cite{Den1989,Oshikawa1992,Totsuka1995},
\begin{equation}
    \begin{aligned}
    \mathcal{O}^{(z)}_{\rm string}(\varphi) = \lim_{|l-k|\rightarrow\infty}\Big\langle &(\hat{S}^z_{2k}+\hat{S}^z_{2k+1})e^{i\varphi\sum_{q=k}^{l-1}(\hat{S}^z_{2q}+\hat{S}^z_{2q+1})}\\
    &\times(\hat{S}^z_{2l}+\hat{S}^z_{2l+1})\Big\rangle,
    \end{aligned}
\end{equation}
remains non-zero in the large-distance limit.
Note that the non-zero string order parameter $\mathcal{O}^{(z)}_{\rm string}(\varphi)$ indicates the existence of the hidden long-range order of antiferromagnetism.
In particular, in the presence of SO(3) symmetry (c.f. Eq.~(\ref{eq:vbs})), the non-zero $\mathcal{O}^{(z)}_{\rm string}(\pi)$ indicates the existence of SPT order \cite{Pollmann2012a}.
We analyze the correlation length $L_{\rm corr}$ and the string order parameter $\mathcal{O}^{(z)}_{\rm string}(\varphi)$ using the transfer-matrix technique of infinite MPS \cite{Cirac2021} (see App.~\ref{sec:largemps} for technical details).

We first consider the range of states generated with $S=1/2$ (see Eq.~(\ref{eq:mpsspin})) as we tune system parameters $\Delta_e$ and $\Delta_b$ (see Fig.~\ref{fig:spinchain1}(a)). 
In the left panel of Fig.~\ref{fig:spinchain1}(a), we see that the correlation length $L_{\rm corr}$ in the pure steady state can be continuously tuned all the way from $0$ to arbitrarily large values.  
The lines with $L_{\rm corr}=0$ are known as disorder lines, and correspond to the condition $|f_0|\;=|f_1|$ on the MPS matrices (c.f.~Eq.~(\ref{eq:mpsspin}) and 
App.~\ref{sec:largemps}).  
These lines can be realized with system parameters $\Delta_b^2=\Delta_e^2$ or $\Delta_b^2=\Delta_e^2+8\chi^2$.
In contrast, having arbitrarily large correlation lengths
$L_{\rm corr}$ can be achieved in the regime $\Delta_e\gg \Delta_b,\chi$.
The string order parameter for $S=1/2$ is given by
\begin{equation}
    \mathcal{O}^{(z)}_{\rm string}(\varphi) = \frac{1}{4}|f_1|^4\sin^2\Big(\frac{\varphi}{2}\Big),
    \label{eq:string}
\end{equation}
indicating that maximal string order is achieved for an angle 
$\varphi=\pi$.
As shown in the right panel of Fig.~\ref{fig:spinchain1}(a), $\mathcal{O}^{(z)}_{\rm string}(\pi)$ reaches its maximum value $4/9$ for parameters corresponding to the spin-$1$ AKLT state ($\Delta_e=0$ and $\Delta_b=\sqrt{2}\,\chi$).

Note, crucially, that for any finite value of $\Delta_e/\chi$ and $\Delta_b/\chi$,  $L_{\rm corr}$ remains finite and $\mathcal{O}^{(z)}_{\rm string}(\pi)$ is always non-zero (see Fig.~\ref{fig:spinchain1}(a)).   
This result indicates that {\it all} the states in Eq.~(\ref{eq:mpsspin}) have SPT order and belong to the same phase as the spin-1 AKLT state.  
This agrees with Ref.~\onlinecite{Kolezhuk1997}, which argued that the parameterized MPS in Eq.~(\ref{eq:mpsspin}) with $S=1/2$ serves as a path smoothly connecting spin-$1/2$ dimer and spin-$1$ AKLT states.
This type of SPT order is due to the emergent SO(3) symmetry in the large-$\Omega$ limit indicated by Eq.~(\ref{eq:vbs}).
For a finite Rabi drive $\Omega$, strictly speaking, there is no true SPT order, as the string order parameter will decay to zero at infinite distance.  
However, for any finite system size, one can always increase $\Omega$ to realize an approximation of SPT order, where the the string order parameter remains non-zero over a distance much larger than system size (see App.~\ref{sec:largemps}).

As we have emphasized, a key virtue of our setup is its extreme flexibility. We now show that the above construction with $S=1/2$ directly generalizes to arbitrary spin-$S$ ensembles on each site.
One can show that the correlation length $L_{\rm corr}$ remains finite for the general case, i.e., all the states in the large-$\Omega$ limit lie in the same phase.
We then perform an analytical calculation of the string order parameter for arbitrary spin $S$, similar to Refs.~\onlinecite{Oshikawa1992,Totsuka1995}. 
As shown in App.~\ref{sec:largemps}, it is possible to separate the dependence of system parameters $\Delta_e,\Delta_b,\chi$ and the dependence of the angle $\varphi$, 
\begin{equation}
    \mathcal{O}^{(z)}_{\rm string}(\varphi) = \bigg(\sum_{J=0}^{2S}\frac{J(J+1)(2J+1)}{2S(S+1)(2S+1)^3}|f_J|^2\bigg)^2[h(\varphi)]^2,
    \label{eq:generalstring}
\end{equation} 
where $h(\varphi)$ is a universal function for spin-$S$ ensembles, 
\begin{equation}
    h(\varphi) = \sum_{\alpha=-S}^{S} \alpha \sin(\alpha\varphi).
\end{equation}
Notice that we have $\mathcal{O}^{(z)}_{\rm string}(\pi)=0$ for integer $S$ (no SPT order), and  $\mathcal{O}^{(z)}_{\rm string}(\pi)>0$ for half-integer $S$ (SPT order).
This result agrees with the discussions of spin-$2S$ AKLT states in Ref.~\onlinecite{Pollmann2012}.
We show the distinct dependence of the angle $\varphi$ for different spin $S$ in Fig.~\ref{fig:spinchain1}(b). 

As we mentioned previously, except for the case of $S=1/2$, the tunability of $\Delta_e/\chi$ and $\Delta_b/\chi$ is not sufficient to exactly realize the spin-$2S$ AKLT state. 
To provide a sense of how close one can approach a spin-$2S$ AKLT state, in Fig.~\ref{fig:spinchain1}(c) we plot the maximum achievable string order parameter $[\mathcal{O}^{(z)}_{\rm string}(\varphi)]_{\rm max}$ by optimizing the detuning parameters $\Delta_e/\chi$ and $\Delta_b/\chi$ and the angle $\varphi$.
Note that, if one considers $f_J$ as arbitrary parameters, spin-$2S$ AKLT states have the largest string order parameter among all the states of the form of Eq.~(\ref{eq:vbs}).
We find that for small spin $S$, $[\mathcal{O}^{(z)}_{\rm string}(\varphi)]_{\rm max}$ is fairly close to the value of spin-$2S$ AKLT states, and it saturates to roughly a factor of $2$ smaller as we increase $S$.

\section{Conclusion and outlook}

We proposed a reconfigurable dissipative platform capable of stabilizing a broad class of exactly solvable pure entangled states between multiple spin ensembles. 
The scheme employs a single collective decay process and permutation-symmetry-breaking Hamiltonian terms, which is easily tunable via the detuning pattern of Rabi drives.
This general approach opens exciting new prospects for robust quantum sensing and many-body state engineering, including Heisenberg-limited differential and curvature sensing, and the stabilization of 1D SPT states such as the spin-1 AKLT state. 

The results presented here open several avenues for further exploration. 
On the metrological side, distributed quantum sensing \cite{Zhang2021} represents a particularly appealing application. 
While explicitly analyzed for two- and four-ensemble configurations, our approach can be scaled to larger networks, potentially enabling spatially resolved field detection and improved sensitivity scaling with system size.
On the state-engineering side, it is intriguing to explore dissipative stabilization schemes for more complex many-body states.
For example, our approach suggests a natural route toward realizing higher-spin AKLT states by further increasing the tunability of the setup by adding additional Hamiltonian terms that have the form of rank-$1$ spherical tensors.
Another interesting direction is to increase complexity by having multiple spin ensembles now with different sizes (see Ref.~\onlinecite{Mivehvar2024} for a related mean-field study).
In addition, the stabilized states in our scheme bear a clear resemblance to quantum spin liquids \cite{Zhou2017}, both consisting of superpositions of singlets with zero total angular momentum. 
This analogy suggests that our framework might be extendable to the dissipative preparation of spin-liquid states.

\begin{acknowledgments}
We thank Ruben Verresen for many useful discussions.
This work was primarily supported by the DOE Q-NEXT Center (Grant No.~DOE 1F-60579). A. A. C. acknowledges support from the Simons Foundation (Grant No. 669487, A. A. C.).
\end{acknowledgments}

\appendix
\section{Analytic calculation in the total angular momentum basis}
\label{sec:analytic}

Here we provide detailed analytical calculations in the total angular momentum basis for the following quantities.
\begin{equation}
    \begin{aligned}
    B_{J,m} &= \langle J+1,m|\hat{S}^{z}_1-\hat{S}^{z}_2|J,m\rangle, \\
    C_{J,m} &= \langle J+1,m|\hat{S}_1^{+}\hat{S}_2^{-}-\hat{S}_1^{-}\hat{S}_2^{+}|J,m\rangle, \\
    B_{J,m}^{+} &= \langle J+1,m+1|\hat{S}^{+}_1-\hat{S}^{+}_2|J,m\rangle, \\
    B_{J,m}^{-} &= \langle J+1,m-1|\hat{S}^{-}_1-\hat{S}^{-}_2|J,m\rangle.
    \end{aligned}
    \label{eq:eva}
\end{equation}

We consider spherical tensors \cite{Brink1968} of rank $k$, which are a set of $2k+1$ operators written as $\hat{T}^{(k)}_{q}$, with $q=-k,-k+1,\cdots, k$. These operators transform under rotation with exactly the same matrix of coefficients as angular momentum eigenstates $|k,q\rangle$.
Notice that $\hat{S}^{z}_1$, $-\hat{S}^{+}_1/\sqrt{2}$, $\hat{S}^{-}_1/\sqrt{2}$ are $k=1$ spherical tensors acting on subsystem 1 (with $q=0,1,-1$, respectively), $\hat{S}^{z}_2$, $-\hat{S}^{+}_2/\sqrt{2}$, $\hat{S}^{-}_2/\sqrt{2}$ are $k=1$ spherical tensors acting on subsystem 2 (with $q=0,1,-1$, respectively).
Therefore, we can conclude that $\hat{S}^{z}_1-\hat{S}^{z}_2$ is a spherical tensor with $k=1$ and $q=0$, $-(\hat{S}^{+}_1-\hat{S}^{+}_2)/\sqrt{2}$ is a spherical tensor with $k=1$ and $q=1$, $(\hat{S}^{-}_1-\hat{S}^{-}_2)/\sqrt{2}$ is a spherical tensor with $k=1$ and $q=-1$.

For $\hat{S}_1^{+}\hat{S}_2^{-}-\hat{S}_1^{-}\hat{S}_2^{+}$, we consider the combination of spherical tensors,
\begin{equation}
    \hat{T}^{(k)}_q = \sum_{q_1q_2}\langle k_1q_1;k_2q_2|kq\rangle\hat{U}^{(k_1)}_{q_1}\hat{V}^{(k_2)}_{q_2},
    \label{eq:a2}
\end{equation}
where $\hat{U}^{(k_1)}_{q_1}$ and $\hat{V}^{(k_2)}_{q_2}$ are arbitrary spherical tensors. This equation leads to
\begin{equation}
    \hat{T}^{(1)}_0(\hat{\bf S}_1\hat{\bf S}_2) = -\frac{1}{2\sqrt{2}}(\hat{S}_1^{+}\hat{S}_2^{-}-\hat{S}_1^{-}\hat{S}_2^{+}).
    \label{eq:comb}
\end{equation}
Therefore, $-(\hat{S}_1^{+}\hat{S}_2^{-}-\hat{S}_1^{-}\hat{S}_2^{+})/(2\sqrt{2})$ is a spherical tensor with $k=1$ and $q=0$.

The Wigner-Eckart theorem for the matrix elements of spherical tensors is given by \cite{Brink1968}
\begin{equation}
    \langle Jm|\hat{T}^{(k)}_{q}|J'm'\rangle = (-1)^{2k}\langle J'm';kq|Jm\rangle\langle J||\hat{\bf T}^{(k)}||J'\rangle,
    \label{eq:wignereckart}
\end{equation}
where $\langle J||\hat{\bf T}^{(k)}||J'\rangle$ is the reduced matrix element. 
For a product of spherical tensors in the form of Eq.~(\ref{eq:a2}), if $\hat{U}^{(k_1)}_{q_1}$ acts on subsystem 1, and $\hat{V}^{(k_2)}_{q_2}$ acts on subsystem 2, the relation between the reduced matrix elements are given by \cite{Brink1968}
\begin{equation}
    \begin{aligned}
    \langle J||\hat{\bf T}^{(k)}||J'\rangle &= \sqrt{(2J'+1)(2k+1)(2j_1+1)(2j_2+1)}\\
    &\times\begin{Bmatrix}
    J & J' & k\\
    j_1 & j'_1 & k_1\\
    j_2 & j'_2 & k_2\\
    \end{Bmatrix} \langle j_1||\hat{\bf U}^{(k_1)}||j'_1\rangle\langle j_2||\hat{\bf V}^{(k_2)}||j'_2\rangle,
    \end{aligned}
    \label{eq:reduced}
\end{equation}
where the curly bracket with $9$ elements is the so-called Wigner's 9-j symbol. A 9-j symbol is invariant under even permutations of rows or columns, while odd permutations yield a phase factor $(-1)^{P}$, where $P$ is the sum of all the $9$ elements.
This symmetry property of the 9-j symbol as well as the Clebsch-Gordan coefficient $\langle J'm';kq|Jm\rangle$ ensures that Eq.~(\ref{eq:eva}) and their complex conjugates are all the non-zero matrix elements in the total angular momentum basis.

Plugging in Eq.~(\ref{eq:wignereckart}) and Eq.~(\ref{eq:reduced}), we have
\begin{equation}
    \begin{aligned}
    B_{J,m} = 2&\times \langle J,m;1,0|J+1,m\rangle (2S+1)\sqrt{3(2J+1)}\\
    &\times\begin{Bmatrix}
    J+1 & J & 1\\
    S & S & 1\\
    S & S & 0\\
    \end{Bmatrix} \langle S||\hat{\bf S}||S\rangle.
    \end{aligned}
\end{equation}
The factor of $2$ is generated by odd permutation of the 9-j symbol, making $-\langle J+1,m|\hat{S}^z_2|J,m\rangle=\langle J+1,m|\hat{S}^z_1|J,m\rangle$. 
Similarly, we have
\begin{equation}
    \begin{aligned}
    C_{J,m} = -2\sqrt{2}&\times \langle J,m;1,0|J+1,m\rangle (2S+1)\sqrt{3(2J+1)}\\
    &\times\begin{Bmatrix}
    J+1 & J & 1\\
    S & S & 1\\
    S & S & 1\\
    \end{Bmatrix} \langle S||\hat{\bf S}||S\rangle\langle S||\hat{\bf S}||S\rangle,\\
    \end{aligned}
\end{equation}
where the factor $-2\sqrt{2}$ is generated by Eq.~(\ref{eq:comb}). Using $\langle S||\hat{\bf S}||S\rangle=\sqrt{S(S+1)}$, and the following property of 9-j symbols \cite{Brink1968},
\begin{equation}
    \begin{Bmatrix}
    J+1 & J & 1\\
    S & S & 1\\
    S & S & 1\\
    \end{Bmatrix} = \frac{J+1}{\sqrt{2S(S+1)}}\begin{Bmatrix}
    J+1 & J & 1\\
    S & S & 1\\
    S & S & 0\\
    \end{Bmatrix},
\end{equation}
we have
\begin{equation}
    C_{J,m} = -(J+1)B_{J,m}.
\end{equation}
This result is directly relevant for the existence of the SU(2)-symmetric unitary transformation in Eq.~(\ref{eq:gate}). Since $B_{J,m}$ and $C_{J,m}$ are both matrix elements of spherical tensors with $k=1$ and $q=0$, they share the same $m$ dependence due to the Wigner-Eckart theorem. This is the reason why the phases $\theta_J$ in Eq.~(\ref{eq:gate}) have no $m$ dependence.   

One can further simplify the formula of $B_{J,m}$ using analytical expressions of Clebsch-Gordan coefficients and 9-j symbols, leading to
\begin{equation}
    B_{J,m} = \sqrt{\frac{(2S-J)(2S+J+2)(J+m+1)(J-m+1)}{(2J+1)(2J+3)}}.
\end{equation}
Following the same procedure, we can also provide analytical calculations for $B_{Jm}^{+}$ and $B_{Jm}^{-}$,
\begin{equation}
    B_{J,m}^{+} = -\sqrt{\frac{(2S-J)(2S+J+2)(J+m+1)(J+m+2)}{(2J+1)(2J+3)}},
\end{equation}
\begin{equation}
    B_{J,m}^{-} = \sqrt{\frac{(2S-J)(2S+J+2)(J-m+1)(J-m+2)}{(2J+1)(2J+3)}}.
\end{equation}

\section{Relaxation time scale}
\label{sec:relax}

\subsection{Lower bound for two spin ensembles}
Here we derive a lower bound of the relaxation time scale to a pure steady state $|\psi_{\rm ss}^{(L)}\rangle$ based on Refs.~\onlinecite{Pocklington2024,Pocklington2025}. 
We define relaxation time $t_{\rm ss}$ such that the fidelity with the steady state $F(t)=\mathrm{tr}\Big(\hat{\rho}_{\rm ss}\,\hat{\rho}(t)\Big)$ satisfies $F(t)>1-\epsilon$ for $t>t_{\rm ss}$, where $\epsilon$ is a small positive number, and $\hat{\rho}_{\rm ss} = |\psi_{\rm ss}^{(L)}\rangle\langle\psi_{\rm ss}^{(L)}|$.
A lower bound of $t_{\rm ss}$ is given as follows,
\begin{equation}
    t_{\rm ss} \geq \frac{1-\epsilon-F(0)}{\max_t |\partial_t F(t)|}.
\end{equation}
We now provide an upper bound for $\max_t |\partial_t F(t)|$. Notice that
\begin{equation}
    \begin{aligned}
    |\partial_t F(t)| &= \Big|\mathrm{tr}\Big(\hat{\rho}_{\rm ss}\partial_t\hat{\rho}(t)\Big)\Big| = \Big|\mathrm{tr}\Big(\hat{\rho}_{\rm ss}\mathcal{L}\hat{\rho}(t)\Big)\Big| \\
    &= \Big|\mathrm{tr}\Big((\mathcal{L}^{\dag}\hat{\rho}_{\rm ss})\hat{\rho}(t)\Big)\Big|,
    \end{aligned}
\end{equation}
where $\mathcal{L}$ is the Lindbladian defined as $\mathcal{L}(\cdot) = -i[\hat{H},(\cdot)] + \sum_{\mu}\Big(\hat{K}_{\mu}(\cdot)\hat{K}^{\dag}_{\mu} - \{\hat{K}^{\dag}_{\mu}\hat{K}_{\mu},(\cdot)\}/2\Big)$, and $\mathcal{L}^{\dag}$ is the adjoint Lindbladian defined as $\mathcal{L}^{\dag}(\cdot) = i[\hat{H},(\cdot)] + \sum_{\mu}\Big(\hat{K}^{\dag}_{\mu}(\cdot)\hat{K}_{\mu} - \{\hat{K}^{\dag}_{\mu}\hat{K}_{\mu},(\cdot)\}/2\Big)$, with $\hat{K}_{\mu}$ Lindblad jump operators. 
We now apply the Cauchy-Schwarz inequality for operators, $|\mathrm{tr}(\hat{A}^{\dag}\hat{B})|^2\leq \mathrm{tr}(\hat{A}^{\dag}\hat{A})\mathrm{tr}(\hat{B}^{\dag}\hat{B})$, leading to
\begin{equation}
    \begin{aligned}
    |\partial_t F(t)| &\leq \sqrt{\mathrm{tr}\Big((\mathcal{L}^{\dag}\hat{\rho}_{\rm ss})^{\dag}(\mathcal{L}^{\dag}\hat{\rho}_{\rm ss})\Big)\mathrm{tr}\Big(\hat{\rho}_t^2\Big)}\\
    &\leq \sqrt{\mathrm{tr}\Big((\mathcal{L}^{\dag}\hat{\rho}_{\rm ss})^{\dag}(\mathcal{L}^{\dag}\hat{\rho}_{\rm ss})\Big)}.
    \end{aligned}
\end{equation}
Using the fact that a pure steady state satisfies $[\hat{H},\hat{\rho}_{\rm ss}] \equiv \hat{L}_{\mu}\hat{\rho}_{\rm ss}=0$, we have $\mathcal{L}^{\dag}\hat{\rho}_{\rm ss} = \sum_{\mu}\hat{K}^{\dag}_{\mu}\hat{\rho}_{\rm ss}\hat{K}_{\mu}$, leading to 
\begin{equation}
    |\partial_t F(t)|\; \leq \sqrt{\sum_{\mu\nu}|\langle\psi_{\rm ss}^{(L)}|[\hat{K}_{\mu},\hat{K}_{\nu}^{\dag}]|\psi_{\rm ss}^{(L)}\rangle|^2}.
\end{equation}
In our case with a single jump operator $\sqrt{\Gamma}\sum_j\hat{S}^{-}_j$, we have
\begin{equation}
    \Gamma t_{\rm ss}\geq \frac{1-\epsilon-F(0)}{\Big|2\langle\sum_j\hat{S}^z_j\rangle_{\rm ss}\Big|}.
\end{equation}

In principle the lower bound derived above can be applied to any number of spin ensembles.
For two spin ensembles ($L=2$), this lower bound is relatively tight, as we shown in Fig.~\ref{fig:twomodesqueezing}(b), the optimal $t_{\rm ss}$ obtained numerically is only a constant factor away.
In this case, we can further relate the lower bound of $t_{\rm ss}$ to the steady-state spin squeezing, since $\langle\hat{S}_1^{z}+\hat{S}_2^{z}\rangle_{\rm ss} = \sum_{J=0}^{2S} |c_{J,-J}|^2 (-J)$, and $\mathrm{Var}(\hat{S}^y_1+\hat{S}^y_2)_{\rm ss} = \frac{1}{2}\sum_{J=0}^{2S} |c_{J,-J}|^2 J$.
So we have
\begin{equation}
    \Gamma t_{\rm ss} \geq \frac{1-\epsilon-F(0)}{4\mathrm{Var}(\hat{S}^y_1+\hat{S}^y_2)_{\rm ss}}.
\end{equation}

\begin{figure}[t]
    \centering
    \includegraphics[width=1.0\columnwidth]{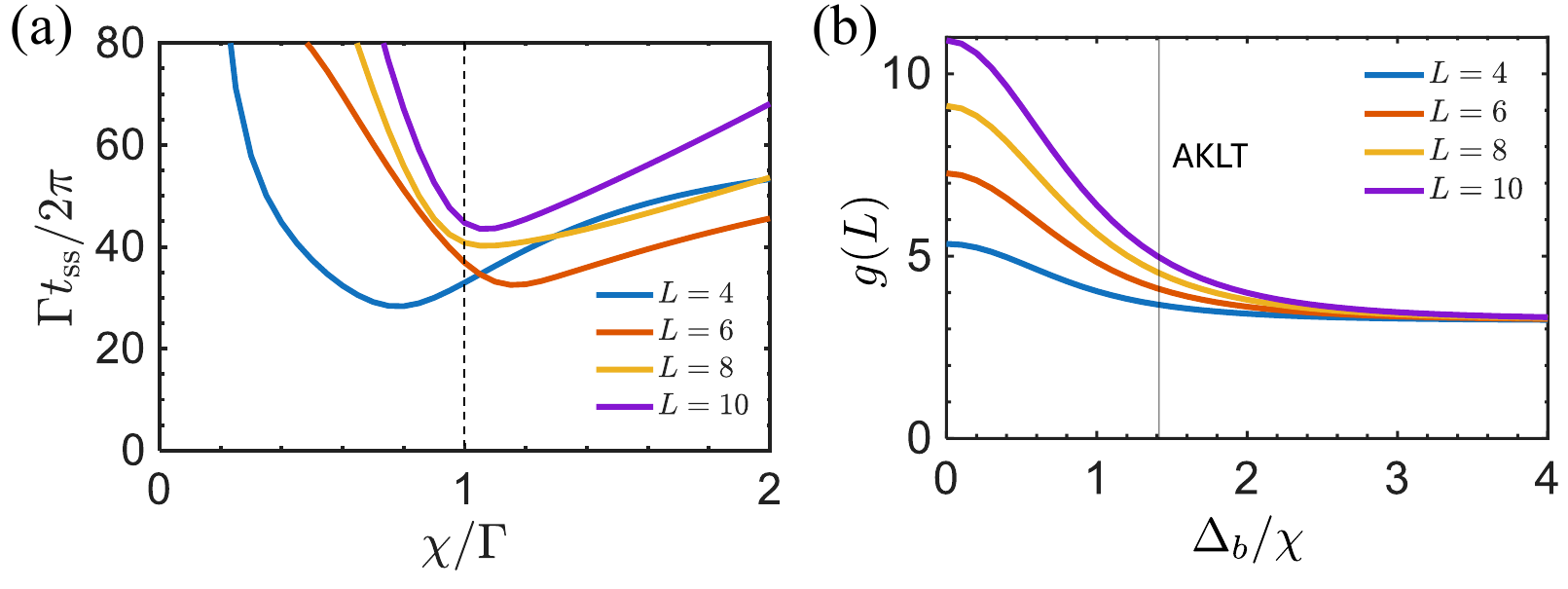}
    \caption{(a) Relaxation time scale $t_{\rm ss}$ for a spin-$1/2$ chain as a function of $\chi/\Gamma$ and system size $L$. We consider the initial state as all spins pointing down, set detunings to $\Delta_e=0$, $\Delta_b/\chi=\sqrt{2}$ (the AKLT case), set Rabi drives to $\Omega/\chi=3$. Optimal $t_{\rm ss}$ can be achieved near $\chi=\Gamma$ (unidirectional propagation) as we increase $L$. (b) At $\chi=\Gamma$, one can approximate the relaxation time scale by $\Gamma t_{\rm ss}/(2\pi)\approx (\Omega/\chi)^2 g(L)$. Except for the regime $\Delta_b\ll\chi$, $g(L)$ depends weakly on $L$.}
    \label{fig:respon1}
\end{figure}

\subsection{Further discussions in 1D chain}
We have already shown the relaxation time scale $t_{\rm ss}$ for the AKLT state (spin-$1/2$ case) in Fig.~\ref{fig:spinchain}(c). 
Here we would like to provide further numerical calculations for the relaxation time scale in a 1D chain.
In Fig.~\ref{fig:respon1}(a), we show that the relaxation time scale $t_{\rm ss}$ under the AKLT parameters ($\Delta_e=0,\Delta_b=\sqrt{2}\chi$) as a function of the ratio $\chi/\Gamma$.
As we increase the system size $L$, the optimal $t_{\rm ss}$ can be achieved near $\chi=\Gamma$ (unidirectional propagation). 
In this case, we expect sequential relaxation, i.e. sub-ensemble $k$ will relax in a faster rate compare to sub-ensemble $l$ if $l>k$, because the information only flows in a single direction.

However, the existence of sequential relaxation in dissipative dynamics does not mean that the relaxation time scale $t_{\rm ss}$ will simply proportional to system size $L$. 
As shown in Fig.~\ref{fig:respon1}(b), we keep the same detuning pattern (see Fig.~\ref{fig:spinchain}(a)) as the AKLT case and vary the bulk detuning $\Delta_b$ (we set the edge detuning $\Delta_e=0$).
At $\chi=\Gamma$, the relaxation time scale $t_{\rm ss}$ (an upper bound for the optimal $t_{\rm ss}$) can be approximated by
\begin{equation}
    \Gamma t_{\rm ss}/(2\pi) \approx \bigg(\frac{\Omega}{\chi}\bigg)^2g(L).
\end{equation}
In the regime $\Delta_b\ll \chi$, where the chiral coupling dominates, we find $g(L)$ depends linearly in $L$. 
While in the regime $\Delta_b\gg \chi$, where the chiral coupling is not playing the dominant role, we find $g(L)$ is independent of $L$.
The AKLT case lies in between these two regimes and $g(L)$ depends weakly in $L$.
We believe this occurs due to the collective nature of the chiral spin exchange coupling, which means that the spins can interact regardless of their distance.

It is worth to mention that the relaxation time scale is nearly the same before and after detuning permutation in a 1D chain, up to transient physics due to different overlap between the initial state and the Liouvillian eigenstates. This is because the detuning permutation is equivalent to a unitary transformation to the Lindblad master equation (see Eq.~(\ref{eq:gate})) and the Liouvillian spectrum is conserved.

\section{Analytic calculation for four spin ensembles}
\label{sec:largedrive}

In the main text, we discuss quantum sensing using four spin ensembles with detunings $\delta_{1}=\Delta_A/2$, $\delta_{2}=\Delta_B/2$,  $\delta_{3}=-\Delta_B/2$, $\delta_{4}=-\Delta_A/2$. Here we analytically calculate the QFI matrix in the large Rabi drive limit ($\Omega\rightarrow \infty$).
In this case, the steady-state solution becomes
\begin{equation}
    |\psi_{\rm ss}^{(4)}\rangle = \hat{U}_{43}\hat{U}_{23}|j_{12}=0,j_{43}=0;j_{\rm tot}=0,m_{\rm tot}=0\rangle,
\end{equation}
where $j_{ab}$ means the combined angular momentum for ensemble $a$ and $b$, $a,b\in\{1,2,3,4\}$. The sequential order of $a$ and $b$ matters because Clebsch-Gordan coefficients might change sign if swapping $a$ and $b$. $j_{\rm tot}$ and $m_{\rm tot}$ are the total angular momentum and magnetic number for the four ensembles. The unitary transformations are given by $\hat{U}_{23}=\exp\Big(i\sum_{jm}\theta_{j}|j,m\rangle\langle j,m|_{23} \Big)$, $\hat{U}_{43}=\exp\Big(i\sum_{jm}\tilde{\theta}_{j}|j,m\rangle\langle j,m|_{43} \Big)$, with
\begin{equation}
    \begin{aligned}
    e^{i(\theta_{j+1}-\theta_{j})} = \frac{(\Delta_A+\Delta_B)/2 -i\chi (j+1)}{-(\Delta_A+\Delta_B)/2 -i\chi (j+1)}, \\
    e^{i(\tilde{\theta}_{j+1}-\tilde{\theta}_{j})} = \frac{(\Delta_A-\Delta_B)/2 -i\chi (j+1)}{-(\Delta_A-\Delta_B)/2 -i\chi (j+1)}.
    \end{aligned}
\end{equation}

Using the Wigner 9-j symbol \cite{Brink1968}, 
\begin{equation}
    \begin{aligned}
    &\langle j_{12},j_{43}; j_{\rm tot}, m_{\rm tot}|j_{14},j_{23};j_{\rm tot},m_{\rm tot}\rangle = \sqrt{(2j_{12}+1)}\\
    &\times\sqrt{(2j_{43}+1)(2j_{14}+1)(2j_{23}+1)}\begin{Bmatrix}
    j_1 & j_2 & j_{12}\\
    j_4 & j_3 & j_{43}\\
    j_{14} & j_{23} & j_{\rm tot}\\
    \end{Bmatrix},
    \end{aligned}
\end{equation}
one can switch between different angular momentum bases ($|j_{12},j_{43};j_{\rm tot},m_{\rm tot}\rangle\leftrightarrow |j_{14},j_{23};j_{\rm tot},m_{\rm tot}\rangle$) of the four ensembles.
Note that, since $\hat{U}_{43}$ is diagonal in the first basis and $\hat{U}_{23}$ is diagonal in the second basis, we obtain
\begin{equation}
    \begin{aligned}
    |\psi_{\rm ss}\rangle = &\frac{1}{2S+1}\sum_{j'=0}^{2S} \sqrt{2j'+1}\;g_{j'}(\Delta_A,\Delta_B,\chi)\\
    &\times|j_{12}=j',j_{43}=j';j_{\rm tot}=0,m_{\rm tot}=0\rangle,
    \end{aligned}
    \label{eq:fourss}
\end{equation}
where
\begin{equation}
    \begin{aligned}
    &g_{j'}(\Delta_A,\Delta_B,\chi) \\
    &= \sum_{j=0}^{2S} (-1)^{2S+j+j'}e^{i\theta_j}e^{i\tilde{\theta}_{j'}}(2j+1) \begin{Bmatrix}
    S & S & j'\\
    S & S & j\\
    \end{Bmatrix}.
    \end{aligned}
    \label{eq:gjp}
\end{equation}
Here the curly bracket with $6$ elements is the so-called Wigner 6-j symbol.

We now calculate the QFI matrix for operators $\hat{S}^z_l$ based on Eq.~(\ref{eq:fourss}),
\begin{equation}
    \mathcal{F}_{kl} = 4\mathrm{Cov}(\hat{S}^z_k,\hat{S}^z_l)_{\rm ss}.
\end{equation}
Notice that $|\psi_{\rm ss}^{(4)}\rangle$ is constructed by linear combination of different $j_{\rm tot}=0$ states, i.e., $|j_{12}=j',j_{43}=j';j_{\rm tot}=0,m_{\rm tot}=0\rangle$. 
So we will still stay in $j_{\rm tot}=0$ subspace if we apply the spin operators $\hat{S}^z_1+\hat{S}^z_2$ and $\hat{S}^z_3+\hat{S}^z_4$, while we will move out of $j_{\rm tot}=0$ subspace if we apply the spin operators $\hat{S}^z_1-\hat{S}^z_2$ and $\hat{S}^z_3-\hat{S}^z_4$.
This allows us to simplify the QFI matrix into a block-diagonal form.
In the first block with spin operators $\hat{S}^z_1+\hat{S}^z_2$ and $\hat{S}^z_3+\hat{S}^z_4$, based on the fact that $(\hat{S}^z_1+\hat{S}^z_2+\hat{S}^z_3+\hat{S}^z_4)|\psi_{\rm ss}^{(4)}\rangle=0$, one can show that $\hat{S}^z_1+\hat{S}^z_2+\hat{S}^z_3+\hat{S}^z_4$ and $\hat{S}^z_1+\hat{S}^z_2-\hat{S}^z_3-\hat{S}^z_4$ form an orthogonal basis. 
In the second block with spin operators $\hat{S}^z_1-\hat{S}^z_2$ and $\hat{S}^z_3-\hat{S}^z_4$, based on the fact that $B_{J,m}=B_{J,-m}$, one can obtain $\mathrm{Var}(\hat{S}^z_1-\hat{S}^z_2)_{\rm ss}=\mathrm{Var}(\hat{S}^z_3-\hat{S}^z_4)_{\rm ss}$, which means that $\hat{S}^z_1-\hat{S}^z_2+\hat{S}^z_3-\hat{S}^z_4$ and $\hat{S}^z_1-\hat{S}^z_2-\hat{S}^z_3+\hat{S}^z_4$ form an orthogonal basis.
Therefore, the diagonalization of the QFI matrix in the large Rabi drive limit leads to 
\begin{equation}
    \mathcal{F}_{++++} \equiv 4\mathrm{Var}(\hat{S}^z_1+\hat{S}^z_2+\hat{S}^z_3+\hat{S}^z_4)_{\rm ss} = 0,
\end{equation}
\begin{equation}
    \begin{aligned}
    \mathcal{F}_{++--} &\equiv 4\mathrm{Var}(\hat{S}^z_1+\hat{S}^z_2-\hat{S}^z_3-\hat{S}^z_4)_{\rm ss}\\
    &= \frac{16}{3(2S+1)^2}\sum_{j'=0}^{2S}j'(j'+1)(2j'+1)|g_{j'}|^2,
    \end{aligned}
\end{equation}
\begin{equation}
    \begin{aligned}
    \mathcal{F}_{+-+-} &\equiv 4\mathrm{Var}(\hat{S}^z_1-\hat{S}^z_2+\hat{S}^z_3-\hat{S}^z_4)_{\rm ss}\\
    &= \frac{8}{3(2S+1)^2}\sum_{j'=0}^{2S}(2S-j')(2S+j'+2)(j'+1) \\
    &\times\Big|g_{j'}+g_{j'+1}\Big|^2,
    \end{aligned}
\end{equation}
\begin{equation}
    \begin{aligned}
    \mathcal{F}_{+--+} &\equiv 4\mathrm{Var}(\hat{S}^z_1-\hat{S}^z_2-\hat{S}^z_3+\hat{S}^z_4)_{\rm ss}\\
    &= \frac{8}{3(2S+1)^2}\sum_{j'=0}^{2S}(2S-j')(2S+j'+2)(j'+1)\\
    &\times \Big|g_{j'}-g_{j'+1}\Big|^2.
    \end{aligned}
\end{equation}
For the sake of brevity, we omitted the arguments of $g_{j'}$ defined in Eq.~\eqref{eq:gjp}.
One can also show that in the large Rabi drive limit,
\begin{equation}
    \mathcal{F}_{++++} + \mathcal{F}_{++--} + \mathcal{F}_{+-+-} +\mathcal{F}_{+--+} = \frac{64}{3}S(S+1).
\end{equation}
This result is based on the orthogonality relation for the Wigner 6-j symbol \cite{Brink1968}, 
\begin{equation}
    \begin{aligned}
    &\sum_{j_3}(2j_3+1)\begin{Bmatrix}
    j_1 & j_2 & j_3\\
    j_4 & j_5 & j_6\\
    \end{Bmatrix}\begin{Bmatrix}
    j_1 & j_2 & j_3\\
    j_4 & j_5 & j'_6\\
    \end{Bmatrix} \\
    &= \frac{\delta_{j_6j'_6}}{2j_6+1}\{j_1\; j_5\; j_6\}\{j_4\; j_2\; j_6\},
    \end{aligned}
    \label{eq:ortho}
\end{equation}
where $\{j_1\; j_5\; j_6\}$ is the triangular delta, which is equal to 1 when the triad $(j_1, j_5, j_6)$ satisfies the triangle conditions, and is zero otherwise.

\section{Analytic calculation for 1D spin chain}
\label{sec:largemps}

\subsection{MPS representation of 1D chain}
In the main text, we consider an even number $L$ of spin-$S$ ensembles placed in a 1D chain with detuning pattern $\delta_1=\Delta_e/2$, $\delta_L=-\Delta_e/2$ on the edges, and $\delta_{2k}=\Delta_b/2$, $\delta_{2k+1}=-\Delta_b/2$ in the bulk, with $k=1,2,\cdots,L/2-1$.
In this case, we can express the steady-state solution using a matrix product state (MPS), 
\begin{equation}
    |\psi_{\rm ss}^{(L)}\rangle = \sum_{\mathbf{m}}(\mathbf{v}_{\rm left}^{m_1})^{T}\bigg(\prod_{k=1}^{L/2-1} \mathbf{A}^{m_{2k}m_{2k+1}} \bigg) \mathbf{v}_{\rm right}^{m_L}|\mathbf{m}\rangle.
    \label{eq:d1}
\end{equation}
Here, $|\mathbf{m}\rangle\equiv\otimes_{j=1}^L|S,m_j\rangle$, $\mathbf{A}^{m_{2k}m_{2k+1}}$ are the matrices for the translationally invariant structure with bond dimension $2S+1$, which is a unit cell with two spin ensembles. $\mathbf{v}_{\rm left}^{m_1}$ and $\mathbf{v}_{\rm right}^{m_L}$ are boundary vectors.
In the following, we calculate $\mathbf{v}_{\rm left}^{m_1}$, $\mathbf{v}_{\rm right}^{m_L}$ and $\mathbf{A}^{m_{2k}m_{2k+1}}$ analytically in the large-Rabi-drive limit ($\Omega\rightarrow\infty$).

In this case, we can simplify Eq.~(\ref{eq:steadyfinal}) into a sequential quantum circuit, 
\begin{equation}
    |\psi_{\rm ss}^{(L)}\rangle = \prod_{p=1}^{L/2-1}\hat{U}_{2p+1,2p+2}\hat{U}_{2p,2p+1} \bigg[\bigotimes_{k=1}^{L/2} |\psi_{\rm ss}^{(2)}\rangle_{2k-1,2k}\bigg],
    \label{eq:seq}
\end{equation}
where, in the large-drive limit, the two-ensemble steady state is given by $|\psi_{\rm ss}^{(2)}\rangle_{2k-1,2k}=|J=0,m=0\rangle_{2k-1,2k}$. In Eq.~(\ref{eq:seq}), we alternately apply the following two quantum gates, $\hat{U}_{2p,2p+1} = \exp\Big(i\sum_{jm}\theta_j|j,m\rangle\langle j,m|_{2p,2p+1} \Big)$, $\hat{U}_{2p+1,2p+2} = \exp\Big(i\sum_{jm}\tilde{\theta}_j|j,m\rangle\langle j,m|_{2p+1,2p+2} \Big)$, with
\begin{equation}
    \begin{gathered}
    e^{i(\theta_{j+1}-\theta_j)} = \frac{(\Delta_{e}+\Delta_{b})/2 -i\chi (j+1)}{-(\Delta_{e}+\Delta_{b})/2 -i\chi (j+1)},\\
    e^{i(\tilde{\theta}_{j+1}-\tilde{\theta}_j)} = \frac{(\Delta_{e}-\Delta_{b})/2 -i\chi (j+1)}{-(\Delta_{e}-\Delta_{b})/2 -i\chi (j+1)}.
    \end{gathered}
\end{equation}

To derive the the column vector $\mathbf{v}_{\rm left}^{m_1}$, we consider the decomposition
\begin{equation}
    \begin{aligned}
    &|J=0,m=0\rangle_{12} \\
    &= \frac{1}{\sqrt{2S+1}}\sum_{m_1} (-1)^{S-m_1} |S,m_1\rangle|S,-m_2\rangle.
    \end{aligned}
\end{equation}
Interpreting the first spin as physical dimension, and the second spin as bond dimension, we have 
\begin{equation}
    [\mathbf{v}_{\rm left}^{m_1}]_{\alpha} = \frac{(-1)^{S-m_1}}{\sqrt{2S+1}}\delta_{\alpha,-m_1},
\end{equation}
where the subscripts $\alpha = S,S-1,\cdots -S$ are indices for the bond dimension.
Note that a boundary vector only contain a single spin ensemble, in contrast to the two-ensemble unit cell in the bulk.

For the column vector $\mathbf{v}_{\rm right}^{m_L}$, we consider the row dimension of the identity matrix as bond dimension, and the column dimension of the identity matrix as physical dimension,
\begin{equation}
    [\mathbf{v}_{\rm right}^{m_L}]_{\alpha} = \delta_{\alpha,m_L}.
\end{equation}

For the matrix $\mathbf{A}^{m_{2k}m_{2k+1}}$, we transform to the total angular momentum basis first,
\begin{equation}
     \mathbf{A}^{Jm} = \sum_{m_{2k}m_{2k+1}} \langle S,m_{2k}; S,m_{2k+1}|J,m\rangle \mathbf{A}^{m_{2k}m_{2k+1}}.
\end{equation}
Then we perform analytical calculation for $\mathbf{A}^{Jm}$. One can either transform back to obtain an analytical expression for $\mathbf{A}^{m_{2k}m_{2k+1}}$, or simply change Eq.~\ref{eq:d1} to the $|J,m\rangle$ basis for each two-ensemble unit cell. The procedure for $\mathbf{A}^{Jm}$ calculation is as follow.
\begin{equation}
    \begin{aligned}
    [\mathbf{A}^{Jm}]_{\alpha\beta} = &\Big(\langle j_{ab}=J,m_{ab}=m|\langle S,m_c=\beta|\Big)\hat{U}_{bc}\hat{U}_{ab}\\
    &\cdot\Big(|S,m_a=\alpha\rangle| j_{bc}=0,m_{bc}=0\rangle\Big),
    \end{aligned}
\end{equation}
where we relabeled $2k\rightarrow a$, $2k+1\rightarrow b$, $2k+2\rightarrow c$.
Considering the definition of the Wigner 6-j symbol \cite{Brink1968}, 
\begin{equation}
    \begin{aligned}
    &|j_{ab},j_c;j_{\rm tot},m_{\rm tot}\rangle = \sum_{j_{bc}} |j_a,j_{bc};j_{\rm tot},m_{\rm tot}\rangle \begin{Bmatrix}
        j_a & j_b & j_{ab}\\
        j_c & j_{\rm tot} & j_{bc}\\
    \end{Bmatrix}\\
    &\times (-1)^{j_a+j_b+j_c+j_{\rm tot}}\sqrt{(2j_{ab}+1)(2j_{bc}+1)},
    \end{aligned}
\end{equation}
one can switch between different angular-momentum bases ($|j_{ab},j_c;j_{\rm tot},m_{\rm tot}\rangle\leftrightarrow |j_a,j_{bc};j_{\rm tot},m_{\rm tot}\rangle$) of the three ensembles.
Note that $\hat{U}_{ab}$ is diagonal in the first basis, and $\hat{U}_{bc}$ is diagonal in the second basis.
We thus obtain
\begin{equation}
    \begin{aligned}
    [\mathbf{A}^{Jm}]_{\alpha\beta} &= \frac{(-1)^{S+\beta}}{\sqrt{2S+1}}\langle S,\alpha;S,-\beta|J,m\rangle f_J(\Delta_e,\Delta_b,\chi),
    \end{aligned}
    \label{eq:mpss}
\end{equation}
where
\begin{equation}
    \begin{aligned}
    f_J(\Delta_e,\Delta_b,\chi) = \sum_{j=0}^{2S}\sum_{j'=0}^{2S}&(-1)^{j-J}e^{i\theta_j}e^{i\tilde{\theta}_j'}(2j+1)(2j'+1)\\
    &\times\begin{Bmatrix}
        S & S & j\\
        S & S & j'\\
    \end{Bmatrix}\begin{Bmatrix}
        S & S & J\\
        S & S & j'\\
    \end{Bmatrix}.
    \end{aligned}
\end{equation}
The orthogonality relation for $f_J$ is based on
Eq.~(\ref{eq:ortho}), and we have
\begin{equation}
    \sum_J (2J+1)|f_J|^2 = (2S+1)^2.
\end{equation}
One can also conclude that the matrix $\mathbf{A}^{Jm}$ is a spherical tensor $\hat{T}^{(k)}_q$ with $k=J$ and $q=m$ based on the Wigner-Eckart theorem, which is acting on the bond dimension (spin-$S$ particle), i.e., $\sum_{\alpha,\beta} [\mathbf{A}^{Jm}]_{\alpha\beta}|S,\alpha\rangle\langle S,\beta|$. The same structure is used in Ref.~\onlinecite{Tu2009} to define the valence bond solid (VBS) state.

\subsection{String order parameter}
Similar to the calculation in Refs.~\onlinecite{Oshikawa1992,Totsuka1995}, here we would like to evaluate the string order parameter for Eq.~(\ref{eq:mpss}) in an infinite 1D chain, 
\begin{equation}
    \begin{aligned}
    \mathcal{O}^{(z)}_{\rm string}(\varphi) = \lim_{|l-k|\rightarrow\infty}\Big\langle &(\hat{S}^z_{2k}+\hat{S}^z_{2k+1})e^{i\varphi\sum_{q=k}^{l-1}(\hat{S}^z_{2q}+\hat{S}^z_{2q+1})}\\
    &\times(\hat{S}^z_{2l}+\hat{S}^z_{2l+1})\Big\rangle.
    \end{aligned}
\end{equation}

The general procedure is to construct a transfer matrix with operator $\hat{O}$ based on an infinite MPS \cite{Cirac2021},
\begin{equation}
    \mathcal{T}_{\hat{O}}(\cdot) = \sum_{Jm}\sum_{J'm'}\langle J',m'|\hat{O}|J,m\rangle \mathbf{A}^{Jm}(\cdot)(\mathbf{A}^{Jm})^{\dag}.
\end{equation}
Here we focus on three types of transfer matrices: 
\begin{itemize}
    \item Ordinary transfer matrix $\mathcal{T}_{\hat{I}}$. For a properly normalized MPS, the largest eigenvalue of $\mathcal{T}_{\hat{I}}$ should be $1$, with corresponding left and right eigenvectors labelled by $V_{\hat{I},L}$ and $V_{\hat{I},R}$. The left and right eigenvectors are normalized by $\mathrm{tr}(V_{\hat{I},L}^{\dag}V_{\hat{I},R})=1$.

    \item String transfer matrix $\mathcal{T}_{\hat{G}}$, with $\hat{G}=\exp\Big(i\varphi(\hat{S}^z_a+\hat{S}^z_b)\Big)$. We denote the largest eigenvalue of $\mathcal{T}_{\hat{G}}$ as $\lambda_{\hat{G}}$, with corresponding left and right eigenvectors labelled by $V_{\hat{G},L}$ and $V_{\hat{G},R}$. The left and right eigenvectors are normalized by $\mathrm{tr}(V_{\hat{G},L}^{\dag}V_{\hat{G},R})=1$.

    \item Endpoint transfer matrices $\mathcal{T}_{\hat{G}_L}$ and $\mathcal{T}_{\hat{G}_R}$, with $\hat{G}_L=(\hat{S}^z_a+\hat{S}^z_b)\exp\Big(i\varphi(\hat{S}^z_a+\hat{S}^z_b)\Big)$ and $\hat{G}_R=\hat{S}^z_a+\hat{S}^z_b$.
\end{itemize}
Using these transfer matrices, the string order parameter can be written as
\begin{equation}
    \mathcal{O}^{(z)}_{\rm string}(\varphi) = \lim_{|l-k|\rightarrow\infty} \mathrm{tr}\Big(V_{\hat{I},L}^{\dag}\mathcal{T}_{\hat{G}_L}\mathcal{T}_{\hat{G}}^{|l-k|-1}\mathcal{T}_{\hat{G}_R}V_{\hat{I},R}\Big).
\end{equation}
As $|l-k|\;\rightarrow\infty$, we only need to focus on the largest eigenvalue of $\mathcal{T}_{\hat{G}}$. If $|\lambda_{\hat{G}}|\;<1$, we have $\mathcal{O}^{(z)}_{\rm string}(\varphi)=0$. If $|\lambda_{\hat{G}}|\;=1$, we have
\begin{equation}
    \mathcal{O}^{(z)}_{\rm string}(\varphi) = \mathrm{tr}\Big(V_{\hat{I},L}^{\dag}\mathcal{T}_{\hat{G}_L}V_{\hat{G},R}\Big)\mathrm{tr}\Big(V_{\hat{G},L}^{\dag}\mathcal{T}_{\hat{G}_R}V_{\hat{I},R}\Big).
\end{equation}

We now estimate the left and right eigenvector of these transfer matrices based on the analytical form of $[\mathbf{A}^{Jm}]_{\alpha\beta}$. For $\mathcal{T}_{\hat{I}}$, one can prove that
\begin{equation}
    [V_{\hat{I},L}]_{\alpha\beta} = [V_{\hat{I},R}]_{\alpha\beta} = \frac{1}{\sqrt{2S+1}}\delta_{\alpha,\beta}.
\end{equation}
For $\mathcal{T}_{\hat{G}}$, one can prove that
\begin{equation}
    [V_{\hat{G},L}]_{\alpha\beta} = [V_{\hat{G},R}]_{\alpha\beta} = \frac{e^{i\alpha\varphi}}{\sqrt{2S+1}}\delta_{\alpha,\beta}.
\end{equation}
Using these results, we have
\begin{equation}
    \mathcal{O}^{(z)}_{\rm string}(\varphi) = \bigg|\sum_{Jm\beta\alpha} \frac{m e^{i\alpha\varphi}}{(2S+1)^2}\;|\langle S,\alpha;S,-\beta|J,m\rangle|^2\;  |f_J|^2\bigg|^2.
\end{equation}
Notice that
\begin{equation}
    \sum_{m\beta} m \;|\langle S,\alpha;S,-\beta|J,m\rangle|^2 = \frac{2J+1}{2S+1} \langle S,\alpha|\hat{J}^z|S,\alpha\rangle.
\end{equation}
Following the same procedure as in App.~\ref{sec:analytic} for a spherical tensor with $k=1$ and $q=0$, we have
\begin{equation}
    \langle S,\alpha|\hat{J}^z|S,\alpha\rangle = \alpha \frac{J(J+1)}{2S(S+1)}.
\end{equation}
So we can further simplify $\mathcal{O}^{(z)}_{\rm string}(\varphi)$ to
\begin{equation}
    \mathcal{O}^{(z)}_{\rm string}(\varphi) = \bigg(\sum_{J=0}^{2S}\frac{J(J+1)(2J+1)}{2S(S+1)(2S+1)^3}|f_J|^2\bigg)^2[h(\varphi)]^2,
\end{equation} 
where
\begin{equation}
    h(\varphi) = \sum_{\alpha=-S}^{S} \alpha \sin(\alpha\varphi).
\end{equation}
One can also perform the summation in $h(\varphi)$ analytically,
\begin{equation}
    h(\varphi) = \frac{(S+1)\sin(S\varphi)-S\sin\Big((S+1)\varphi\Big)}{2\sin^2(\varphi/2)}.
\end{equation}

\subsection{Correlation length}

\begin{figure}[t]
    \centering
    \includegraphics[width=1.0\columnwidth]{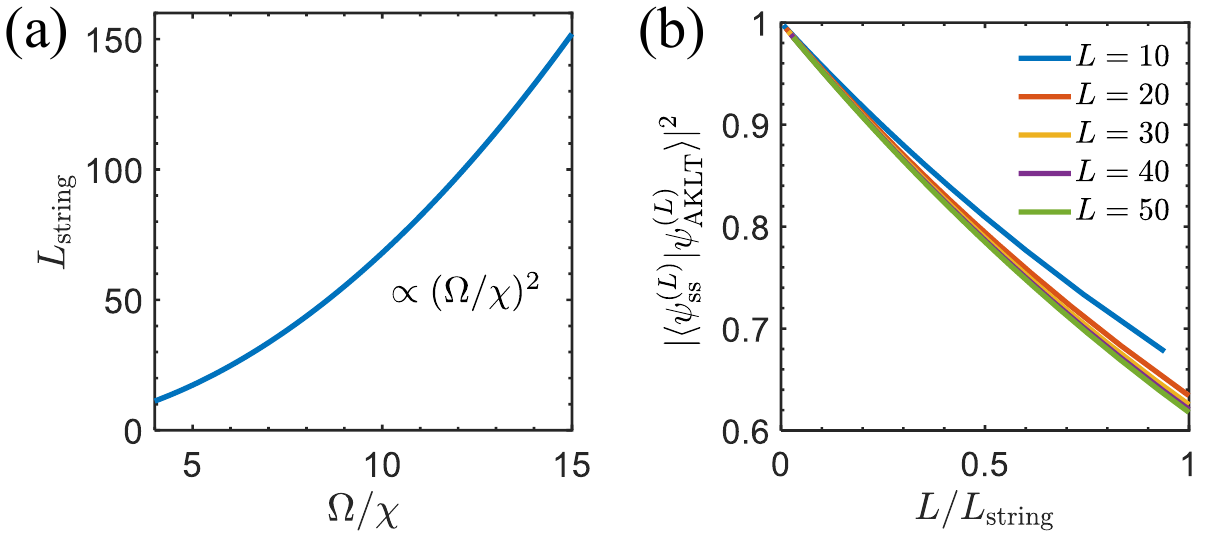}
    \caption{(a) String order correlation length $L_{\rm string}$ as a function of $\Omega/\chi$ for the case of $S=1/2$. We set $\Delta_e=0$ and $\Delta_b=\sqrt{2}\,\chi$ (parameters for spin-$1$ AKLT state in the large Rabi drive limit) for both (a) and (b). We also set the angle $\varphi=\pi$ in the string order parameter. (b) String order correlation length $L_{\rm string}$ as a measure of the fidelity between the steady-state solution and the spin-$1$ AKLT state.}
    \label{fig:corr}
\end{figure}

One can also evaluate the spin-spin correlation length $L_{\rm corr}$ (defined by Eq.~(\ref{eq:corr})) using the ordinary transfer matrix $\mathcal{T}_{\hat{I}}$. We sort the eigenvalues $\lambda_j$ of $\mathcal{T}_{\hat{I}}$ by their absolute values, $|\lambda_0|\;>|\lambda_1|\;\geq|\lambda_2|\;\geq \cdots$, with $j=0,1,\cdots,(2S+1)^2-1$. 
A properly normalized MPS should have $\lambda_0=1$.
Since we are focusing on the connected part of spin-spin correlations, $\langle \hat{S}^z_k\hat{S}^z_l\rangle - \langle \hat{S}^z_k\rangle\langle\hat{S}^z_l\rangle$, the contribution of $\lambda_0$ is canceled, and and the spin-spin correlation length is determined by the second largest eigenvalue $\lambda_1$,
\begin{equation}
    L_{\rm corr} = -\frac{2}{\ln|\lambda_1|},
    \label{eq:corr1}
\end{equation}
where the factor of $2$ in Eq.~(\ref{eq:corr1}) is to take account of the two-ensemble unit cell.
In the case of $S=1/2$, one can evaluate $\lambda_1$ analytically,
\begin{equation}
    \lambda_1 = \frac{|f_0|^2-|f_1|^2}{4}.
\end{equation}
In the left panel of Fig.~\ref{fig:spinchain1}(a), we calculate $L_{\rm corr}$ based on the procedure above. 

As we mention in the main text, in the case of a finite Rabi drive $\Omega$, we break the symmetry required for the SPT phase and the string order parameter vanishes in an infinite chain. In this case, one can define the string order correlation length $L_{\rm string}$ as follows,
\begin{equation}
    \begin{aligned}
        &\Big\langle(\hat{S}^z_{2k}+\hat{S}^z_{2k+1})e^{i\varphi\sum_{q=k}^{l-1}(\hat{S}^z_{2q}+\hat{S}^z_{2q+1})}(\hat{S}^z_{2l}+\hat{S}^z_{2l+1})\Big\rangle\\
        &\sim e^{-2|l-k|/L_{\rm string}},
    \end{aligned}
\end{equation}
where the factor of $2$ is to account for the two-ensemble unit cell. 
Similarly, we evaluate $L_{\rm string}$ using the string transfer matrix $\mathcal{T}_{\hat{G}}$. 
For finite Rabi drive $\Omega$, the largest eigenvalue $\lambda_{\hat{G}}$ has absolute value smaller than $1$, so $L_{\rm string}$ is given by
\begin{equation}
    L_{\rm string} = -\frac{2}{\ln|\lambda_{\hat{G}}|}.
\end{equation}

Here, we consider the case of $S=1/2$ and set $\varphi=\pi$. We also set $\Delta_e=0$ and $\Delta_b=\sqrt{2}\,\chi$ (parameters for spin-$1$ AKLT state in the large-Rabi-drive limit).
In Fig.~\ref{fig:corr}(a), a numerical calculation of $L_{\rm string}$ shows that $L_{\rm string}\propto (\Omega/\chi)^2$.
When $L_{\rm string}\gg L$, we effectively realize the SPT phase since we cannot tell the difference in a finite-size system.
One can also interpret $L/L_{\rm string}$ as a measure of the fidelity to the spin-$1$ AKLT state, as shown in Fig.~\ref{fig:corr}(b). 
As we increase $L$, we find that $L/L_{\rm string}$ and the fidelity collapse into a single line.
For example, $90\%$ fidelity corresponds to $L/L_{\rm string}\approx 0.2$.
Based on the discussions in the main text, we have the relaxation time scale $t_{\rm ss}\propto (\Omega/\chi)^2g(L)$, leading to  the AKLT relaxation time scale $t_{\rm AKLT}\propto Lg(L)$ if fixing the same fidelity to the spin-$1$ AKLT state. 

\section{Second-order cumulant expansion for two spin ensembles}
\label{sec:cumulant}

\begin{figure}[t]
    \centering
    \includegraphics[width=1.0\columnwidth]{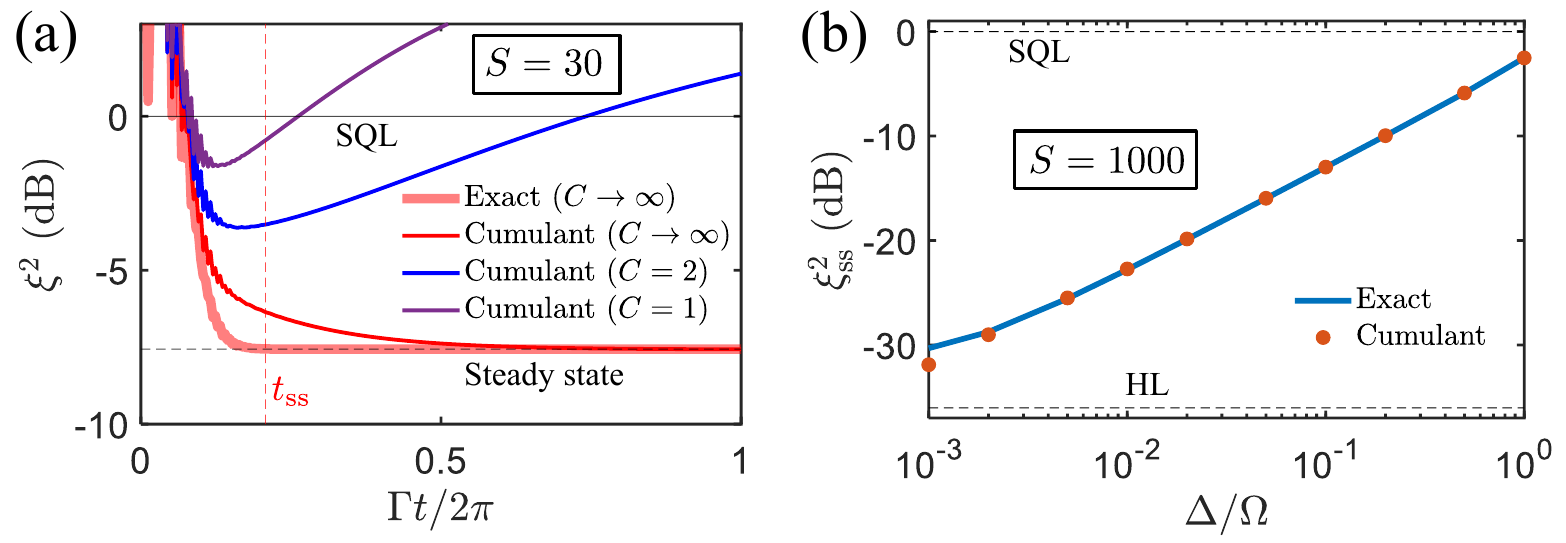}
    \caption{(a) Evolution of the Wineland spin-squeezing parameter $\xi^2$ in the case of $S=30$, $S\Delta/\Omega = 10$, $S\Gamma/\Omega = 10$, with initial state as all spins pointing down. We compare numerical results based on the quantum-jump method (exact) and the second-order cumulant expansion. We apply the cumulant expansion to include single-particle decay processes with rate $\gamma = \Gamma/C$, with $C$ the single-atom cooperativity. (b) Wineland spin squeezing parameter for the steady state $|\psi_{\rm ss}^{(2)}\rangle$ of two spin ensembles with $S=1000$ and $\gamma=0$. Second-order cumulant expansion correctly captures the steady-state spin squeezing up to $\Delta/\Omega\sim 1/S$, which is the parameter regime for Heisenberg-scaling quantum enhancement.}
    \label{fig:cumulant}
\end{figure}

Here we provide details of the second-order cumulant expansion for the case of two spin ensembles. Note that this approach can in principle be generalized to the case of many spin ensembles. We consider the following Lindblad master equation.
\begin{equation}
    \begin{gathered}
    \begin{aligned}
        \frac{d}{dt}\hat{\rho} &= -i[\hat{H},\hat{\rho}] + \Gamma\mathcal{D}[\hat{S}_1^{-}+\hat{S}_2^{-}]\hat{\rho}\\
        &+\gamma\sum_{l=1}^{2S}\bigg(\mathcal{D}[\hat{s}_{1,l}^{-}]\hat{\rho}+\mathcal{D}[\hat{s}_{2,l}^{-}]\hat{\rho}\bigg),
    \end{aligned}\\
    \hat{H} = \Omega (\hat{S}^x_1+\hat{S}^x_2) +\frac{\Delta}{2}(\hat{S}^z_1-\hat{S}^z_2),
    \end{gathered}
    \label{eq:culind}
\end{equation}
where $\hat{s}_{j,l}^{-}$ are spin-$1/2$ operators. The collective spin operators are constructed by the addition of $2S$ spin-$1/2$ operators in the same spin ensemble, for example, $\hat{S}_j^{-}=\sum_{l=1}^{2S}\hat{s}_{j,l}^{-}$.

The general procedure of second-order cumulant expansion is as follows: 1) Derive Heisenberg equations of motion for spin operators and two-operator products of spin operators. 2) Take their expectation values and then split the expectation values for products of three or more operators to obtain a closed set of equations of motion. For example, expectation values for three-operator product can be split in the following way \cite{Kubo1962},   
\begin{equation}
    \langle \hat{A}\hat{B}\hat{C}\rangle\approx \langle \hat{A}\hat{B}\rangle\langle\hat{C}\rangle + \langle \hat{A}\hat{C}\rangle\langle\hat{B}\rangle + \langle \hat{B}\hat{C}\rangle\langle\hat{A}\rangle - 2\langle \hat{A}\rangle\langle\hat{B}\rangle\langle\hat{C}\rangle.
\end{equation}
In the procedure above, we need to specify a convention for counting the number of operators in an operator product. To avoid any ambiguity, we rewrite Eq.~(\ref{eq:culind}) in terms of Pauli matrices $\hat{\sigma}^x_{j,l}$, $\hat{\sigma}^y_{j,l}$, $\hat{\sigma}^z_{j,l}$, with $j=1,2$ and $l=1,\cdots,2S$. We then derive Heisenberg equations of motion for these Pauli matrices. Since all the spin-$1/2$ operators can be expressed in the basis of Pauli matrices and identity, in this case the number of operators in an operator product can be naturally defined as the number of spin-$1/2$ particles involved.
We also apply permutation invariance for each of the spin ensembles to reduce the number of independent expectation values, e.g., $\langle \hat{\sigma}^\alpha_{j,l} 
\rangle = \langle \hat{\sigma}^\alpha_{j,1} \rangle$ for all $l=1, \dots, 2S$ and $\alpha \in \{x,y,z\}$.
As a result, we find $27$ independent expectation values:
\begin{itemize}
    \item One-body operators in ensemble 1:
    \[\langle\hat{\sigma}_{1,1}^x\rangle,\quad \langle\hat{\sigma}_{1,1}^y\rangle, \quad \langle\hat{\sigma}_{1,1}^z\rangle\]
    
    \item One-body operators in ensemble 2:
    \[\langle\hat{\sigma}_{2,1}^x\rangle,\quad \langle\hat{\sigma}_{2,1}^y\rangle, \quad \langle\hat{\sigma}_{2,1}^z\rangle\]
    
    \item Two-body operators in ensemble 1:
    \[\begin{gathered}
        \langle\hat{\sigma}_{1,1}^x\hat{\sigma}_{1,2}^x\rangle,\quad \langle\hat{\sigma}_{1,1}^y\hat{\sigma}_{1,2}^y\rangle, \quad \langle\hat{\sigma}_{1,1}^z\hat{\sigma}_{1,2}^z\rangle\\
        \langle\hat{\sigma}_{1,1}^x\hat{\sigma}_{1,2}^y\rangle,\quad \langle\hat{\sigma}_{1,1}^x\hat{\sigma}_{1,2}^z\rangle, \quad \langle\hat{\sigma}_{1,1}^y\hat{\sigma}_{1,2}^z\rangle\\
    \end{gathered}\]
    
    \item Two-body operators in ensemble 2:
    \[\begin{gathered}
        \langle\hat{\sigma}_{2,1}^x\hat{\sigma}_{2,2}^x\rangle,\quad \langle\hat{\sigma}_{2,1}^y\hat{\sigma}_{2,2}^y\rangle, \quad \langle\hat{\sigma}_{2,1}^z\hat{\sigma}_{2,2}^z\rangle\\
        \langle\hat{\sigma}_{2,1}^x\hat{\sigma}_{2,2}^y\rangle,\quad \langle\hat{\sigma}_{2,1}^x\hat{\sigma}_{2,2}^z\rangle, \quad \langle\hat{\sigma}_{2,1}^y\hat{\sigma}_{2,2}^z\rangle\\
    \end{gathered}\]
    
    \item Two-body operators between two ensembles:
    \[\begin{gathered}
        \langle\hat{\sigma}_{1,1}^x\hat{\sigma}_{2,1}^x\rangle,\quad \langle\hat{\sigma}_{1,1}^x\hat{\sigma}_{2,1}^y\rangle, \quad \langle\hat{\sigma}_{1,1}^x\hat{\sigma}_{2,1}^z\rangle\\
        \langle\hat{\sigma}_{1,1}^y\hat{\sigma}_{2,1}^x\rangle,\quad \langle\hat{\sigma}_{1,1}^y\hat{\sigma}_{2,1}^y\rangle, \quad \langle\hat{\sigma}_{1,1}^y\hat{\sigma}_{2,1}^z\rangle\\
        \langle\hat{\sigma}_{1,1}^z\hat{\sigma}_{2,1}^x\rangle,\quad \langle\hat{\sigma}_{1,1}^z\hat{\sigma}_{2,1}^y\rangle, \quad \langle\hat{\sigma}_{1,1}^z\hat{\sigma}_{2,1}^z\rangle\\
    \end{gathered}\]
    
\end{itemize}

After numerically solving the corresponding equations of motion, we then use these expectation values of Pauli matrices to calculate the expectation values of collective spin operators. For example,
\begin{equation}
    \langle \hat{S}^x_1-\hat{S}^x_2\rangle = S\Big(\langle\hat{\sigma}_{1,1}^x\rangle-\langle\hat{\sigma}_{2,1}^x\rangle\Big),
\end{equation}
\begin{equation}
    \langle \hat{S}^y_1+\hat{S}^y_2\rangle = S\Big(\langle\hat{\sigma}_{1,1}^y\rangle+\langle\hat{\sigma}_{2,1}^y\rangle\Big),
\end{equation}
\begin{equation}
    \begin{aligned}
    &\langle (\hat{S}^y_1+\hat{S}^y_2)^2\rangle \\
    &= S + S\Big(S-\frac{1}{2}\Big)\Big(\langle\hat{\sigma}_{1,1}^y\hat{\sigma}_{1,2}^y\rangle+\langle\hat{\sigma}_{2,1}^y\hat{\sigma}_{2,2}^y\rangle\Big)+ 2S^2\langle\hat{\sigma}_{1,1}^y\hat{\sigma}_{2,1}^y\rangle\\
    &\approx S + S^2\Big(\langle\hat{\sigma}_{1,1}^y\hat{\sigma}_{1,2}^y\rangle+\langle\hat{\sigma}_{2,1}^y\hat{\sigma}_{2,2}^y\rangle+2\langle\hat{\sigma}_{1,1}^y\hat{\sigma}_{2,1}^y\rangle\Big).
    \end{aligned}
    \label{eq:largeS}
\end{equation}
Based on these results, we can numerically calculate the two-mode generalization of the Wineland spin-squeezing parameter,
\begin{equation}
    \xi^2 = \frac{4S\mathrm{Var}(\hat{S}^y_1+\hat{S}^y_2)}{\langle\hat{S}^x_1-\hat{S}^x_2\rangle^2}.
\end{equation}
We find numerically that the large-$S$ approximation in Eq.~(\ref{eq:largeS}) can lead to a better agreement of the steady-state spin squeezing.

In Fig.~\ref{fig:cumulant}(a), we first benchmark the second-order cumulant expansion with exact calculation via the quantum jump method in the case without single-particle decay ($C=\Gamma/\gamma \rightarrow\infty$). 
We conclude that the second-order cumulant expansion correctly captures the steady-state spin squeezing, while predicting a slower relaxation time scale.
In Fig.~\ref{fig:cumulant}(b), we further demonstrate that the second-order cumulant expansion captures the steady-state spin squeezing up to Heisenberg scaling.
These findings ensure the reliability of second-order cumulant expansion in our case, at least it is possible to provide a qualitative prediction in the regime beyond the reach of exact numerical methods.
We then use second-order cumulant expansion to calculate the case with finite $C$ as shown in Fig.~\ref{fig:cumulant}(a).
In this case the optimal squeezing is achieved at a finite time before the relaxation time scale $t_{\rm ss}$.

\section{Uniqueness of steady-state solution}
\label{sec:unique}

\subsection{Two spin ensembles}
Here we discuss the uniqueness of the steady-state solution for two spin ensembles (see Eq.~(\ref{eq:twospin})).
We can show the uniqueness of the steady-state solution in the limit $\Omega \gg S\Gamma$ and $\Omega \gg S\Delta$ using perturbation theory in the Liouvillian space.

First, we apply the rotation $\hat R(\theta) = e^{-i\theta(\hat S_1^y-\hat S_2^y)/2}$ with $\theta=\arctan\Big(\Delta/(2\Omega)\Big)$, such that the Hamiltonian becomes
\begin{equation}
    \hat{H}_R = \hat R(\theta) \hat H \hat R^\dagger(\theta) = \sqrt{\Omega^2+(\Delta/2)^2}(\hat S_1^x + \hat S_2^x),
\end{equation}
and the jump operator becomes
\begin{equation}
    \begin{aligned}
    \hat S^-_R = \cos^2\frac{\theta}{2}\bigg[ \hat S^- - \tan^2\frac{\theta}{2} \hat S^+ - 2\tan\frac{\theta}{2}(\hat S_1^z - \hat S_2^z) \bigg],
    \end{aligned}
\end{equation}
where $\hat{S}^{\pm} = \hat{S}^{\pm}_1+\hat{S}^{\pm}_2$.
Based on the condition $\Omega \gg S\Delta$, one can separate the Lindbladian into zeroth, first and second-order terms. The zeroth-order Lindbladian is given by 
\begin{equation}
    \mathcal{L}_0\hat \rho = -i[\hat H_R, \hat \rho] + \Gamma_R\mathcal{D}[\hat S^-]\hat\rho,
\end{equation}
where $\Gamma_R=\Gamma\cos^4(\theta/2)$. 
The first-order Lindbladian is given by
\begin{equation}
    \begin{aligned}
    \mathcal{L}_1\hat \rho &= -2\tan\bigg(\frac{\theta}{2}\bigg)\Gamma_R \bigg(\hat{S}^{-}\hat{\rho}(\hat{S}^z_1-\hat{S}^z_2) + (\hat{S}^z_1-\hat{S}^z_2)\hat{\rho}\hat{S}^{+}\\
    &-\frac{1}{2}\{\hat{S}^{+}(\hat{S}^z_1-\hat{S}^z_2) + (\hat{S}^z_1-\hat{S}^z_2)\hat{S}^{-},\hat{\rho}\}\bigg),
    \end{aligned}
\end{equation}
and the second-order Lindbladian is given by
\begin{equation}
    \mathcal{L}_2\hat \rho = 4\tan^2\bigg(\frac{\theta}{2}\bigg)\Gamma_R \mathcal{D}[\hat{S}^z_1-\hat{S}^z_2]\hat{\rho}.
\end{equation}
Here we drop the $\hat{S}^{+}$ term in $\hat{S}_R^{-}$ since this term does not play a role in lifting the degeneracy of steady states of $\mathcal{L}_0$.

Second, we focus on the spectrum of $\mathcal{L}_0$. Note that both $\hat{H}_R$ and $\hat{S}^{-}$ do not couple different total angular momentum $J$ sectors, such that one can separate $\mathcal{L}_0$ into different subspaces labeled by $(J,J')$,
\begin{equation}
    \mathcal{L}_0 = \bigoplus_{J,J'} \mathcal{L}_0^{J,J'}.
\end{equation}
In the limit $\Omega\gg S\Gamma$, $\mathcal{L}_0$ is dominated by the Hamiltonian $\hat{H}_R$, so its eigen-operator to the leading order takes the following form,
\begin{equation}
    \hat{O}^{J,J'} \approx \sum_{m} c_m |J,m\rangle_x\langle J',m+m_0|,
\end{equation}
where $\ket{J,m}_x=e^{-i\pi(\hat S^y_1 + \hat S^y_2)/2}\ket{J, m}$ is the total angular momentum state in the $x$ basis. Note that, in this case, the left and right eigen-operators are the same to leading order. To reach the eigenvalue closest to $0$ in each $(J,J')$ subspace, we have $m_0=0$. The eigenvalues and coefficients $c_m$ can be determined by the dissipative part, and here we simply list the results. 
For $\mathcal{L}_0^{J,J}$, we have a steady state solution with eigenvalue $0$, and the corresponding eigen-operator is given by  
\begin{equation}
    \hat{O}^{J,J} \approx \sum_{m=-J}^{J} \ket{J,m}_x\!\bra{J,m},
\end{equation}
For $\mathcal{L}_0^{J,J+1}$, the eigenvalue closest to $0$ is $-\Gamma/2$, and the corresponding eigen-operator is given by 
\begin{equation}
    \hat{O}^{J,J+1} \approx \sum_{m=-J}^{J} \frac{\sqrt{(J+1)^2-m^2}}{J+1}\ket{J,m}_x\!\bra{J+1,m},
\end{equation}
Similarly, for $\mathcal{L}_0^{J,J-1}$, we have $\hat{O}^{J,J-1}=(\hat{O}^{J-1,J})^{\dag}$ with eigenvalue $-\Gamma/2$.

We apply perturbation theory to lift the $(2S+1)$-fold degeneracy ($\hat{O}^{J,J}$ for each $J$) of the steady states of $\mathcal{L}_0$. We define a projection superoperator $\mathcal{P}$ for all the $\hat{O}^{J,J}$ (ground manifold), and a projection superoperator $\mathcal{Q}$ for all the $\hat{O}^{J,J\pm 1}$ (excited manifold). Notice that $\mathcal{L}_1$ only couples between the ground and excited manifold, while $\mathcal{L}_2$ can couple within the ground manifold. So the effective Lindbladian in the ground manifold is given by 
\begin{equation}
    \mathcal{L}_{\rm eff} = \frac{\mathcal{P}\mathcal{L}_1\mathcal{Q}\mathcal{L}_1\mathcal{P}}{\Gamma/2} + \mathcal{P}\mathcal{L}_2\mathcal{P}.
\end{equation}
Analytical calculation shows that $\mathcal{L}_{\rm eff}$ is a tri-diagonal matrix with dimension $(2S+1)$,
\begin{equation}
    \mathcal{L}_{\rm eff} = \Gamma\sin^2(\theta)\begin{pmatrix}
 a_0 & c_0 & & & \\
 b_1 & a_1 & c_1 & & \\
 & b_2 & \ddots & \ddots &\\
 & & \ddots & \ddots & c_{2S-1} \\
 & &  & b_{2S} & a_{2S} \\
\end{pmatrix}.
\label{eq:perturb}
\end{equation}
The matrix elements are given by
\begin{equation}
    \begin{aligned}
    a_J &= -\frac{1}{3}J(J+1)\Big((2S+1)^2-J(J+1)\Big),\\
    b_J &= \frac{1}{3(2J+1)}(J-1)^2J\Big((2S+1)^2-J^2\Big),\\
    c_J &= \frac{1}{3(2J+1)}(J+2)^2(J+1)\Big((2S+1)^2-(J+1)^2\Big),
    \end{aligned}    
\end{equation} 
where $J=0,\cdots, 2S$. Since $a_{J=0}=b_{J=1}=0$, one can show that $\hat{O}^{J=0,J=0}$ is still a steady-state solution.

\begin{figure}[t]
    \centering
    \includegraphics[width=1.0\columnwidth]{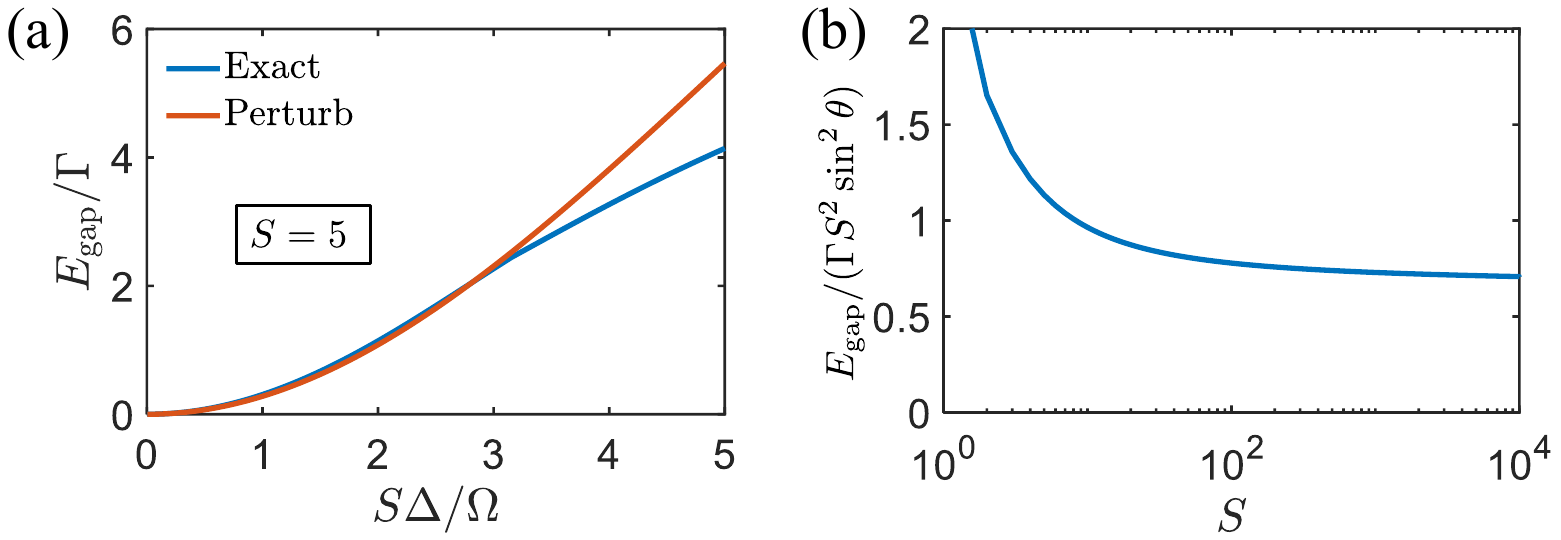}
    \caption{(a) Comparison of dissipative gap $E_{\rm gap}$ between exact diagonalization and perturbative calculation in the case of two spin ensembles with $S=5$. The perturbative calculation can capture the exact result up to $S\Delta/\Omega \sim 3$. (b) Perturbative calculation of the dissipative gap $E_{\rm gap}$ up to $S=10^4$. In the perturbative regime, we have $E_{\rm gap}\propto S^2$ for $S\gg 1$.}
    \label{fig:perturb}
\end{figure}

To show the uniqueness of the steady-state solution, we numerically diagonalize the matrix in Eq.~(\ref{eq:perturb}) to obtain the dissipative gap $E_{\rm gap}$. In Fig.~\ref{fig:perturb}(a), we compare the perturbative calculation with exact diagonalization for $E_{\rm gap}$ at $S=5$, which shows that the perturbative calculation can capture the exact result up to $S\Delta/\Omega \sim 3$. In Fig.~\ref{fig:perturb}(b), we perform the perturbative calculation for $E_{\rm gap}$ to confirm the uniqueness of the steady state up to $S=10^4$. 
In the perturbative regime, one can obtain $E_{\rm gap}\propto \Gamma(S\Delta/\Omega)^2$. 

Assuming the Lindbladian is diagonalizable, we can extend the uniqueness of the steady-state solution by increasing $\Delta$ beyond the perturbative regime.
We also numerically confirm the uniqueness of the steady state beyond the perturbative regime up to $S=30$, as shown in Fig.~\ref{fig:twomodesqueezing}(b).
In this regime we have $E_{\rm gap}\propto \Gamma(S\Delta/\Omega)$.
The uniqueness of the steady state solution in the case of $\chi\neq 0$ can be discussed in a similar way.

\subsection{Many spin ensembles}
In the case of many spin ensembles, due to the existence of the unitary transformation (see Eq.~(\ref{eq:gate})) to swap detunings, we only need to discuss the uniqueness of the steady-state solution for a specific detuning pattern, $\delta_{2k-1}=\Delta_k/2$, $\delta_{2k}=-\Delta_k/2$ with $k=1,2,\cdots,L/2$.
With this detuning pattern, the steady state is simply a tensor product of ensemble pairs, $|\psi_{\rm ss}^{(L)}\rangle = \bigotimes_{k} |\psi_{\rm ss}^{(2)}\rangle_{2k-1,2k}$.

It is possible to show the uniqueness of the steady-state solution in the case of $\chi=\Gamma$, where the Lindblad master equation becomes a cascaded master equation.
Similar to Refs.~\cite{Stannigel2012,Agusti2023}, the procedure is based on induction: 1) Show the unique steady state for the case of two spin ensembles. 2) Suppose the steady state is unique for $L$ spin ensembles ($L$ is an even number), and prove that the steady state is unique for $L+2$ spin ensembles. 
The first step is discussed in the previous subsection, here we discuss the second step.

We use subsystem $A$ to label the first $L$ ensembles, and subsystem $B$ to label the last two ensembles. When $\chi=\Gamma$, the Lindbladian of the system (see Eq.~(\ref{eq:model})) can be written as
\begin{equation}
    \mathcal{L}\hat{\rho} = \mathcal{L}_A\hat{\rho} + \mathcal{L}_B\hat{\rho} - \Gamma\Big([\hat{S}_B^{+},\hat{S}_A^{-}\hat{\rho}]+[\hat{\rho}\hat{S}_A^{+},\hat{S}_B^{-}]\Big), 
    \label{eq:cascade}
\end{equation}
where $\hat{S}_A^{\pm}=\sum_{k=1}^L\hat{S}^{\pm}_k$, and $\hat{S}_B^{\pm}=\hat{S}^{\pm}_{L+1}+\hat{S}^{\pm}_{L+2}$.
Based on Eq.~(\ref{eq:cascade}), one can show that
\begin{equation}
    \mathrm{tr}_B(\mathcal{L}\hat{\rho}) = \mathcal{L}_A \mathrm{tr}_B(\hat{\rho}).
    \label{eq:tracecas}
\end{equation}
Apply Eq.~(\ref{eq:tracecas}) to the steady state solution, and assuming that we have a unique pure steady state $\hat{\rho}_{A,\rm ss}=|\psi_{\rm ss}^{(L)}\rangle_A\langle \psi_{\rm ss}^{(L)}|$ for subsystem $A$, the steady-state solution for the whole system should take the following form, $\hat{\rho}_{\rm ss} = \hat{\rho}_{A,\rm ss}\otimes\hat{\rho}'$. Based on Eq.~(\ref{eq:cascade}), we have $\mathcal{L}\hat{\rho}_{\rm ss} = \mathcal{L}_B\hat{\rho}'=0$, reducing the problem to the case of two spin ensembles discussed in the previous subsection.
This result indicates the unique steady state solution in the case of $\chi=\Gamma$.

Assuming the Lindbladian is diagonalizable, we can extend the uniqueness of the steady-state solution to the case of $\chi\neq \Gamma$ if the permutation symmetry between ensembles is completely broken. 
We also numerically confirm the uniqueness for the case of $\chi\neq \Gamma$ for small system sizes (up to $L=10$ for $S=1/2$).  

\section{Experimental implementation of chiral spin-exchange couplings}
\label{sec:experiment}

\subsection{Raman-coupled spin-exchange interactions}

\begin{figure}[t]
    \centering
    \includegraphics[width=1.0\columnwidth]{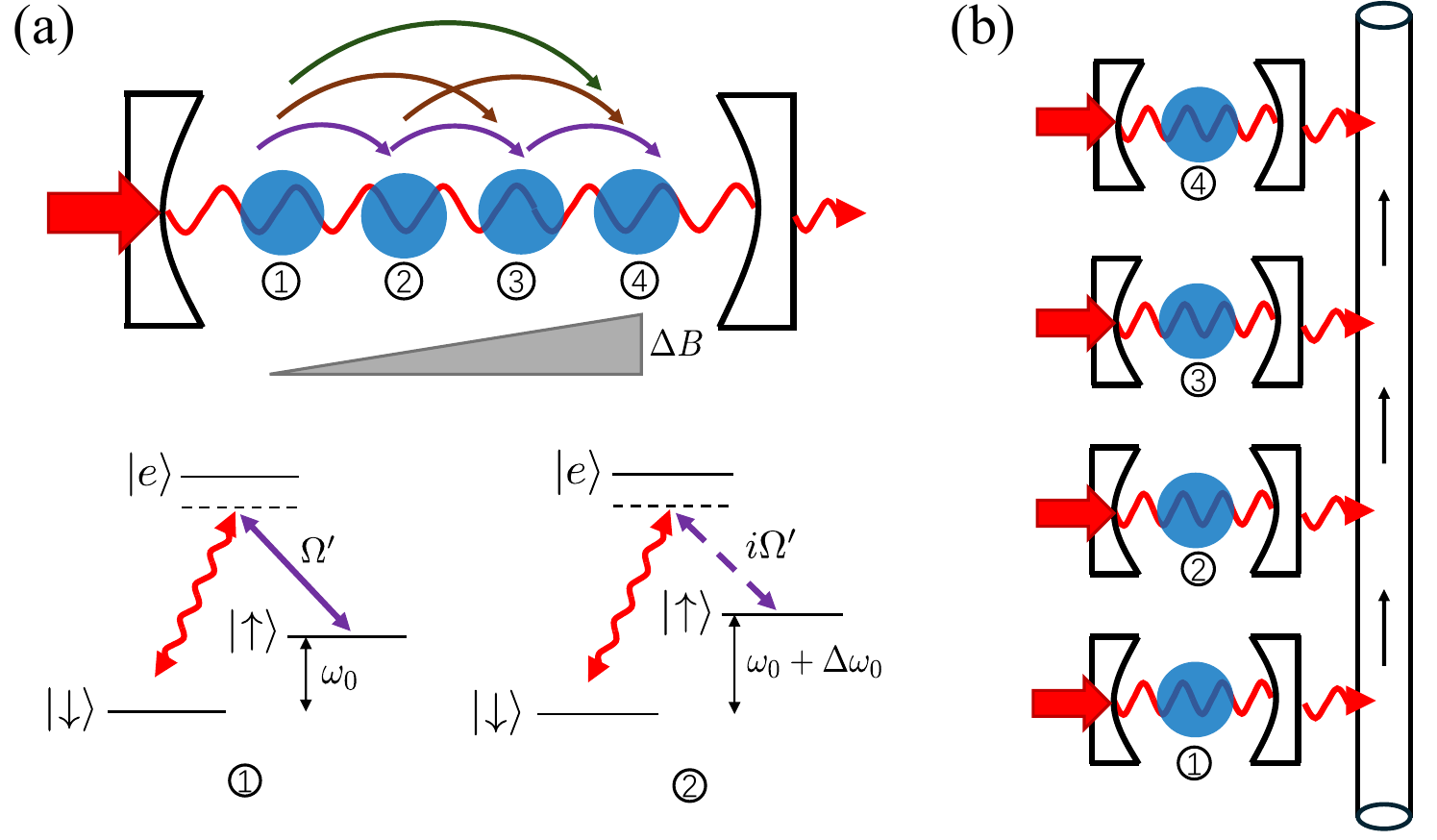}
    \caption{(a) Schematic of engineering chiral spin-exchange couplings via Raman transitions. We place all the spin ensembles in a single optical cavity, and apply a magnetic-field gradient along the cavity axis. Spin-exchange couplings between ensembles separated by different distances can be independently tuned via frequency selection. (b) Schematic of engineering chiral spin-exchange couplings via chiral quantum networks. We couple each spin ensemble with an optical cavity, and then connect all these cavities by a single-directional waveguide.}
    \label{fig:exp}
\end{figure}

One way to implement the chiral spin-exchange couplings is to engineer Raman-coupled spin-exchange interactions similar to the setup in Ref.~\cite{Periwal2021}. As shown in Fig.~\ref{fig:exp}(a), we define $|\uparrow\rangle$ and $|\downarrow\rangle$ states in the ground manifold with transition frequency $\omega_0$ (generated by either hyperfine shift or Zeeman shift from a uniform magnetic field depending on specific platforms), and excited state $|e\rangle$ with transition frequency $\omega_e$ with respect to $|\uparrow\rangle$ state.
We then apply a magnetic-field gradient along the cavity such that the transition frequency between $|\uparrow\rangle$ and $|\downarrow\rangle$ states becomes $\omega_{0,l} = \omega_0 + (l-1)(\Delta\omega_0)$.
To engineer chiral spin-exchange couplings between nearest-neighbor ensembles (purple arrows), we apply a two-tone drive to transition from $|\uparrow\rangle$ to $|e\rangle$ off-resonantly (solid and dashed arrows). The two tones have Rabi frequency $\Omega'$ and $i\Omega'$, and drive frequency $\omega_d$ and $\omega_d-\Delta\omega_0$ respectively. We still use the same detuning $\delta_e=\omega_d-\omega_e$ to the excited state $|e\rangle$ assuming $\delta_e$ much larger than Zeeman shifts. The cavity mode couples the transition from $|\downarrow\rangle$ to $|e\rangle$ with cavity resonant frequency $\omega_c$. After adiabatic elimination of the excited states ($\delta_e\gg \Omega'$), the Hamiltonian is given by
\begin{equation}
    \begin{aligned}
    \hat{H} &= \omega_c \hat{a}^{\dag}\hat{a} + \sum_l \omega_{0,l}\hat{S}^z_l +\sum_l\bigg(\frac{\Omega'\mathcal{G}}{\delta_e}\hat{a}\hat{S}^{+}_l e^{-i\omega_dt} + h.c.\bigg)\\
    &+\sum_l\bigg(\frac{i\Omega'\mathcal{G}}{\delta_e}\hat{a}\hat{S}^{+}_l e^{-i(\omega_d-\Delta\omega_0)t} + h.c.\bigg), \\
    \end{aligned}
\end{equation}
where $2\mathcal{G}$ is the single-atom vacuum Rabi splitting of the cavity. Using the following unitary transformation to the rotating frame, $\hat{U} = e^{-i(\omega_d+\omega_0) \hat{a}^{\dag}\hat{a}t-i\sum_l \omega_{0,l}\hat{S}^z_l t}$, we have
\begin{equation}
    \begin{aligned}
    \hat{H}' &= -\delta_c \hat{a}^{\dag}\hat{a} + \sum_l\bigg(\frac{\Omega'\mathcal{G}}{\delta_e}\hat{a}\hat{S}^{+}_l e^{-i(l-1)(\Delta\omega_0)t} + h.c.\bigg) \\
    &+\sum_l\bigg(\frac{i\Omega'\mathcal{G}}{\delta_e}\hat{a}\hat{S}^{+}_l e^{-il(\Delta\omega_0)t} + h.c.\bigg),
    \end{aligned}
\end{equation}
where $\delta_c=\omega_d+\omega_0-\omega_c$. Assuming $\delta_c\gg \Omega'\mathcal{G}\sqrt{S}/\delta_e$, we can further adiabatically eliminate the cavity mode.
If we also apply the rotating wave approximation and only keep the time-independent terms, the effective Hamiltonian becomes
\begin{equation}
    \hat{H}_{\rm eff} = \frac{1}{\delta_c}\bigg(\frac{\Omega'\mathcal{G}}{\delta_e}\bigg)^2\bigg(\sum_l \hat{S}^{+}_l\hat{S}^{-}_l + \sum_l i(\hat{S}^{+}_l\hat{S}^{-}_{l+1}-\hat{S}^{-}_l\hat{S}^{+}_{l+1})\bigg).
    \label{eq:efftwotone}
\end{equation}
The second term in Eq.~(\ref{eq:efftwotone}) is the chiral spin-exchange coupling between nearest-neighbor ensembles, which is due to frequency selection via the two-tone frequency difference $\Delta\omega_0$.
Other chiral spin-exchange coupling terms can be engineered by two-tone frequency differences
$2\Delta\omega_0,3\Delta\omega_0,\cdots$.
One can also cancel the first term in Eq.~(\ref{eq:efftwotone}) using a single-tone drive that generates $\sum_l \hat{S}^{+}_l\hat{S}^{-}_l$ with opposite $\delta_c$.

\subsection{Chiral quantum networks}
Another way to implement the chiral spin-exchange couplings is to couple each spin ensemble with an optical cavity, and then connect all these cavities by a unidirectional waveguide as shown in Fig.~\ref{fig:exp}(b). Based on Refs.~\cite{Stannigel2012,Pichler2015}, the Lindblad master equation of this system can be written as
\begin{equation}
    \begin{gathered}
    \frac{d}{dt}\hat{\rho} = -i[\hat{H},\hat{\rho}] + \kappa \mathcal{D}\bigg[\sum_le^{-i\phi_l}\hat{a}_{l}\bigg]\hat{\rho},\\
    \hat{H}=\sum_{l}\hat{H}_l + i\frac{\kappa}{2}\sum_{l>k}\Big(e^{i(\phi_k-\phi_l)}\hat{a}^{\dag}_k\hat{a}_l-e^{-i(\phi_k-\phi_l)}\hat{a}_k\hat{a}^{\dag}_l\Big),
    \end{gathered}
    \label{eq:chiral1}
\end{equation}
where $\kappa$ is the linewidth of the optical cavity, $\hat{a}_l$ is the bosonic annihilation operator of the $l$-th cavity, $\phi_l$ is the phase due to the running-wave mode of the single-directional waveguide. The Hamiltonian within the $l$-th cavity is based on the Tavis-Cumming model, and the frequency of the drive is resonant with the cavity mode,
\begin{equation}
    \hat{H}_l = \frac{\Omega}{2}(e^{i\phi'_l}\hat{S}^+_l+e^{-i\phi'_l}\hat{S}^-_l) + \delta_{l}\hat{S}^z_l + \mathcal{G}(\hat{a}^{\dag}_l\hat{S}^-_l + \hat{S}^+_l\hat{a}_l),
    \label{eq:chiral2}
\end{equation}
where $2\mathcal{G}$ is the single-atom vacuum Rabi splitting of the cavity, $\delta_l$ are the drive detunings to the optical transition of the $l$-th sub-ensemble, and $\phi'_l$ are the phases of Rabi drives for each sub-ensemble.
We perform the following gauge transformation, $\hat{S}^{+}_l\rightarrow \hat{S}^{+}_le^{-i\phi'_l}$, $\hat{S}^{-}_l\rightarrow \hat{S}^{-}_le^{i\phi'_l}$, $\hat{a}_l\rightarrow \hat{a}_le^{i\phi'_l}$, $\hat{a}^{\dag}_l\rightarrow \hat{a}^{\dag}_le^{-i\phi'_l}$, Eq.~(\ref{eq:chiral1}) and (\ref{eq:chiral2}) become
\begin{equation}
    \begin{gathered}
    \frac{d}{dt}\hat{\rho} = -i[\hat{H}',\hat{\rho}] + \kappa \mathcal{D}\bigg[\sum_le^{-i\phi_l}\hat{a}_{l}\bigg]\hat{\rho},\\
    \hat{H}'=\sum_{l}\hat{H}'_l + i\frac{\kappa}{2}\sum_{l>k}\Big(e^{i(\phi_k-\phi_l)}\hat{a}^{\dag}_k\hat{a}_l-e^{-i(\phi_k-\phi_l)}\hat{a}_k\hat{a}^{\dag}_l\Big),\\
    \hat{H}'_l=\Omega\hat{S}^x_l + \delta_{l}\hat{S}^z_l + \mathcal{G}(\hat{a}^{\dag}_l\hat{S}^-_l + \hat{S}^+_l\hat{a}_l),
    \end{gathered}
\end{equation}
where $\tilde{\phi}_l = \phi_l - \phi'_l$.
When $\kappa\gg \mathcal{G}\sqrt{S}$, it is possible to adiabatically eliminate all these cavity modes and obtain an effective spin-only Lindblad master equation. When $\tilde{\phi}_{l+1}-\tilde{\phi}_{l}=\pi$, one can obtain Eq.~(\ref{eq:model}) with $\chi=\Gamma=4\mathcal{G}^2/\kappa$.
This phase requirement can be satisfied by controlling either the separation distance between cavities holding the sub-ensembles or the phases of the Rabi drives.

\section{Discussions of experimental imperfections}
\label{sec:imperfection}

\begin{figure}[t]
    \centering
    \includegraphics[width=1.0\columnwidth]{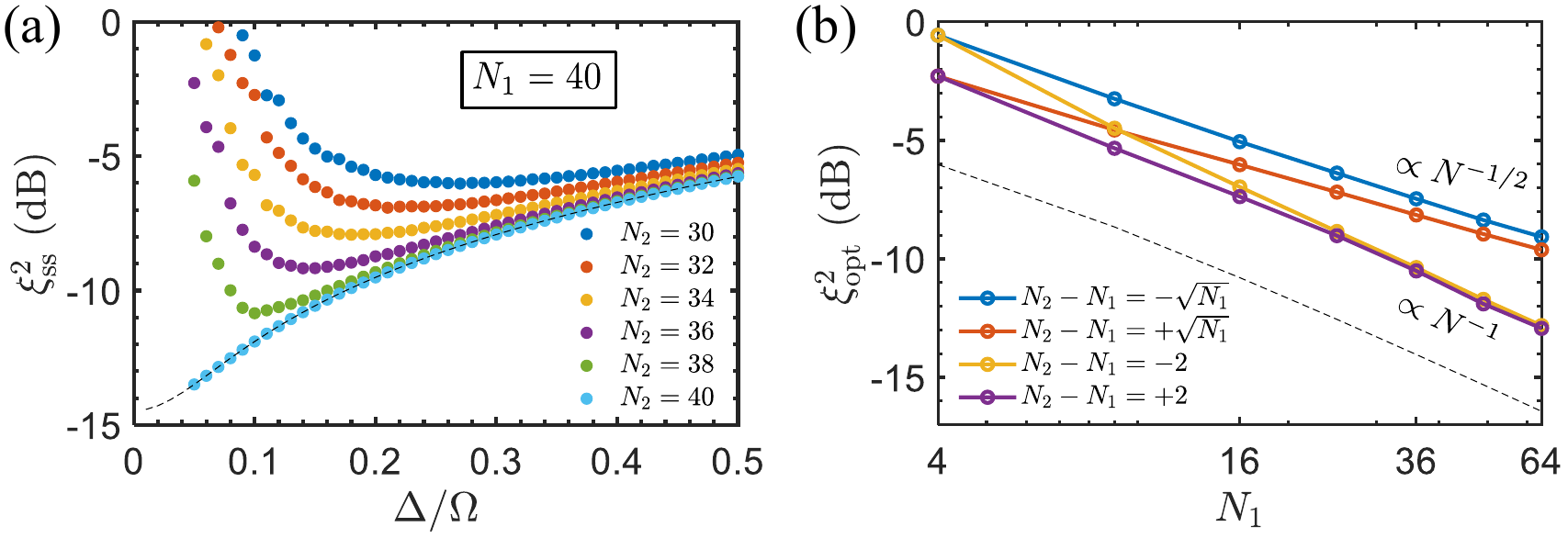}
    \caption{(a) Steady state spin squeezing $\xi^2_{\rm ss}$ considering atom number imbalance between two spin ensembles. We fix the atom number in the first ensemble $N_1=40$, and vary the atom number in the second ensemble $N_2$. When $N_1\neq N_2$, $\xi^2_{\rm ss}$ reaches a optimal value at non-zero $\Delta/\Omega$ ratio. $\xi^2_{\rm ss}$ degrades as we increases the atom number imbalance. (b) Scaling of optimal steady state spin squeezing $\xi^2_{\rm opt}$ (optimized over $\Delta/\Omega$). We compare the cases with atom number imbalance scales as a constant value and those with imbalance scales like $\sqrt{N}$.}
    \label{fig:imbalance}
\end{figure}

\subsection{Population imbalance between spin ensembles}
In the main text, we assume each spin ensemble has the same number of atoms.
Here we analyze the effects on the steady-state spin squeezing generated by atom-number imbalance between two spin ensembles, as shown in Fig.~\ref{fig:imbalance}.
We define the atom number in the first ensemble and the second ensemble as $N_1$ and $N_2$ respectively, and define averaged atom number $N=(N_1+N_2)/2$ and averaged spin value as $S=N/2$.
In the presence of atom-number imbalance, the lowest total angular momentum is $J_{\rm min} = |N_1-N_2|/2>0$.
As shown in Fig.~\ref{fig:twoensemble}(c), the distribution of the wave function across the different total-angular-momentum sectors depends on the ratio $\Delta/\Omega$.
This effect determines the behavior of the steady state spin squeezing parameter $\xi_{\rm ss}^2$ (defined in Eq.~(\ref{eq:squeeze})):
\begin{itemize}
    \item In the regime $S\Delta/\Omega \gg J_{\rm min}$, the steady state is close to a pure state since the relevant angular-momentum structure is very similar to balanced case $N_1=N_2$: 
    The state populates subspaces of large total angular momentum and is not sensitive to the fact that the lowest total-angular-momentum sector has been moved from $J_{\mathrm{min}}=0$ to $J_{\mathrm{min}} = \vert N_1 - N_2\vert/2$.
    In this regime, spin squeezing enhances as we decrease the ratio $\Delta/\Omega$, similar to Fig.~\ref{fig:twoensemble}(d).
    \item In the regime $S\Delta/\Omega \ll J_{\rm min}$, the steady state mainly occupies low total-angular-momentum subspaces and becomes sensitive to the effect of population imbalance. 
    Now, the steady state becomes a highly mixed state limited by the population imbalance, since we cannot find angular momentum smaller than $J_{\rm min}$. 
    This causes the squeezing to degrade.
\end{itemize}
The optimal spin squeezing is typically achieved when $S\Delta/\Omega \sim J_{\rm min}$, where the steady state is roughly a pure state centered near total angular momentum $J_{\rm min}$.
The qualitative understanding above agrees with our numerical findings in Fig.~\ref{fig:imbalance}(a), in which the steady state spin squeezing $\xi^2_{\rm ss}$ reaches an optimal value at non-zero $\Delta/\Omega$ ratio when $N_1\neq N_2$.
In Fig.~\ref{fig:imbalance}(b), we show that the steady-state spin squeezing is mainly set by $J_{\rm min}$, i.e. $\xi^2_{\rm opt}\sim J_{\rm min}/N$. When $J_{\rm min}\sim \sqrt{N}$, we have $\xi^2_{\rm opt}\propto 1/\sqrt{N}$; When $J_{\rm min}$ is a constant value, we have $\xi^2_{\rm opt}\propto 1/N$.

\subsection{Measurement noise}

\begin{figure}[t]
    \centering
    \includegraphics[width=1.0\columnwidth]{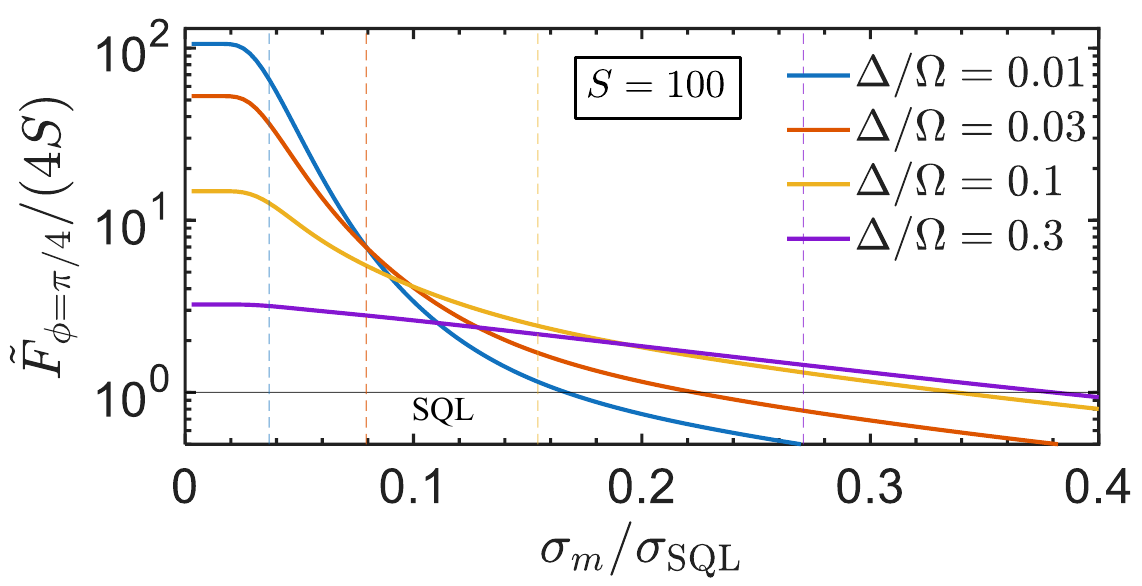}
    \caption{Classical Fisher information (CFI) $\tilde{F}_{\phi}$ for differential phase $\phi$ (near $\phi=\pi/4$) including measurement noise $\sigma_m$ (see text). $\sigma_{\rm SQL}=1/\sqrt{8S}$ is the SQL-level noise for excitation fractions. The dashed lines mark the spin-projection noise for excitation fractions $\sigma_{\rm QPN}$ (see text) with colors indicating the value of $\Delta/\Omega$.}
    \label{fig:measnoise}
\end{figure}

Here we examine the effects of measurement noise in the case of ellipse fitting for differential phase $\phi$. 
The measurement noise turns the projective measurement of each spin ensemble in the basis $|S,m_1\rangle\otimes|S,m_2\rangle$ into the following POVM-type measurements,
\begin{equation}
    \hat{E}_{\tilde{p}_1\tilde{p}_2} = \bigotimes_{j=1,2} \bigg[\sum_{m_j} |S,m_j\rangle\langle S,m_j|\,\frac{e^{-(\tilde{p}_j-p_j)^2/(2\sigma_m^2)}}{\sqrt{2\pi\sigma_m^2}}\bigg],
\end{equation}
with $\int d\tilde{p}_1d\tilde{p}_2 \hat{E}_{\tilde{p}_1\tilde{p}_2} = 1$. 
Here, $p_j=m_j/(2S)+1/2$ are the excitation fractions without measurement noise, $\tilde{p}_j$ are the excitation fractions in the presence of measurement noise, and $\sigma_m$ controls the level of measurement noise. 
Without measurement noise ($\sigma_m \to 0$), we use the conditional probability $P(p_1,p_2|\phi)$ defined in Eq.~(\ref{eq:condprob}) to estimate the classical Fisher information (CFI) $F_{\phi}$ for the differential phase $\phi$ (see Eq.~(\ref{eq:cfi})).
In the presence of measurement noise, we calculate the conditional probability using the above POVM and find
\begin{equation}
    P(\tilde{p}_1,\tilde{p}_2|\phi) = \sum_{p_1,p_2}P(p_1,p_2|\phi) \frac{1}{2\pi \sigma_m^2}e^{-\sum_{j}(\tilde{p}_j-p_j)^2/(2\sigma_m^2)}.
\end{equation}
Using $P(\tilde{p}_1,\tilde{p}_2|\phi)$, one can now estimate the CFI $\tilde{F}_{\phi}$ including measurement noise. In Fig.~\ref{fig:measnoise}, we show $\tilde{F}_{\phi}$ as a function of the measurement noise $\sigma_m$. The full power of the quantum enhancement can only obtained in the regime $\sigma_m \ll \sigma_{\rm QPN}$, i.e., if the level of measurement noise is well below the spin-projection noise. Here the spin-projection noise is given by $\sigma_{\rm QPN} = \sqrt{\mathrm{Var}(\hat{S}^y_1 + \hat{S}^y_2)_{\rm ss}}/(4S)$, which can be interpreted as the half-thickness of the ellipse.

Note that this result is expected for all spin-squeezing protocols that do not use signal amplification before readout.
If the measurement noise in an experiment is larger than the spin-projection noise, the signal-to-noise ratio is dominated by the ratio between signal and measurement noise. 
In this case, amplification can be used to increase the signal and the spin-projection noise until measurement noise and spin-projection noise become comparable, which improves the overall signal-to-noise ratio.
For unitary OAT and dissipative single-mode spin squeezing protocols, simple signal amplification protocols have been proposed to mitigate measurement noise \cite{Davis2016,Martin2023}.
It is an interesting topic to study how these concepts can be generalized to multimode spin squeezing and ellipse-fitting measurements.

\bibliography{reference}

\end{document}